\documentclass[12pt]{article}
\usepackage{graphicx}
\usepackage{latexsym,amsmath,amsfonts,amssymb}
\usepackage{bbm}
\usepackage{tikz}
\usetikzlibrary{shapes.misc,arrows,positioning,fit,backgrounds,calc,decorations.markings,cd}
\usepackage{pgfplots}
\usepackage[latin1]{inputenc}
\usepackage[american]{babel}
\usepackage{bbm}
\usepackage{adjustbox}
\usepackage{pdfpages}
\usepackage{array}
\usepackage{float}
\usepackage{tabu}
\usepackage{comment}
\usepackage{cancel}
\usepackage{authblk}
\usepackage{jheppub}

\newcommand{\ket}[1]{\ensuremath{| #1 \rangle}}
\newcommand{\bra}[1]{\ensuremath{\langle #1 |}}

\newcommand{\wt}{\widetilde}

\numberwithin{equation}{section}

\numberwithin{equation}{section}

\newcommand{\nn}{\nonumber}

\newcommand{\be}{\begin{equation}} \newcommand{\ee}{\end{equation}}
\newcommand{\bea}{\begin{equation} \begin{aligned}} \newcommand{\eea}{\end{aligned} \end{equation}}

\newcommand{\cD}{\mathcal{D}}
\newcommand{\cE}{\mathcal{E}}
\newcommand{\cF}{\mathcal{F}}

\newcommand{\cH}{\mathcal{H}}

\newcommand{\cJ}{\mathcal{J}}

\newcommand{\cO}{\mathcal{O}}

\newcommand{\cQ}{\mathcal{Q}}

\newcommand{\cW}{\mathcal{W}}

\newcommand{\bC}{\mathbb{C}}

\newcommand{\bN}{\mathbb{N}}

\newcommand{\bR}{\mathbb{R}}

\newcommand{\bZ}{\mathbb{Z}}

\def\U{\mathrm{U}}

\def\bal#1\eal{\begin{align}#1\end{align}}
\def\repa{\raise4pt\hbox{$\square$}\mkern-14mu\raise-4pt\hbox{$\square$}}
\def\repab{\overline{\raise4pt\hbox{$\square$}\mkern-14mu\raise-4pt\hbox{$\square$}\mkern-1mu}}

\def\smileface{\ensuremath{\hbox{\large$\bigcirc$}\mkern-15mu\raise-1pt\hbox{\scriptsize$\smallsmile$}%
		\mkern-10mu\raise4pt\hbox{..}\mkern4mu}}
\def\frownface{\ensuremath{\hbox{\large$\bigcirc$}\mkern-15mu\raise-1pt\hbox{\scriptsize$\smallfrown$}%
		\mkern-10mu\raise4pt\hbox{..}\mkern4mu}}
\allowdisplaybreaks

\DeclareMathOperator{\Tr}{Tr}

\DeclareMathOperator{\sgn}{sgn}

\DeclareMathOperator{\re}{\mathbb{R}e}
\DeclareMathOperator{\im}{\mathbb{I}m}

\usepackage{eso-pic}
\newcommand{\reportnum}[2]{
  \AddToShipoutPictureBG*{%
    \AtPageUpperLeft{%
      \hspace{0.75\paperwidth}%
      \raisebox{#1\baselineskip}{%
        \makebox[0pt][l]{\textnormal{#2}}
  }}}%
}

\reportnum{-6}{USTC-ICTS/PCFT-25-06}

\title{\textbf{Gauging the complex SYK model}}

\author[a]{Ziruo Zhang}

\author[a,b]{Cheng Peng}

\emailAdd{zhangziruo@ucas.ac.cn} \emailAdd{pengcheng@ucas.ac.cn}

\affiliation[a]{\small Kavli Institute for Theoretical Sciences (KITS), University of the Chinese Academy of Sciences, Beijing 100190, China}
\affiliation[b]{\small Peng Huanwu Center for Fundamental Theory, Hefei, Anhui 230026, China}

\abstract{Motivated by SYK-like models describing near-BPS black holes in string/M-theory, we consider gauging the $\U(1)$ symmetry of the complex SYK model in the presence of a Wilson line with charge $k$. At a fixed background gauge field, solutions to the Schwinger-Dyson equations display vastly different properties from those at a fixed real chemical potential. In the partition function and the two-point function, the integral over the gauge field is performed either directly or via a large $N$ saddle point approximation, and both results are consistent with exact diagonalization data. From the behaviour of the two-point function at large $N$, we deduce that the conformal symmetry at low energies is preserved at fixed $\kappa = k/N = 0$, but broken at $\kappa\neq 0$. In addition, we find that there is maximal chaos for all $k$.}


\arxivnumber{2502.18595}

\pgfplotsset{compat=1.18}
\begin{document}
\maketitle

\setcounter{page}{1}

\tableofcontents

\section{Introduction}

Recently, there has been considerable progress in understanding the physics of near-extremal black holes, through analysing the quantum effects coming from a universal subsector of gravity on near-AdS$_2$ \cite{Maldacena:2016upp, Iliesiu:2020qvm}. This subsector consists of the Schwarzian mode, $\U(1)$ gauge fields if the black hole is charged, and their superpartners if the theory is supersymmetric \cite{Heydeman:2020hhw,Boruch:2022tno,Lin:2022zxd}. It is universal in the sense that its presence is guaranteed by the symmetry breaking pattern arising in the AdS$_2$ near-horizon region. On the other hand, the same subsector arises in the low energy limit of the SYK model~\cite{Sachdev:1992fk, Parcollet:1999itf, Kitaevtalk:2015}, and its charged and supersymmetric generalizations~\cite{Gu:2019jub,Fu:2016vas,Peng:2020euz}. This observation constitutes a series of so-called near-AdS$_2$/near-CFT$_1$ dualities~\cite{Maldacena:2016hyu, Jevicki:2016bwu, Engelsoy:2016xyb}. 

Given a near-extremal black hole, an interesting question to ask is whether one can \emph{derive} the SYK-like quantum mechanics that is dual to gravity in the AdS$_2$ near-horizon region. A natural setting to answer this question is that of asymptotically AdS$_4$ near-BPS black holes in M-theory or massive Type IIA string theory~\cite{Cacciatori:2009iz,Guarino:2017eag,Guarino:2017pkw,Benini:2015eyy,Bobev:2020pjk,Heydeman:2024fgk}, which are dual to near-BPS states of Chern-Simons matter theories on a Riemann surface $\Sigma$, possibly with a topological twist~\cite{Guarino:2015jca}. The SYK-like model could be obtained in principle by dimensionally reducing the Chern-Simons matter theory on $\Sigma$, and taking the low energy limit. This procedure was implemented concretely in~\cite{Benini:2022bwa} for the case of spherically symmetric dyonic black holes in massive Type IIA string theory. A simplified model based on the result of \cite{Benini:2022bwa} can be found in \cite{Benini:2024cpf}. The feature that we would like to emphasize in the resulting SYK-like model is the fact that it is \emph{gauged}, with gauge group $\U(1)^N$, where $N$ is the rank of the gauge group in the higher-dimensional Chern-Simons matter theory. In addition, the Chern-Simons level $k$ descends to a Wilson line of charge $k$ in the quantum mechanics. These features are generic when reducing any Chern-Simons matter theory on $\Sigma$, and are therefore expected for any SYK-like model that describes asymptotically AdS$_4$ near-BPS black holes in M-theory or Type IIA string theory. With this motivation in mind, we would like to study the effects of gauging a SYK-like model in the presence of a Wilson line. In this work we focus on the simplest SYK-like model that possesses a continuous symmetry which can be gauged, namely the complex SYK model of \cite{Gu:2019jub}. It has random interactions of order $q\in 2\bN$ between $N$ complex fermions. In light of the aforementioned motivations, we will also add a Wilson line of charge $k$.\footnote{We notice that the work~\cite{Murugan:2023vhk} has similar motivations, but focuses on the free, namely $q = 2$, model. On the other hand, we shall be interested in the interacting $q\geq 4$ models, and focus on $q = 4$ in particular. More importantly, their theory is not gauged since they do not integrate over the gauge field in the path integral; it is instead treated as a non-dynamical background field. }       


%

Unsurprisingly, the physical interpretation of gauging in the Hamiltonian formalism is to restrict observables such as the partition function or correlation functions to the sector with charge $k$. In other words, a non-trivial Wilson line with charge $k\neq 0$ will shift the Gauss' law constraint by $k$. Although it was not presented from this perspective, \cite{Gu:2019jub} essentially studies the same problem by passing (at large $N$) from the grand canonical ensemble at fixed chemical potential to the canonical ensemble at fixed $\kappa = k/N$. Note that in \cite{Gu:2019jub}, the normalized charge $\kappa$ is known as $\cQ$. As we will discuss, this connection becomes explicit when the integral over the gauge field $u$ is performed in the large $N$ saddle point approximation. 

To compute observables, we first gauge fix the gauge field to a constant holonomy $u\in (-\pi,\pi)$. We then proceed in the usual fashion by taking the large N limit in the annealed approximation, where the theory boils down to solving a set of Schwinger-Dyson equations for the bilocal fields $(G,\Sigma)$. Here, we discover that there are drastic differences between the numerical solutions of the Schwinger-Dyson equations in the presence of a gauge field, and those in the presence of a chemical potential, which was analysed in~\cite{Gu:2019jub}. For instance, the solutions for $G(\tau)$ are not real, and this can be traced to the non-hermiticity of the Hamiltonian $\hat{H}$. As an aside, we also solve the equations analytically at large $q$. Importantly, there are \emph{infinitely} many solutions at any given $u\in (-\pi,\pi)$, labelled by integers $n\in\bZ$. These solutions are not gauge equivalent and their contributions should be summed in the path integral. We show that summing these contributions is gauge equivalent to working with a single solution at each $u$, with the price that $u$ must be extended to the range $(-\infty,\infty)$. It will also be shown that the sum over infinitely many solutions is essential to ensuring that the fermion partition function has the required periodicity as a function of $u$. Furthermore, the labels $n$ can be identified with the winding numbers of the $\U(1)$ axion in the low energy effective action of the complex SYK model derived in \cite{Davison:2016ngz,Gu:2019jub}.   

Each observable contains an integral in $u$ coming from the gauge fixed path integral, and this can be done in two ways. The first is to approximate the integrand by a relatively simple fitted function, which can then be integrated analytically over $u\in(-\infty,\infty)$. The second method is to apply a large $N$ saddle point approximation to the $u$ integral, where one finds that the saddle points lie on the imaginary axis of the complex $u$ plane. The result is then obtained by deforming the contour from the real axis to the imaginary axis, and picking up the contributions of the saddles. The integral over imaginary $u$ can be identified with an integral over chemical potentials $\beta\mu = -iu$, i.e. a Laplace transform of the grand canonical partition function. The large $N$ saddle point approximation subsequently simplifies this integral into the Legendre transform relating the grand canonical and canonical ensembles. This can be seen from the saddle point equation \eqref{u saddle cmplx u} with respect to $u$, which identifies $\kappa = k/N$ with the expectation value of the charge computed from the bilocal field $G$, defined as $\cQ$ in \cite{Gu:2019jub}. We compute the partition function and the two-point function in a fixed charge sector using these two methods, and find that they agree quite well with results coming from exact diagonalization. In particular, if the large $N$ limit is taken while $\kappa = k/N$ is held fixed, the two-point function is approximately conformal if $\kappa = 0$, but not otherwise. This is consistent with physical expectations that conformality is broken in nonzero charge sectors.

In addition, we compute the spectrum of low-lying operators in the presence of a background gauge field $u\in(-\infty,\infty)$, and examine how the operator dimensions depend on $u$. Although the operator with dimension $h = 1$ always appears in the spectrum, corresponding to the $\U(1)$ axion and gauge transformations acting on the bilocal field, it is lifted since we have gauge fixed the $\U(1)$ gauge field to the constant holonomy $u$, and further gauge transformations are not allowed except for large gauge transformations which wind around $S^1$. Because of this, the only degree of freedom surviving from the $\U(1)$ axion in the low energy effective action is its winding number. Lastly, following the retarded kernel method in \cite{Murugan:2017eto}, we compute the chaos exponent from the late time behaviour of out of time order double commutators. We find that there is maximal chaos regardless of $u$, which means that there is maximal chaos in every charge sector.

The outline of the paper is as follows. In Section \ref{sec: model}, we review the definition of the model, the gauge fixing procedure, and apply the Hamiltonian formalism after gauge fixing. In Section \ref{sec: annealed}, leaving the integration over the gauge field for later, we examine the solutions to the Schwinger-Dyson equations in the presence of a background gauge field, showing that the properties which make them so different from the solutions with a chemical potential can in fact be derived from the Hamiltonian formalism. In sections \ref{sec: partition fn} and \ref{sec: 2 pt fn}, we compute the partition function and the two-point function respectively, treating the integral over the gauge field using the aforementioned two methods: integrating an approximate fitted function, and using the large $N$ saddle point approximation. Each result is checked against data coming from exact diagonalization. Finally, in Sections \ref{sec: op dim} and \ref{sec: chaos exp}, we compute the lowest lying operator dimensions and the chaos exponent respectively.  

\section{The model}\label{sec: model}

We consider gauging the global $\U(1)$ symmetry  of  the complex SYK model~\cite{Sachdev:2015efa,Gu:2019jub} consisting of $N$ complex fermions $\psi_{i=1,\ldots,N}$. The gauging is implemented by introducing a gauge field $A_\tau$ with a 1d Chern-Simons term (or a Wilson line of charge $k$) in the (Euclidean) action
\be\label{Euclidean action}
 \wt S =\int d\tau\left[ikA_\tau+\sum_{i=1}^N\psi_i^\dagger(\partial_\tau-iA_\tau)\psi_i+\sum_{\substack{j_1<\ldots<j_{q/2}\\ k_1<\ldots<k_{q/2}}}J_{j_1\cdots j_{q/2},k_1\cdots k_{q/2}}\psi_{j_1}^\dagger\cdots\psi_{j_{q/2}}^\dagger\psi_{k_1}\cdots\psi_{k_{q/2}}\right],
\ee
where $q\in 2\bN$ and $J_{j_1\cdots j_{q/2},k_1\cdots k_{q/2}}\in\bC$ is antisymmetric in both $j_1\cdots j_{q/2}$ and $k_1\cdots k_{q/2}$. In order for the potential term and the Hamiltonian to be real, we also need 
\be
J^*_{j_1\cdots j_{q/2},\,k_1\cdots k_{q/2}}=J_{k_1\cdots k_{q/2},\,j_1\cdots j_{q/2}}\,.
\ee
The lexicographic ordering of the tuples $(j_1,\ldots,j_{q/2})$ is taken in \eqref{Euclidean action} because the independent couplings are $J_{j_1\cdots j_{q/2},\,k_1\cdots k_{q/2}}\in\bC$ for $(j_1,\ldots,j_{q/2})<(k_1,\ldots,k_{q/2})$, as well as $J_{j_1\cdots j_{q/2},\,j_1\cdots j_{q/2}}\in\bR$. The couplings are Gaussian random variables with zero mean and variance\footnote{This formula applies to both the complex couplings $J_{j_1\cdots j_{q/2},\,k_1\cdots k_{q/2}}$ with $(j_1,\ldots,j_{q/2})< (k_1,\ldots,k_{q/2})$ as well as the real couplings $J_{j_1\cdots j_{q/2},\,j_1\cdots j_{q/2}}$.} 
\be\label{variance couplings}
\overline{|J_{j_1\cdots j_{q/2},k_1\cdots k_{q/2}}|^2}=J^2\frac{(q/2)!(q/2-1)!}{N^{q-1}}\,.
\ee
The scaling dimension of the fields and couplings are $[\psi_i]=0$, $[A_\tau]=1$ , $[J]=1$. 

\subsection{Gauge fixing}

It is convenient to carry out our subsequent analysis in a fixed gauge. The one-dimensional equivalent of the Lorenz gauge, $\partial_\tau A_\tau=0$, can always be reached by the gauge transformation
\be\label{gt to reach Lorenz gauge}
\alpha(\tau)=\frac{\tau u}{\beta}-\int_0^\tau d\tau' A_\tau(\tau')\,,\quad u\equiv\int_0^\beta d\tau'A_\tau(\tau')\,,
\ee
so that the transformed gauge field is a constant
\be\label{gauge fixing}
A_\tau'=A_\tau+\partial_\tau\alpha=\frac{u}{\beta}=\text{const}\,.
\ee
Note that \eqref{gt to reach Lorenz gauge} is a well-defined $\U(1)$ transformation since $\alpha(\tau+\beta)=\alpha(\tau)$. The residual gauge transformations must preserve the Lorenz gauge, which means the gauge transformation parameters satisfy $\partial_\tau^2\alpha=0$. This means that $\alpha$ is at most linear in $\tau$. The only option compatible with $\alpha$ being periodic in $\tau$ modulo $2\pi$ is
\be
\alpha(\tau)=\frac{2\pi n\tau}{\beta}+\alpha_0\,,\quad n\in\bZ\,,\quad \alpha_0=\text{const}\,.\label{large_gauge_transf}
\ee
These are the constant ($n=0$) and large gauge transformations under which $u\mapsto u+2\pi n$. They are fixed by restricting the range of $u$ to be $(-\pi,\pi]$. Different values of $u$ in this range are not gauge-equivalent to each other, since they are not related by a well-defined gauge transformation. Strictly speaking, the gauge fixing procedure introduces ghost fields, but the path integral over ghost fields in an abelian theory is just an overall numerical factor that is independent of $u$. 

The partition function of the theory can be computed by first performing the path integral over the fermions and then path integrating (it is actually only an ordinary integral) over $u$
\bea\label{ferm path integral}
&Z_k = \int_{-\pi}^{\pi} du\,e^{-iku}\wt Z(u)\,,\quad \wt Z(u)\equiv\int \prod_{i=1}^N\cD[\psi_i^\dagger]\cD[\psi_i]e^{- S[u,\psi,\psi^\dagger]}\,,\\
&S[u,\psi,\psi^\dagger]\equiv\int_0^1 d\tau\left[\sum_{i=1}^N\psi_i^\dagger\left(\partial_\tau-iu\right)\psi_i+\sum_{\substack{j_1<\ldots<j_{q/2}\\ k_1<\ldots<k_{q/2}}}\beta J_{j_1\cdots j_{q/2},k_1\cdots k_{q/2}}\psi_{j_1}^\dagger\cdots\psi_{j_{q/2}}^\dagger\psi_{k_1}\cdots\psi_{k_{q/2}}\right]
\eea
In the above, we have changed coordinates to $\tau'=\tau/\beta\in [0,1]$ so that the imaginary time direction is a unit circle, and the theory is characterized by the dimensionless coupling $\beta J_{j_1\cdots j_{q/2},k_1\cdots k_{q/2}}$.

\subsection{Hamiltonian formalism}

In the following, we will study this model~\eqref{Euclidean action}
by exact diagonalization, so let us first introduce its Hamiltonian formulation. It will also be useful when analysing the properties of solutions to the Schwinger-Dyson equations. 

From the kinetic terms of the Lagrangian, the canonical commutation relations are 
\bal
\{\hat{\psi}_j,\hat{\psi}_i^\dagger\}=\delta_{ij}\ .
\eal
Bulding up from the Fock vacuum $\ket{0}$ defined by
\be
\hat{\psi}_i\ket{0}=0,\quad\forall i=1,\ldots,N\,,
\ee
the Hilbert space $\cH$ of dimension $2^N$ is the Fock space
\be\label{Hilbert space}
\cH=\text{span}_{\bC}\big\{\big(\hat{\psi}_N^\dagger\big)^{b_N}\cdots\big(\hat{\psi}_1^\dagger\big)^{b_1}\ket{0}:\; b_{1},\ldots,b_{N}=0,1\big\}\,.
\ee
The Hamiltonian corresponding to~\eqref{Euclidean action}  is then
\bea\label{H op def}
&\hat{H}(u)=-iu\hat{Q}+\hat{H}_0\,,\quad \hat{Q}=\frac{1}{2}\sum_{i=1}^N\big[\hat{\psi}_i^\dagger,\hat{\psi}_i\big]\,,\\
&\hat{H}_0\equiv\sum_{\substack{j_1<\ldots<j_{q/2}\\ k_1<\ldots<k_{q/2}}}\beta J_{j_1\cdots j_{q/2},k_1\cdots k_{q/2}}\hat{\psi}_{[j_1,}^\dagger\cdots\hat{\psi}_{j_{q/2}}^\dagger\hat{\psi}_{k_1}\cdots\hat{\psi}_{,\,k_{q/2}]}\,,
\eea
where the square bracket on the indices of $\hat{H}_0$ denote the antisymmetrized product. Note that $\hat{Q}$ and $\hat{H}_0$ are hermitian but $\hat{H}(u)$ is not.~\footnote{This is a peculiarity of working in Euclidean signature. In real time $t=-i\tau$, time evolution will still be unitary since $A_t=iA_\tau$} The ordering ambiguity in $\hat{H}_0$ has been fixed by requiring that it commutes with the charge conjugation operator $\hat{C}$
\be\label{charge conj op}
\hat{C}\equiv\hat{K}\hat{X}\,,\quad \hat{X}\equiv(\hat{\psi}^\dagger_1+\hat{\psi}_1)\cdots(\hat{\psi}_N^\dagger+\hat{\psi}_N)\,,\quad \hat{X}^\dagger=(-1)^\frac{N(N-1)}{2}\hat{X}\,,\quad \hat{C}^2=(-1)^\frac{N(N-1)}{2}\,,
\ee
where $\hat{K}$ is the anti-linear complex conjugation operator defined with respect to the basis in \eqref{Hilbert space}, and $\hat{X}$ exchanges $0 \leftrightarrow 1$ in all the occupancy numbers $b_{1},\ldots,b_{N}$ labelling the basis states\footnote{There is also an overall sign $(-1)^{\sum_{i=1}^N(1-b_i)(i-1)}$ in the resulting state.}, and we simply have $\hat{K}\hat{X}\hat{K}=\hat{X}$. The combination $\hat{C}$ exchanges $\hat{\psi}_i$ and $\hat{\psi}_i^\dagger$ up to a sign. Specifically,
\be\label{charge conj comm w psi}
\hat{C}\hat{\psi}_i=(-1)^{N-1}\hat{\psi}_i^\dagger\hat{C}\,,\quad \hat{C}\hat{\psi}_i^\dagger=(-1)^{N-1}\hat{\psi}_i\hat{C}\,.
\ee
The operator ambiguity in $\hat{Q}$ is fixed by requiring that $\hat{Q}\hat{C}=-\hat{C}\hat{Q}$, i.e. that $\hat{C}$ flips the sign of the charge.

Note that on the basis states in \eqref{Hilbert space}, $\hat{Q}$ has eigenvalues
\be\label{eq: Q quantization}
Q=N_\uparrow-\frac{N}{2}=-\frac{N}{2},-\frac{N}{2}+1,\ldots,\frac{N}{2}-1,\frac{N}{2}\,,\quad N_\uparrow\equiv\sum_{i=1}^N b_i\,.
\ee

The subspaces of the Hilbert space with fixed charge $Q$ or $N_\uparrow$ are denoted by
\be\label{Q subspace}
\cH_Q = \text{span}_{\bC}\left\{\big(\hat{\psi}_N^\dagger\big)^{b_N}\cdots\big(\hat{\psi}_1^\dagger\big)^{b_1}\ket{0}:\; b_{1},\ldots,b_{N}=0,1,\;N_\uparrow=Q+N/2\right\}\,,\quad \dim\cH_Q=\binom{N}{N_\uparrow}\,.
\ee
The charge conjugation operator $\hat{C}$ maps states with charge $Q$ to those with charge $-Q$, and this is an isomorphism
\be\label{charge conj on Hilb space}
\hat{C}\cH_Q \cong \cH_{-Q}\,.
\ee

\subsection{Anomaly free condition and charge quantisation}\label{subsec: k quantisation}

Since we want to gauge the $\U(1)$ symmetry, the theory must be free of gauge anomalies. In a quantum mechanics on $S^1$, there is a possible anomaly under large gauge transformations with nonzero winding number. This can be seen both in the Hamiltonian and path integral formalisms. In the Hamiltonian formalism, we have chosen an operator ordering in $\hat{Q}$ such that it takes integer spaced values from $-\frac{N}{2}$ to $\frac{N}{2}$. Therefore, the Hilbert space will be a direct sum of projective $\U(1)$ representations if $N$ is odd. Meanwhile, the Wilson line carries charge $k$, which is also a projective representation if $k\notin\bZ$. The total system is therefore anomaly free if
\be\label{anomaly free condition}
k+\frac{N}{2}\in \bZ\,.
\ee
Alternatively, a more direct way to get this condition is by writing the partition function as a sum over charge sectors. Given that $\hat{Q}$ commutes with $\hat{H}_0$,
\bea\label{Zk sum Q}
Z_k&=\int_{-\pi}^{\pi} du\,e^{-iku}\wt Z(u)=\int_{-\pi}^{\pi} du\,e^{-iku}\Tr_\cH e^{iu\hat{Q}-\hat{H}_0}=\sum_{Q=-\frac{N}{2}}^{\frac{N}{2}}\int_{-\pi}^{\pi} du\,e^{iu(Q-k)}\Tr_{\cH_Q}e^{-\hat{H}_0}\,.
\eea
The exponential factor $e^{iu(Q-k)}$ is invariant under large gauge transformations of $u$ if and only if the condition \eqref{anomaly free condition} holds. From the integral over $u$ in this expression, we also learn that the values of $k$ giving nonzero values of $Z_k$ coincide with the allowed values of $Q$, which are
\be\label{allowed k}
k = -\frac{N}{2},-\frac{N}{2}+1,\ldots,\frac{N}{2}-1,\frac{N}{2}\,.
\ee

On the other hand, one can imagine computing the path integral for $\wt Z(u)$. The chosen operator ordering in the Hamiltonian formalism corresponds to a regularisation of the path integral where the fermion partition function $\wt Z(u)$ acquires an overall sign $(-1)^{Nn}$ under a gauge transformation with winding number $n$. The charge $k$ Wilson line transforms as $(-1)^{2kn}$ under the same large gauge transformation. In order for the total system to be gauge invariant, one again needs \eqref{anomaly free condition}. We shall see this explicitly when considering the regularisation of the large $N$ on-shell action in Section \ref{sec: partition fn}. 

These observations are just another manifestation of previous results in~\cite{Dunne:1989hv,Hori:2014tda}, see also~\cite{Redlich:1983kn}. As emphasized in \cite{Dunne:1989hv}, our situation is identical to the ``parity anomaly" in 3d Chern-Simons-matter theories, and \eqref{anomaly free condition} is exactly the same as the quantization condition for the Chern-Simons level. This is not surprising since a theory like \eqref{Euclidean action} can be obtained by dimensionally reducing a 3d Chern-Simons-matter theory on a compact 2d manifold.

\section{Annealed average}\label{sec: annealed}

Assuming that the theory is self-averaging, we can follow the standard steps and perform the disorder average to rewrite the averaged path integral in terms of bilocal fields. By completing the square in $J$, the disorder average gives the action
\bea
S_{\text{eff}}&=\int d\tau\sum_{i=1}^N\psi_i^\dagger\left(\partial_\tau-iu\right)\psi_i-\int d\tau_1d\tau_2\frac{(-1)^{\frac{q}{2}}(\beta J)^2}{N^{q-1}q}\\
&\qquad \times\sum_{\substack{j_1,\ldots,j_{q/2}\\ k_1,\ldots,k_{q/2}}}\psi_{j_1}^\dagger(\tau_2)\psi_{j_1}(\tau_1)\cdots\psi_{j_{\frac{q}{2}}}^\dagger(\tau_2)\psi_{j_{\frac{q}{2}}}(\tau_1)\psi_{k_1}^\dagger(\tau_1)\psi_{k_1}(\tau_2)\cdots \psi_{k_{\frac{q}{2}}}^\dagger(\tau_1)\psi_{k_{\frac{q}{2}}}(\tau_2)\,.
\eea
The bilocal fields $G$, $\Sigma$ are introduced by inserting
\be\label{id bilocal}
1=\int\cD [G]\cD[\Sigma] \exp\left\{ N\int d\tau_1d\tau_2\Sigma(\tau_1,\tau_2)\left[G(\tau_2,\tau_1)-\frac{1}{N}\sum_{i=1}^N\psi_i^\dagger(\tau_1)\psi_i(\tau_2)\right]\right\}\,. 
\ee
For the action to stay invariant under large gauge transformations, we need $\Sigma$ and $G$ to transform as
\be\label{bilocal gt}
G(\tau_1,\tau_2)\mapsto e^{2\pi i n(\tau_1-\tau_2)}G(\tau_1,\tau_2)\,,\quad\Sigma(\tau_1,\tau_2)\mapsto e^{2\pi i n(\tau_1-\tau_2)}\Sigma(\tau_1,\tau_2)\,.
\ee
Integrating over the  fields $\psi$ gives the action
\bea\label{wt S def}
-\wt S[u,G,\Sigma]&\equiv -\frac{S[u,G,\Sigma]}{N}=\log\det\left[-\delta'(\tau_{12})+iu\delta(\tau_{12})-\Sigma(\tau_1,\tau_2)\right]\\
&\qquad +\frac{(\beta J)^2}{q}\int d\tau_1d\tau_2\bigg[\Sigma(\tau_1,\tau_2)G(\tau_2,\tau_1)\left(-G(\tau_1,\tau_2)G(\tau_2,\tau_1)\right)^\frac{q}{2}\bigg]\,.
\eea
With a slight abuse of notation, we denote the bilocal action and the original action in \eqref{ferm path integral} by the same symbol $S$. For convenience, we also define the rescaled $N$-independent $\wt S$. At leading order in large $N$, it is sufficient to solve the saddle point equations, which are the Schwinger-Dyson equations 
\be\label{SD eqn unit beta}
\left(\partial_{\tau_1}-iu\right)G(\tau_{12})+\int_0^1 d\tau_3\,\Sigma(\tau_{13})G(\tau_{32})=-\delta(\tau_{12})\,,\quad\Sigma(\tau)=(\beta J)^2G(\tau)^\frac{q}{2}\left(-G(-\tau)\right)^{\frac{q}{2}-1}\,.
\ee
We have simplified the equations using the fact that the bilocal fields are functions of the time differences $\tau_{12}\equiv\tau_1-\tau_2$ due to time translational invariance. While the equations~\eqref{SD eqn unit beta} are similar to the equations of the complex SYK model~\cite{Gu:2019jub}, the major difference is that the chemical potential $\mu$ in~\cite{Gu:2019jub} is replaced by the purely imaginary $i u/\beta$. It is more convenient to write the first equation of \eqref{SD eqn unit beta} in frequency space, which leads to the following form of the equations~\footnote{Our Fourier transformation convention is
\be\label{Fourier convention}
G(\tau)=\sum_{n\in\bZ}G(\omega_n)e^{i\omega_n\tau}\,,\quad G(\omega_n)=\int_{-\frac{1}{2}}^\frac{1}{2} d\tau e^{-i\omega_n\tau}G(\tau)\,. 
\ee
}
\be\label{full SD eqn}
G(\omega_n)=\frac{i}{\omega_n-u-i\Sigma(\omega_n)}\,,\quad\Sigma(\tau)=(\beta J)^2G(\tau)^\frac{q}{2}\left(-G(-\tau)\right)^{\frac{q}{2}-1}\ .
\ee
The equations~\eqref{SD eqn unit beta} are also explicitly invariant under large gauge transformations~\eqref{large_gauge_transf}. Under the large gauge transformations in \eqref{bilocal gt}, the second equation of \eqref{full SD eqn} is straightforwardly satisfied. 
The first equation of~\eqref{full SD eqn} is also invariant given the fact that the large gauge transformation~\eqref{large_gauge_transf} maps $u\mapsto u+2\pi n$,~$G(\omega_m)\mapsto G(\omega_{m-n})$, and $\Sigma(\omega_m)\mapsto \Sigma(\omega_{m-n})$, coming from~\eqref{bilocal gt} and the Fourier transformation~\eqref{Fourier convention}. In other words, if $(G,\Sigma)$ is a solution of \eqref{full SD eqn} at $u$, then $(e^{2\pi i n\tau} G,e^{2\pi i n\tau} \Sigma)$ is a solution of \eqref{full SD eqn} at $u + 2\pi n$. 

\subsection{Properties of the on-shell action and 2-point function}\label{subsec: 2 pt fn prop}

Since the Hamiltonian $\hat{H}(u)$ in \eqref{H op def} is not hermitian, observables of the theory have rather unusual properties, which we aim to introduce and derive in this section.~\footnote{There are other non-Hermitian SYK-like models such as~\cite{Garcia-Garcia:2023yet} and references therein, and it is interesting to make contact with those models. } These in turn descend to properties of solutions to the Schwinger-Dyson equations which are not commonly seen in the literature. 

We consider the partition function
\be\label{op def wt Z}
\wt Z(u) = \Tr_\cH e^{-\hat{H}(u)}\,,
\ee
and the 2-point function 
\be\label{2pt fn def}
\wt G(\tau_1,\tau_2)\equiv\frac{1}{N}\Tr_\cH\left[e^{-\hat{H}(u)}\text{T}\hat{\psi}_i^\dagger(\tau_2)\hat{\psi}_i(\tau_1)\right]\,,
\ee
where T is the time ordering operator. In the large $N$ saddle point approximation, they are computed by sums over solutions $G_0$ to the Schwinger-Dyson equations  
\bea\label{SD sum Z G}
\wt Z(u) &= \int\cD [G]\cD[\Sigma]e^{-S[u, G, \Sigma]}=\sum_{G_0\,\in\, \text{SD sols}}\frac{1}{\sqrt{\Delta(G_0)}}e^{-S[u, G_0,\Sigma_0]}(1+\cO(N^{-1}))\,,\\
\wt G(\tau_1,\tau_2) &= \int \cD[\psi^\dagger]\cD[\psi]\frac{1}{N}\psi_i^\dagger(\tau_2)\psi_i(\tau_1)e^{-S[u,\psi,\psi^\dagger]} = \int\cD [G]\cD[\Sigma] G(\tau_1,\tau_2)e^{-S[u, G, \Sigma]} \\
&=\sum_{G_0\,\in\, \text{SD sols}}\frac{1}{\sqrt{\Delta(G_0)}}e^{-S[u, G_0,\Sigma_0]}\left[G_0(\tau_1,\tau_2)+\cO(N^{-1})\right]\,,
\eea
where $\Sigma_0$ is determined in terms of $G_0$ via the second equation of \eqref{full SD eqn}, and $\Delta(G_0)$ is the determinant of the kinetic operator appearing at quadratic order in the expansion of $\wt S=S/N$ around $G_0$. Concretely, it is the determinant of the operator sandwiched between two $\delta G$'s in \eqref{S quad expansion}. Notice that $\Delta(G_0)\sim\cO(N^0)$ and it is subleading in $1/N$. Using the operator definitions \eqref{op def wt Z} and \eqref{2pt fn def}, one can derive various properties of $\wt Z$ and $\wt G$. These in turn descend to properties of $G_0$ and the on-shell action, if we assume that they are satisfied independently for each solution of the Schwinger-Dyson equations.    

\paragraph{KMS condition} As usual, $\wt G$ is anti-periodic on the circle and satisfies the KMS condition
\be\label{KMS condition}
\wt G(\tau+1)=-\wt G(\tau)\,,
\ee
where we again use time translational invariance to express $\wt G(\tau_1,\tau_2)$ as a function of $\tau_{12}\equiv\tau_1-\tau_2$. To show this we notice that for $\tau\in(-1,0)$, we have
\bea
&\wt G(\tau+1)=-\frac{1}{N}\Tr_\cH\left[ e^{-\hat{H}}\hat{\psi}_i(\tau+1)\hat{\psi}_i^\dagger(0)\right]=-\frac{1}{N}\Tr_\cH \left[e^{\tau\hat{H}}\hat{\psi}_i(0)e^{-(\tau+1)\hat{H}}\hat{\psi}_i^\dagger(0)\right]\\
&=-\frac{1}{N}\Tr_\cH \left[\hat{\psi}_i(\tau)e^{-\hat{H}}\hat{\psi}_i^\dagger(0)\right]=-\frac{1}{N}\Tr_\cH\left[ e^{-\hat{H}}\hat{\psi}_i^\dagger(0)\hat{\psi}_i(\tau)\right]=-\wt G(\tau)\,.
\eea
A minus sign appears in the first equality because the time ordering includes a sign if two fermionic operators are swapped. The second to last equality uses the cyclicity of the trace. Notice that a cyclic permutation of fermionic operators in the trace does not introduce any sign, as it is clear by representing each operator in the trace as a matrix. 

Consequently, the same should hold for each solution $G_0$, i.e.
\be\label{KMS condition sd}
\wt G_0(\tau+1)=-\wt G_0(\tau)\,.
\ee
We can therefore expand $G_0$ in Matsubara frequencies which are half integer multiples of $2\pi$
\be
\omega_n = 2\pi\left(n+\frac{1}{2}\right)\,.
\ee

\paragraph{Reality of $\wt Z$} Usually, the reality of $\wt Z$ follows directly from the hermiticity of $\hat{H}$. But here this is not obvious since the Hamiltonian $\hat{H}(u)$ in \eqref{H op def} is not hermitian. Nevertheless, $\wt Z(u)$ is real because
\bea\label{Z tilde cc}
\wt Z(u)^*&=\left(\Tr_\cH e^{iu\hat{Q}-\hat{H}_0}\right)^*=\bigg(\sum_{Q=-N/2}^{N/2}e^{iuQ}\Tr_{\cH_Q}e^{-\hat{H}_0}\bigg)^*=\sum_{Q=-N/2}^{N/2}e^{-iu Q}\Tr_{\cH_Q}e^{-\hat{H}_0}=\wt Z(-u)\\
&=\sum_{Q=-N/2}^{N/2}e^{-iuQ}\sum_{n=1}^{\dim\cH_Q}\bra{n}\hat{C}^2e^{-\hat{H}_0}\hat{C}^2\ket{n}=\sum_{Q=-N/2}^{N/2}e^{-iuQ}\sum_{n=1}^{\dim\cH_Q}\bra{n}\hat{X}^\dagger\hat{K}e^{-\hat{H}_0}\hat{C}\ket{n}\\
&=\sum_{Q=-N/2}^{N/2}e^{-iuQ}\sum_{n=1}^{\dim\cH_Q}(\hat{C}\ket{n})^\dagger e^{-\hat{H}_0}\hat{C}\ket{n}=\sum_{Q=-N/2}^{N/2}e^{-iuQ}\Tr_{\cH_{-Q}}e^{-\hat{H}_0}=\wt Z(u)\,.
\eea
In the second line, $\{\ket{n}\}$ is the basis of $\cH_Q$ in \eqref{Q subspace} and $\hat{C}$ is the charge conjugation operator defined in \eqref{charge conj op}. The second equality of the second line uses the fact that $\hat{C}$ commutes with $\hat{H}_0$, like any symmetry generator of the original hamiltonian, and $(-1)^\frac{N(N-1)}{2}\hat{C}=\hat{X}^\dagger\hat{K}$. To get to the last line, note that $\bra{n}\hat{X}^\dagger=(\hat{C}\ket{n})^T$ and $\hat{K}e^{-\beta\hat{H}_0}\hat{C}\ket{n}=(e^{-\beta\hat{H}_0}\hat{C}\ket{n})^{\dagger\; T}=((\hat{C}\ket{n})^\dagger e^{-\beta\hat{H}_0})^T$. The second to last equality uses the fact that $\{\hat{C}\ket{n}\}$ is a basis for $\cH_{-Q}$, as stated in \eqref{charge conj on Hilb space}. The last equality follows from a relabelling of charges $Q\mapsto -Q$. A partial result of the above is that the partition function of $\hat{H}_0$ over the charge $k$ and charge $-k$ sectors are equal, i.e.
\be\label{Z Q eq minus Q}
Z_k=\Tr_{\cH_k}e^{-\hat{H}_0}=\Tr_{\cH_{-k}}e^{-\hat{H}_0}=Z_{-k}\,.
\ee
Assuming that the contribution in \eqref{SD sum Z G} of each Schwinger-Dyson solution to $\wt Z$ is real, one gets that 
\be\label{SD contrib Z real}
\frac{1}{\sqrt{\Delta(G_0)}}e^{-S[u, G_0,\Sigma(G_0)]}\in\bR\,,\quad \text{and is an even function of $u$}\,.
\ee

\paragraph{Complex conjugate of $\wt G$} One important peculiarity of having a nonzero gauge field and a non-hermitian Hamiltonian \eqref{H op def} is that $\wt G$ is not real. Firstly, note that the 2-point function for $\tau\in(0,1)$ can be written as
\bea
\wt G(\tau)&=-\frac{1}{N}\Tr_\cH\left[ e^{-(1-\tau)\hat{H}}\hat{\psi}_i(0)e^{-\tau\hat{H}}\hat{\psi}_i^\dagger(0)\right]=-\frac{1}{N}\Tr_\cH\left[e^{iu(1-\tau)\hat{Q}-(1-\tau)\hat{H}_0}\hat{\psi}_i(0)e^{iu\tau\hat{Q}-\tau\hat{H}_0}\hat{\psi}_i^\dagger(0)\right]\\
&=-\frac{1}{N}\sum_{Q=-N/2}^{N/2-1}e^{iu\tau+iuQ}\Tr_{\cH_Q}\left[e^{-(1-\tau)\hat{H}_0}\hat{\psi}_i(0)e^{-\tau\hat{H}_0}\hat{\psi}_i^\dagger(0)\right]
\eea
To get the last equality, we used the fact that if $\ket{n}$ has charge $Q$, then $\hat{\psi}_i^\dagger(0)\ket{n}$ (when nonzero) has charge $Q+1$. The state with maximum charge $Q=N/2$ does not contribute since it is annihilated by all $\hat{\psi}_i^\dagger(0)$. Similarly, the 2-point function for $\tau\in(-1,0)$ is
\bea\label{G neg tau}
\wt G(\tau)&=\frac{1}{N}\Tr_\cH\left[e^{-(1+\tau)\hat{H}}\hat{\psi}_i^\dagger(0)e^{\tau\hat{H}}\hat{\psi}_i(0)\right]=\frac{1}{N}\Tr_\cH\left[e^{iu(1+\tau)\hat{Q}-(1+\tau)\hat{H_0}}\hat{\psi}_i^\dagger(0)e^{-iu\tau\hat{Q}+\tau\hat{H}_0}\hat{\psi}_i(0)\right]\\
&=\frac{1}{N}\sum_{Q=-N/2+1}^{N/2}e^{iu\tau+iuQ}\Tr_{\cH_Q}\left[e^{-(1+\tau)\hat{H}_0}\hat{\psi}_i^\dagger(0)e^{\tau\hat{H}_0}\hat{\psi}_i(0)\right]\,.
\eea
Now, making use of the charge conjugation symmetry like in \eqref{Z tilde cc}, the complex conjugate of the 2-point function on $\tau\in(0,1)$ can be written as
\bea
&\wt G(\tau ;u)^*=-\frac{1}{N}\sum_{Q=-N/2}^{N/2-1}e^{-iu\tau-iuQ}\Tr_{\cH_Q}\left[\hat{\psi}_i(0) e^{-\tau\hat{H}_0}\hat{\psi}_i^\dagger(0)e^{-(1-\tau)\hat{H}_0}\right]=\wt G(\tau ;-u)\\
&=-\frac{1}{N}\sum_{Q=-N/2}^{N/2-1}e^{-iu\tau-iuQ}\sum_{n=1}^{\dim\cH_Q}\bra{n}\hat{C}^2\hat{\psi}_i(0) e^{-\tau\hat{H}_0}\hat{\psi}_i^\dagger(0)e^{-(1-\tau)\hat{H}_0}\hat{C}^2\ket{n}\\
&=-\frac{1}{N}\sum_{Q=-N/2}^{N/2-1}e^{-iu\tau-iuQ}\sum_{n=1}^{\dim\cH_Q}\bra{n}\hat{X}^\dagger\hat{K}\hat{\psi}_i^\dagger(0) e^{-\tau\hat{H}_0}\hat{\psi}_i(0)e^{-(1-\tau)\hat{H}_0}\hat{C}\ket{n}\\
&=-\frac{1}{N}\sum_{Q=-N/2}^{N/2-1}e^{-iu\tau-iuQ}\sum_{n=1}^{\dim\cH_Q}(\hat{C}\ket{n})^\dagger e^{-(1-\tau)\hat{H}_0}\hat{\psi}_i^\dagger(0)e^{-\tau\hat{H}_0}\hat{\psi}_i(0)\hat{C}\ket{n}\\
&=-\frac{1}{N}\sum_{Q=-N/2}^{N/2-1}e^{-iu\tau-iuQ}\Tr_{\cH_{-Q}}\left[e^{-(1-\tau)\hat{H}_0}\hat{\psi}_i^\dagger(0)e^{-\tau\hat{H}_0}\hat{\psi}_i(0)\right]\\
&=-\frac{1}{N}\sum_{Q=-N/2+1}^{N/2}e^{-iu\tau+iuQ}\Tr_{\cH_{Q}}\left[e^{-(1-\tau)\hat{H}_0}\hat{\psi}_i^\dagger(0)e^{-\tau\hat{H}_0}\hat{\psi}_i(0)\right]=-\wt G(-\tau ; u)=\wt G(1-\tau ; u)\,.
\eea
In the third line, we have used the commutation relations \eqref{charge conj comm w psi}. The second to last equality holds by comparing with \eqref{G neg tau}, and the last equality uses the KMS condition \eqref{KMS condition}.

Assuming that the contribution in \eqref{SD sum Z G} of each Schwinger-Dyson solution to $\wt G$ satisfies the same relations, and using \eqref{SD contrib Z real}, one gets that 
\be\label{G cc relation}
G_0(\tau ; u)^*=G_0(1-\tau; u)=G_0(\tau ; -u)\,,\quad \tau\in (0,1)
\ee
or in terms of the real and imaginary parts,
\bea\label{G cc re and im}
\re G_0(1-\tau;u)&=\re G_0(\tau;u)\,,\quad \im G_0(1-\tau;u)=-\im G_0(\tau;u)\,,\\
\re G_0(\tau;-u)&=\re G_0(\tau; u)\,,\quad \im G_0(\tau;-u)=-\im G_0(\tau; u)\,.
\eea
In words, the real part of $G_0$ is symmetric under reflection about $\tau=\frac{1}{2}$, or $u\mapsto -u$, whereas its imaginary part is antisymmetric. Substituting \eqref{G cc relation} into the complex conjugate of the second equation in \eqref{full SD eqn}, one finds that $\Sigma_0$ also satisfies $\Sigma_0(\tau;u)^*=\Sigma_0(1-\tau;u)=\Sigma_0(\tau,-u)$. In frequency space, \eqref{G cc relation} and the same equation for $\Sigma_0$ are equivalent to
\be\label{G FT imag}
G_0(\omega_n;u)^*=-G_0(\omega_n;u)=G_0(-\omega_n;-u)\,,\quad \Sigma_0(\omega_n;u)^*=-\Sigma_0(\omega_n;u) =\Sigma_0(-\omega_n;-u)
\ee
The Fourier coefficients of $G_0$ and $\Sigma_0$ are purely imaginary, as opposed to $G_0(\omega_n)^*=G_0(-\omega_n)$ and $\Sigma_0(\omega_n)^*=\Sigma_0(-\omega_n)$ for the usual case when $\wt G$ is real. Both relations half the degrees of freedom in the Fourier coefficients, but in a different way.

\paragraph{Boundary values of $\wt G$} The boundary values of $\re \wt G$ at $\tau=0^+,1^-$ is dictated by the canonical commutation relations $\{\psi_i,\psi_i^\dagger\}=1\;\forall i$. Specifically, we must have
\bea\label{ccr from G}
2\re \wt G(0^+)&=2\re \wt G(1^-)=\lim_{\epsilon\rightarrow 0^+}\left(\wt G(\epsilon)+\wt G(1-\epsilon)\right)=\lim_{\epsilon\rightarrow 0^+}\left(\wt G(\epsilon)-\wt G(-\epsilon)\right)\\
&=-\frac{1}{N}\lim_{\epsilon\rightarrow 0^+}\Tr_{\cH}e^{-\hat{H}}\Big[\hat{\psi}_i^\dagger(0)\hat{\psi}_i(-\epsilon) +\hat{\psi}_i(\epsilon)\hat{\psi}_i^\dagger(0)\Big]=-\wt Z\,.
\eea
The first and second equalities follow from \eqref{G cc relation}, and the last equality uses the canonical commutation relation. Assuming that \eqref{ccr from G} holds for the contribution of each Schwinger-Dyson solution to $\wt G$ and $\wt Z$ together with the relation~\eqref{SD sum Z G}, one obtains  
\be\label{reG bdy}
\re G_0(0^+)=\re G_0(1^-)=-\frac{1}{2}\,.
\ee

In a similar fashion, the boundary values of $\im \wt G$ are determined by the expectation value of $\hat{Q}$:
\bea\label{imG bdy}
\im \wt G(0^+)&=-\im \wt G(1^-)=\frac{1}{2i}\lim_{\epsilon\rightarrow 0^+}\left(\wt G(\epsilon)-\wt G(1-\epsilon)\right)=\frac{1}{2i}\lim_{\epsilon\rightarrow 0^+}\left(\wt G(\epsilon)+\wt G(-\epsilon)\right)\\
&=\frac{1}{2iN}\lim_{\epsilon\rightarrow 0^+}\Tr_{\cH}e^{-\hat{H}}\Big[\hat{\psi}_i^\dagger(0)\hat{\psi}_i(-\epsilon) -\hat{\psi}_i(\epsilon)\hat{\psi}_i^\dagger(0)\Big]=\frac{1}{iN}\Tr_{\cH}e^{-\hat{H}}\hat{Q}=-\frac{1}{N}\partial_u\wt Z\,.
\eea

The first and second equalities follow from \eqref{G cc relation}, while the last equality comes from the $u$-dependence of $\hat{H}$ in \eqref{H op def}. Substituting \eqref{SD sum Z G} into \eqref{imG bdy} and equating the contributions from each solution, one obtains
\bea
&\frac{1}{\sqrt{\Delta(G_0)}}e^{-S[u,G_0,\Sigma_0]}\im G_0(0^+)(1+\cO(N^{-1}))=-\frac{1}{N}\partial_u\left[\frac{1}{\sqrt{\Delta(G_0)}}e^{-S[u,G_0,\Sigma_0]}(1+\cO(N^{-1}))\right]\\
&=\frac{1}{\sqrt{\Delta(G_0)}}e^{-S[u,G_0,\Sigma_0]}\left(  \partial_u\wt S +\frac{\partial_u\Delta(G_0)}{2N\Delta(G_0)}+\cO(N^{-1})\right)=\frac{1}{\sqrt{\Delta(G_0)}}e^{-S[u,G_0,\Sigma_0]}\left(\partial_u\wt S +\cO(N^{-1})\right)\,.
\eea
Note that \eqref{SD contrib Z real} has been used in the first equality when taking the imaginary part of $\wt G$. Equating the leading order terms then gives
\be\label{imG S deriv}
\im G_0(0^+) = \partial_u\wt S\,.
\ee

\subsection{Conformal solutions}

Following the same steps as in~\cite{Gu:2019jub}, we redefine $\wt\Sigma(\omega_n)\equiv\Sigma(\omega_n)-iu$ or equivalently $\wt\Sigma(\tau)=\Sigma(\tau)-iu\delta(\tau)$, which simplifies the Schwinger-Dyson equations to
\bea\label{diff inv SD eqn}
G(\omega_n)&=\frac{i}{\omega_n-i\wt\Sigma(\omega_n)}\approx -\frac{1}{\wt\Sigma(\omega_n)}\,,\\
\wt\Sigma(\tau)&=-iu\delta(\tau)+(\beta J)^2G(\tau)^\frac{q}{2}\left(-G(-\tau)\right)^{\frac{q}{2}-1} = (\beta J)^2G(\tau)^\frac{q}{2}\left(-G(-\tau)\right)^{\frac{q}{2}-1}\,.
\eea
The first approximation is valid at frequencies that are small relative to the effective coupling, that is, $|\omega_n|\ll (\beta J)^2$ (or equivalently at time scales $\Delta\tau \gg 1/(\beta J)^2$), while the equality in the second line holds if we are only interested in the domain $\tau\in(0,1)$ away from the singular points $\tau = 0,1$. 
The simplified equations are invariant under diffeomorphisms of the circle, and admit conformal invariant solutions. 

Let us emphasize that we only anticipate this enhanced conformal symmetry to occur in the deep IR $|\omega_n|\ll (\beta J)^2$, where the gauge field could in principle get heavily renormalised, so there is no reason to expect a priori that the expectation value of $u$ is the same as the $u$ we started with in the UV~\eqref{full SD eqn}. From now on, we denote the renormalized gauge field in the IR by $u_\text{IR}$. As derived in Appendix \ref{app: conformal ansatz deriv}, a generic conformal solution in the domain $\tau\in(-1,1)$ satisfying the various properties derived in Section \ref{subsec: 2 pt fn prop} must have the form
\be\label{conf ansatz full}
G_0(\tau)=-ge^{iu_\text{IR}\tau}\bigg|\frac{\pi}{\sin(\pi\tau)}\bigg|^{2\Delta}\left[e^{-\frac{iu_\text{IR}}{2}}\Theta(\tau)-e^\frac{iu_\text{IR}}{2}\Theta(-\tau)\right]\,,\quad g\in\bR\,,\quad \sgn(g)=(-1)^{\lfloor\frac{u_\text{IR}+\pi}{2\pi}\rfloor}\,.
\ee
The Fourier transform of this ansatz is
\be
G_0(\omega_n)=-g(2\pi)^{2\Delta}\frac{i(-1)^n\Gamma(1-2\Delta)}{\Gamma\left(\frac{3}{2}+n-\frac{u_\text{IR}}{2\pi}-\Delta\right)\Gamma\left(\frac{1}{2}-n+\frac{u_\text{IR}}{2\pi}-\Delta\right)}\,.
\ee
To be precise, the Fourier integral is only convergent for $\Delta<\frac{1}{2}$, but we use the analytic continuation of the formula when $\Delta$ is beyond this range. Plugging this into the IR equations \eqref{diff inv SD eqn}, one observes that we must have $\Delta=1/q$ in order for $\Sigma(\omega_n)G(\omega_n)$ to be independent of $n$, and the equations also determine the coefficient $g$ to be
\be
g=(-1)^{ \lfloor\frac{u_\text{IR}+\pi}{2\pi}\rfloor}\left[\frac{\left(1-\frac{2}{q}\right)\sin\frac{2\pi}{q}}{2\pi (\beta J)^2\left(\cos u_\text{IR}+\cos\frac{2\pi}{q}\right)}\right]^\frac{1}{q}\,.
\ee
Since $g$ must be real, and $q$ is even, the quantity in the square bracket must be real and positive, i.e.
\be\label{uIR range}
\cos u_\text{IR}+\cos\frac{2\pi}{q}>0\,,\quad u_\text{IR}\in \left(-\frac{\pi(q-2)}{q},\frac{\pi(q-2)}{q}\right)\mod 2\pi\,.
\ee
Thus the allowed values of $u_\text{IR}$ lie in intervals which are centered around $2\pi n$ for $n\in\bZ$, with a spread of $\left(-\frac{\pi(q-2)}{q},\frac{\pi(q-2)}{q}\right)\subset (-\pi,\pi)$ around the centre $2\pi n$. Explicitly,  we can thus write
\be\label{split u_IR}
u_\text{IR}=\wt u_\text{IR} + 2\pi n\,, \quad \wt u_\text{IR}\in \left(-\frac{\pi(q-2)}{q},\frac{\pi(q-2)}{q}\right)\ .
\ee
In addition, the integer $n$ labelling the centre of each allowed interval of $u_\text{IR}$ also determines the sign of $g$ in \eqref{conf ansatz full} since
\be\label{sign integer eq centre}
 \left\lfloor\frac{u_\text{IR}+\pi}{2\pi}\right\rfloor = \left\lfloor\frac{\wt u_\text{IR}+\pi}{2\pi}\right\rfloor + n = n\,.
\ee
The last equality follows from
\be
\frac{\wt u_\text{IR}+\pi}{2\pi}\in \left(\frac{1}{q},1-\frac{1}{q}\right)\,.
\ee

To summarize, we expect that the solutions to \eqref{full SD eqn} within $(\beta J)^{-2}\ll\tau\ll 1-(\beta J)^{-2}$ are well approximated by the conformal solution
\be\label{conf const fixed}
G_0(\tau)=-(-1)^n\left[\frac{\left(1-\frac{2}{q}\right)\sin\frac{2\pi}{q}}{2\pi (\beta J)^2\left(\cos u_\text{IR}+\cos\frac{2\pi}{q}\right)}\right]^\frac{1}{q}e^{iu_\text{IR}(\tau-\frac{1}{2})}\bigg[\frac{\pi}{\sin(\pi\tau)}\bigg]^{2/q}\,.
\ee
The actual solution to \eqref{full SD eqn} will differ from \eqref{conf const fixed} when $\tau$ is within distance $\sim (\beta J)^{-2}$ to the points $\tau=0,1$ since \eqref{conf const fixed} diverges near $\tau=0,1$, while $G_0(\tau)$ must be finite in order to satisfy \eqref{reG bdy}. 

One can follow the same steps as in~\cite{Gu:2019jub} to derive the Luttinger-Ward identity expressing the difference between the boundary values of $G_0$ in terms of $u_\text{IR}$. Since \eqref{conf const fixed} is identical to the conformal solution in the presence of a chemical potential under the identification $2\pi\cE=-iu_\text{IR}$, where $\cE$ is the spectral asymmetry parameter used in \cite{Gu:2019jub}, the result is simply an analytic continuation of the formula in \cite{Gu:2019jub}:
\be\label{LW identity}
\frac{1}{2i}(G_0(0^+)-G_0(1^-)) = \left(\frac{1}{2}-\frac{1}{q}\right)\frac{\sin u_\text{IR}}{\cos u_\text{IR}+\cos\big(\frac{2\pi}{q}\big)}+\frac{1}{2\pi}\log\left[\frac{\cos\big(\frac{\pi}{q}-\frac{u_\text{IR}}{2}\big)}{\cos\big(\frac{\pi}{q}+\frac{u_\text{IR}}{2}\big)}\right]\,.
\ee
Due to the range of $u_\text{IR}$ in \eqref{uIR range}, the $\log$ term is real, which is consistent with the expectation from \eqref{G cc relation} that the left hand side is real.

\subsection{Numerical solution to Schwinger-Dyson equations}\label{sec:num}

We can solve the Schwinger-Dyson equations \eqref{full SD eqn} numerically using the standard iterative method in \cite{Maldacena:2016hyu}. Specifically, we sample the interval $(0,1)$ at $M$ points 
\be\label{discrete tau}
\tau_r = \frac{r-1/2}{M}\,,\quad r = 1\,,\ldots\,,M
\ee
where $M$ is a large even positive integer, after which $G$ and $\Sigma$ are represented by the lists of their values at these points $\{G(\tau_r)\}$, $\{\Sigma(\tau_r)\}$. Alternatively, via the discrete Fourier transform, one can represent the same data using lists of Fourier coefficients $\{G(\omega_r)\}$ and $\{\Sigma(\omega_r)\}$, where the sample frequencies are
\be
\omega_r = 2\pi\left(r-\frac{1}{2}-\frac{M}{2}\right)\,,\quad r = 1\,,\ldots\,,M\,.
\ee
Since $M$ is even, the frequencies are symmetrically distributed around zero, with the same number of positive and negative frequencies. The algorithm to solve \eqref{full SD eqn} begins with a set of initial data $\{G_0(\omega_r)\}$ that is preferably a good approximation to the exact solution. In this discussion, the subscript $k$ on $G_k$ labels the number of iterations, and $G_k$ is the result after $k$ iterations. In the $k^\text{th}$ iteration, $\{G_{k-1}(\omega_r)\}$ is converted to $\{G_{k-1}(\tau_r)\}$ using the inverse Fourier transform and plugged into the second equation of \eqref{full SD eqn} to give $\{\Sigma_{k-1}(\tau_r)\}$. After another Fourier transform to $\{\Sigma_{k-1}(\omega_r)\}$, we update the result from $\{G_{k-1}(\omega_r)\}$ to $\{G_k(\omega_r)\}$ using the rule 
\be
G_k(\omega_r)=(1-x)G_{k-1}(\omega_r)+x\frac{i}{\omega_r-u_\text{uv}-i\Sigma_{k-1}(\omega_r)}\,,
\ee
where $x\in(0,1)$ is an arbitrary weighting factor. Note that if $\{G_{k-1}(\omega_r)\}$ and $\{G_k(\omega_r)\}$ are equal, it means that $\{G_k(\omega_r)\}$ satisfies \eqref{full SD eqn} and the solution has been found. To test this, we compute an error estimate
\be
\Delta G_k = \frac{1}{M}\left[\sum_{r=1}^M\big|G_k(\omega_r)-G_{k-1}(\omega_r)\big|^2\right]^\frac{1}{2}
\ee
after each iteration. If $\Delta G_k < \epsilon$ for some chosen precision tolerance $\epsilon$, the algorithm terminates and the result is taken as an approximate solution. If $\Delta G_k > \Delta G_{k-1}$, that is if the updated data is a worse approximation to the solution than the result of the previous iteration, the weighting factor $x$ is halved. This is a numerical trick in \cite{Maldacena:2016hyu} that seems to improve convergence. Unless otherwise stated, we shall consider the model with $q=4$ and the initial value of $x=1/2$.

\begin{figure}[htbp]
\centering
\includegraphics[width=\textwidth]{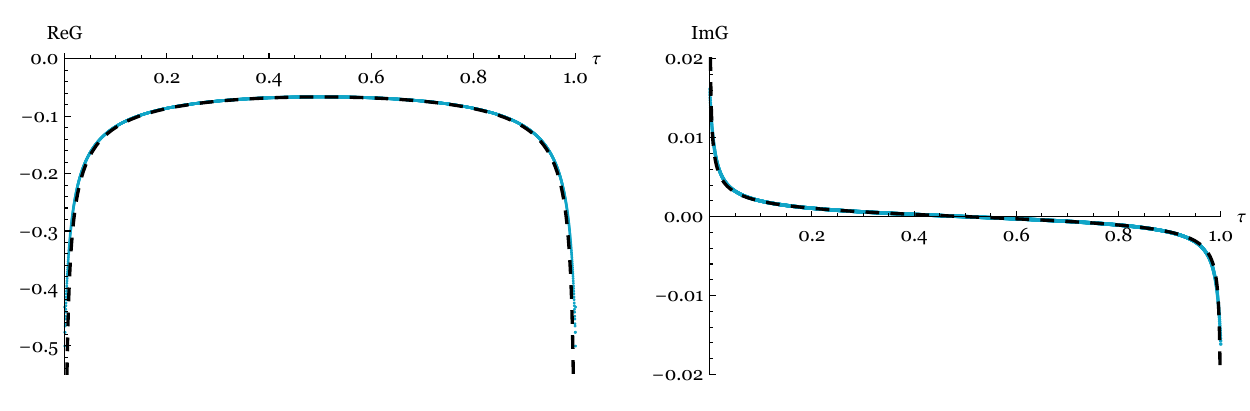}
\caption{The left and right panels of this plot show the real and imaginary parts of $G(\tau)$ respectively. The points in blue give the numerical solution in the case $q=4$, $\beta J=200$, $u=\pi$, computed with $M=2^{12}$ discretised points and tolerance $\epsilon=10^{-10}$. They are displayed together with plots of the analytic conformal solution \eqref{conf const fixed} in the dotted black lines. The parameter $u_\text{IR} = 0.04077$ is determined by fitting against the numerical solution. In other words, the numerical solution gives the explicit renormalization of $u$. Notice that there is excellent agreement between the numerical and conformal solutions except for a small window near $\tau = 0, 1$ as we expected. }
\label{fig: eg num sol}
\end{figure}

Using the algorithm above, we want to find all possible solutions for $G(\tau)$ as $u\in (-\pi,\pi)$ is varied. Firstly, all of the solutions that we found at large $\beta J$ agree with the conformal solution for $(\beta J)^{-1}\ll\tau\ll 1-(\beta J)^{-1}$, regardless of the value of $u$. See Figure \ref{fig: eg num sol} for an illustrative example. This is in contrast with the case of a chemical potential $\mu$, where there is a gapped phase when $|\mu|\gg J$. One can understand this difference from the Hamiltonian in \eqref{H op def}. The magnitudes of contributions to the partition function depend on the real part $\hat{H}_0$ of $\hat{H}$ and are not affected by $u$, which only controls the imaginary part of $\hat{H}$. In the case of a chemical potential, the magnitudes of contributions vary with $\mu$ because there is a competition between the terms $\mu\hat{Q}$ and $\hat{H}_0$, which are both real. 

In addition, we observe that the solutions (such as Figure \ref{fig: eg num sol}) satisfy the relations \eqref{G cc relation}, \eqref{G cc re and im}, \eqref{reG bdy}, and \eqref{LW identity}. Some of these should only be true when charge conjugation symmetry is preserved by $\hat{H}_0$. This is an important motivation for choosing the charge conjugation invariant operator ordering of $\hat{H}_0$ in \eqref{H op def}. This would later allow us to compare results from exact diagonalization with those from the large $N$ saddle point analysis.    

\begin{figure}[htbp]
    \centering
    \includegraphics[width=\textwidth]{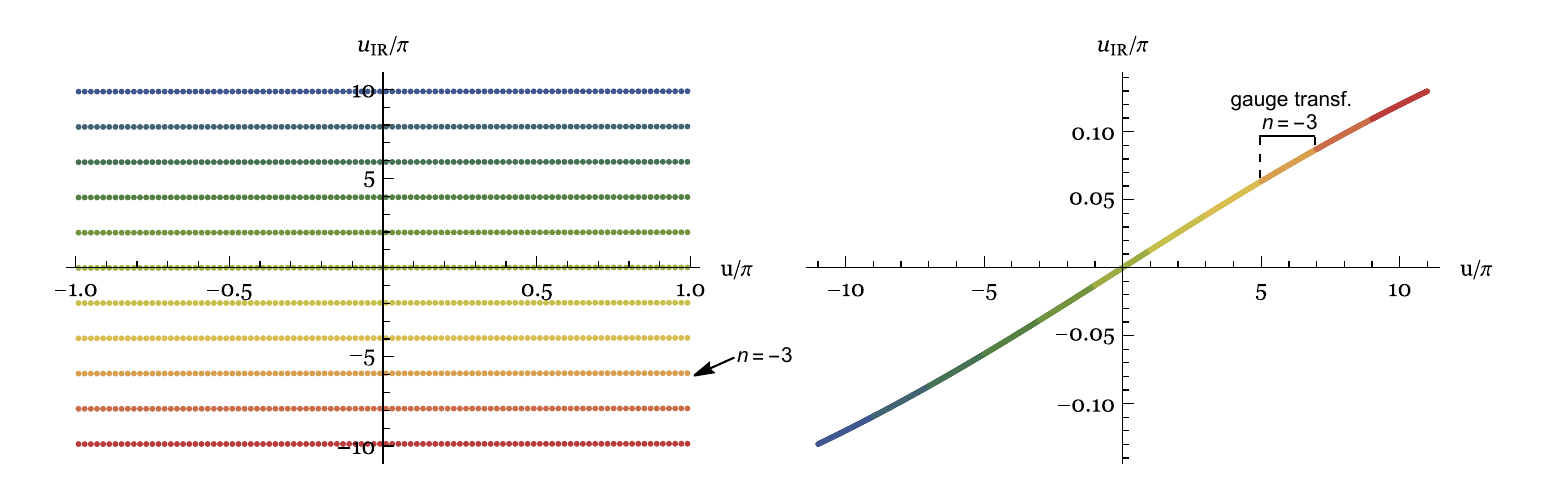}
    \caption{The left panel above plots $u_\text{IR}$, obtained by fitting numerical solutions $G_0(\tau;u,n)$ to the conformal ansatz, against $u$. The solutions on different branches $n$ are coloured differently, but distinct solutions on the same branch have the same colour. The right panel plots the fitted values of $u_\text{IR}$ for the gauge equivalent solutions $G_0(\tau;u-2\pi n,0)= e^{-2\pi in\tau}G_0(\tau;u,n)$. Each segment with a uniform colour on the right panel plots the solutions which are gauge equivalent to those in a particular branch with the same colour on the left panel. For instance, in the left panel, we have labelled the $n=-3$ branch of solutions, which are all coloured orange. Their gauge equivalent counterparts lie on the orange coloured segment in the right panel, which is also labelled.}
    \label{fig:uIR vs uUV}
\end{figure}

It turns out that the numerics in this case are more subtle than the case of a chemical potential; there are multiple fixed points of the iterative algorithm and therefore the numerics is very  sensitive to the initial data provided at the beginning of the iterations. To better handle this sensitivity, we scan over $u$ in small increments. Namely, we start with $u = \Delta u/2$ where $\Delta u=\pi/M$ for some large integer $M$ and take a constant $G(\tau)=-\frac{1}{2}$ as initial data. The solution found is then fed as initial data to find the solution at a slightly larger value of $u = 3\Delta u/2$. This is then iterated, where the solution at each value of $u$ is used as initial data for $u+\Delta u$. When $u = \pi-\Delta u/2$ is reached at the $M$-th step, the next point $u=\pi +\Delta u/2$ is not in the range $(-\pi,\pi)$, but it can be brought back to $u=-\pi+\Delta u/2$ via a large gauge transformation. We thus take the solution for $u = \pi-\Delta u/2$, apply the gauge transformation $G\mapsto e^{-2\pi i\tau}G$, and use this as initial data to solve for the solution at $u =-\pi + \Delta u/2$. Repeating this process allows us to scan over $u\in(-\pi,\pi)$ repeatedly. We also applied this method to scan over $u$ in negative increments. Each numerical solution at a fixed $u$ is fitted against the conformal solution \eqref{conf const fixed} to determine $u_\text{IR}$, and the left panel of Figure \ref{fig:uIR vs uUV} shows the plot of $u_\text{IR}$ as a function of $u$ obtained in this way.  

Importantly, there are infinitely many solutions with different values of $u_\text{IR}$ which solve \eqref{full SD eqn} for the same value of $u\in (-\pi,\pi)$. For each solution, the best-fit value of $u_\text{IR}$ satisfies \eqref{uIR range} and is centred around $2\pi n$ for some $n\in\bZ$ defined in \eqref{split u_IR}, \eqref{sign integer eq centre}. The different solutions at a fixed $u$ have different values of $n$, and we denote them by $G_0(\tau;u,n)$. In the left panel of Figure \ref{fig:uIR vs uUV}, each ``branch'' of solutions with a distinct value of $n$ is coloured differently. As we will explicitly show in Section \ref{sec: partition fn}, the presence of infinitely many solutions is essential to ensure that the fermion partition function $\wt Z(u)$ has the required periodicity $\wt Z(u+2\pi n)=(-1)^{Nn}\wt Z(u)$, which is in turn equivalent to the charge quantization \eqref{eq: Q quantization}.

Alternatively, the situation is gauge equivalent to working with the \emph{single} branch of solutions with $n=0$, but spread across $u\in\bR$ instead of $u\in(-\pi,\pi)$, as we will explain in the following. Consider $G_0(\tau;u_0,n)$, the solution at $u=u_0\in(-\pi,\pi)$ on the branch $n$. Due to the invariance of \eqref{full SD eqn} under large gauge transformations as discussed below \eqref{Fourier convention}, $e^{-2\pi in\tau}G_0(\tau;u_0,n)$ must be a solution at $u=u_0-2\pi n$, and we checked this numerically on every solution found in the left panel of Figure \ref{fig:uIR vs uUV}. The values of $u_\text{IR}$ obtained from the gauge transformed solutions are plotted in the right panel of Figure \ref{fig:uIR vs uUV}, where points which are related by a gauge transformation are displayed in the same colour. In addition, it is simple to check that multiplying the conformal ansatz with parameter $u_\text{IR}=u_{\text{IR},0}$ in \eqref{conf const fixed} by $e^{-2\pi in\tau}$ shifts the parameter to $u_\text{IR}=u_{\text{IR},0}-2\pi n$. Therefore $e^{-2\pi in\tau}G_0(\tau;u_0,n)$ will have a fitted $u_\text{IR}$ that is on the branch $n = 0$, according to the definition \eqref{split u_IR}, \eqref{sign integer eq centre}. This is numerically confirmed by the range of $u_\text{IR}$ in the right panel of Figure \ref{fig:uIR vs uUV}. We thus denote the gauge transformed solutions as
\be\label{gt sols branch 0}
G_0(\tau;u-2\pi n,0)\equiv e^{-2\pi in\tau}G_0(\tau;u,n)\,,\quad u\in (-\pi,\pi)\,.
\ee
Since $u-2\pi n\in (-2\pi n-\pi,-2\pi n+\pi)$, working with the gauge transformed solutions for every $n\in\bZ$ will extend the range of $u$ from $(-\pi,\pi)$ to $(-\infty,\infty)$. As $u\rightarrow \pm\infty$, we find that the fitted values of $u_\text{IR}\rightarrow \pm \frac{\pi}{2}$, which are the maximum or minimum values allowed by the consistency condition \eqref{uIR range} when $q=4$. 


\subsection{Large q analysis}

As an aside, we note that the Schwinger-Dyson equations \eqref{full SD eqn} can be solved analytically in the large $q$ limit, by adapting Appendix A.1 of \cite{Louw:2022njq}. This method of taking the large $q$ limit is distinct from that in Appendix C of \cite{Davison:2016ngz}. Firstly, we redefine
\be\label{redef rem mu}
G(\tau)\rightarrow e^{iu\tau}G(\tau)\,,\quad \Sigma(\tau)\rightarrow e^{iu\tau}\Sigma(\tau)\,,
\ee
so that in terms of the new variables, $u$ is absent from the equations 
\be\label{sd wo mu}
\partial_{\tau_1}G(\tau_{12})+\int_0^1 d\tau_3 \Sigma(\tau_{13})G(\tau_{32}) = -\delta(\tau_{12})\,,\quad \Sigma(\tau) = (\beta J)^2 G(\tau)\left[-G(\tau)G(-\tau)\right]^{\frac{q}{2}-1}\,.
\ee
In exchange, the KMS condition is modified to
\be\label{modified KMS}
G(\tau\pm 1) = -e^{\mp iu}G(\tau)\,,\quad \Sigma(\tau\pm 1) = -e^{\mp iu}\Sigma(\tau)\,.
\ee
We consider the ansatz
\be\label{large q ansatz}
G(\tau)=\left(\cQ - \frac{1}{2}\sgn(\tau)\right)e^{\frac{1}{q}g(\tau)}\,,\quad \cQ\equiv i\im G(0^\pm)\,,
\ee
where $\cQ$ and $g$ are assumed to be $\cO(1)$, and we want to solve for $g(\tau)$ in the range $\tau\in(0,1)$. For the moment, $\cQ$ is an undetermined parameter which specifies the boundary condition of $G$. It turns out that the equations determining $g(\tau)$ are most conveniently written in terms of the combinations
\be
g_\pm(\tau)\equiv\frac{1}{2}\left(g(\tau) \pm g(-\tau)\right)\,.
\ee
Before deriving these equations however, we first examine the boundary conditions for $g_\pm$ because they will be used in the process.

\paragraph{Boundary conditions} 
From the properties of solutions to the Schwinger-Dyson equations in  \eqref{reG bdy} and \eqref{imG bdy}, we know that $G(0^\pm)=\cQ\mp\frac{1}{2}$. On the ansatz \eqref{large q ansatz}, this implies
\be
g(0^\pm) = 2\pi i qn_\pm\,,\qquad n_\pm\in\bZ\,,
\ee
which in turn implies
\be
g_\pm(0^+) = \pi i q(n_+\pm n_-)\,.
\ee
The reality property $G(\tau)^* = -G(-\tau)$ still holds after the redefinition \eqref{redef rem mu}. This implies $g(\tau)^* = g(-\tau) + 2\pi i q n_\text{cc}$. Evaluating on $\tau = 0^+$, we see that the integer ambiguity is determined to be $n_\text{cc} = -n_+ - n_-$, i.e.
\be\label{cc wt G}
g(\tau)^* - g(-\tau) = - 2\pi i q(n_+ + n_-)\,.
\ee
In addition, the modified KMS condition \eqref{modified KMS} implies
\bea\label{KMS on wt G}
g(1-\tau) &= g(-\tau) + q\log\left(\frac{1+2\cQ}{1-2\cQ}\right) - iqu + 2\pi i q l\,,\\
g(\tau - 1) &= g(\tau) - q\log\left(\frac{1+2\cQ}{1-2\cQ}\right) + iqu + 2\pi i q m\,,\qquad l,m\in\bZ\,.
\eea
Subtracting the second equation from the complex conjugate of the first equation gives $m = -l$ upon using \eqref{cc wt G}. One also needs to note that $\cQ$ is imaginary by its definition \eqref{large q ansatz}. Taking the sum of the two equations in \eqref{KMS on wt G} then gives
\be\label{wt G+ symm}
g_+(1-\tau) = g_+(\tau)\,.
\ee
In other words, $g_+$ is symmetric or even under the reflection $\tau\mapsto 1-\tau$ about the axis $\tau = \frac{1}{2}$. Evaluating this at $\tau = 0^+$ gives the boundary condition $g_+(1^-) = \pi i q(n_+ + n_-)$. Evaluating the difference of the two equations in \eqref{KMS on wt G} at $\tau = 0^+$ then determines $g_-(1^-)$. In summary, $g_\pm$ need to satisfy \eqref{wt G+ symm} and the boundary conditions
\bea\label{large q bc}
g_+(0^+) &= g_+(1^-) = \pi i q(n_+ + n_-)\,, \quad g_-(0^+) = \pi i q(n_+ - n_-)\,,\\
g_-(1^-) &= \pi i q(n_- - n_+) + q\log\left(\frac{1+2\cQ}{1-2\cQ}\right) - iqu + 2\pi i l\,.
\eea

\paragraph{Equations for $g_\pm$}
We proceed to derive the differential equations satisfied by $g_\pm$. Substituting the ansatz \eqref{large q ansatz} into the equation for $\Sigma$ in \eqref{sd wo mu} gives
\be
\Sigma(\tau) = (\beta J)^2\left(\frac{1}{4}-\cQ^2\right)^{\frac{q}{2}-1}\left(\cQ - \frac{1}{2}\sgn(\tau)\right)e^{\frac{1}{q}g(\tau) + \left(1-\frac{2}{q}\right)g_+(\tau)}\,.
\ee
Using $\partial_{\tau_1}G(\tau_{12}) = -\partial_{\tau_2}G(\tau_{12})$, setting $\tau_1 = 0$, and relabelling $\tau = -\tau_2$, $\tau' = \tau_3$, the first equation of \eqref{sd wo mu} becomes
\be\label{interm sd pos}
\partial_\tau G(\tau) + \int_0^1 d\tau' \Sigma(-\tau')G(\tau'+\tau) = -\delta(\tau)\,.
\ee
Despite the additional exponential factors in \eqref{modified KMS}, we still have $\Sigma(-\tau')G(\tau'+\tau) = \Sigma(1-\tau')G(\tau'-1+\tau)$. Using this rewriting in \eqref{interm sd pos} and redefining $\tau' - 1\rightarrow \tau'$ then gives
\be\label{interm 2 sd pos}
\partial_\tau G(\tau) + \int_{-1}^0 d\tau' \Sigma(-\tau')G(\tau'+\tau) = -\delta(\tau)\,.
\ee
Notice that the range of $\tau'$ has been shifted with respect to \eqref{interm sd pos}. Let $\tau > 0$. After substituting the ansatz \eqref{large q ansatz} into \eqref{interm 2 sd pos} and multiplying it by $qG(\tau)^{-1}$, one obtains
\bea\label{first deriv pos}
\partial_\tau g(\tau)+(\beta\cJ)^2\int_{-1}^0 d\tau' &\left(\cQ - \frac{1}{2}\sgn(\tau'+\tau)\right)e^{\frac{1}{q}g(-\tau')+\left(1-\frac{2}{q}\right)g_+(-\tau')+\frac{1}{q}g(\tau +\tau')-\frac{1}{q}g(\tau)} = 0\,,\\
(\beta\cJ)^2&\equiv q(\beta J)^2\left(\frac{1}{4}-\cQ^2\right)^{\frac{q}{2}-1}\,.
\eea
We assume that $\beta\cJ$ is kept constant as $q\rightarrow \infty$. On the other hand, \eqref{interm sd pos} relabeled using $\tau\rightarrow -\tau$ is
\be\label{interm sd neg}
\partial_\tau G(-\tau) - \int_0^1 d\tau'\,\Sigma(-\tau')G(\tau'-\tau) = \delta(-\tau)\,.
\ee
Like before, we consider $\tau > 0$, substitute \eqref{large q ansatz} into \eqref{interm sd neg}, and multiply the equation by $qG(-\tau)^{-1}$. This gives
\be\label{first deriv neg}
\partial_\tau g(-\tau)-(\beta\cJ)^2\int_0^1 d\tau'\left(\cQ + \frac{1}{2}\sgn(\tau - \tau')\right)e^{\frac{1}{q}g(-\tau')+\left(1-\frac{2}{q}\right)g_+(-\tau')+\frac{1}{q}g(\tau'-\tau) - \frac{1}{q}g(-\tau)} = 0\,.
\ee
If we redefine the dummy integration variable in \eqref{first deriv pos} as $\tau'\rightarrow -\tau'$  and add or subtract \eqref{first deriv neg}, we have 
\bea\label{first deriv comb}
\partial_\tau g_+(\tau) &= \frac{(\beta\cJ)^2}{2}\int_0^1 d\tau' \sgn(\tau - \tau')e^{\left(1-\frac{2}{q}\right)g_+(\tau')} + \cO(q^{-1})\,,\\
\partial_\tau g_-(\tau) &= -\cQ (\beta\cJ)^2\int_0^1 d\tau' e^{\left(1-\frac{2}{q}\right)g_+(\tau')} + \cO(q^{-1})\,.
\eea
Taking the $\tau$ derivative of \eqref{first deriv comb} then gives
\be\label{sec deriv pm}
\partial_\tau^2 g_+(\tau) = (\beta\cJ)^2e^{\left(1-\frac{2}{q}\right)g_+(\tau)}+\cO(q^{-1})\,,\quad \partial_\tau^2 g_-(\tau) = \cO(q^{-1})\,,\quad \tau > 0\,.
\ee
Going back to the equation for $g_-$ in \eqref{first deriv comb}, we see from the expression for $\partial_\tau^2g_+$ in \eqref{sec deriv pm} that
\be\label{deriv Gm}
\partial_\tau g_-(\tau) = -\cQ\int_0^1 d\tau'\partial_{\tau'}^2g_+(\tau') = -\cQ\,\partial_{\tau'}g_+(1^-) + \cQ\,\partial_{\tau'}g_+(0^+)= 2\cQ\,\partial_{\tau'}g_+(0^+)\,.
\ee
In the last equality, we have used the symmetry \eqref{wt G+ symm} to rewrite $\partial_{\tau'}g_+(1^-)=\lim_{\epsilon\rightarrow 0^+}\frac{g_+(1)-g_+(1-\epsilon)}{\epsilon}=\lim_{\epsilon\rightarrow 0^+}\frac{g_+(0)-g_+(\epsilon)}{\epsilon}=-\partial_{\tau'}g_+(0^+)$.

\paragraph{Solutions for $g_\pm$} Let us first focus on the solution for $g_+$. The general solution to the differential equation in \eqref{sec deriv pm} is
\be
e^{\left(1-\frac{2}{q}\right)g_+(\tau)} = \frac{2v^2}{(\beta\cJ)^2\left(1-\frac{2}{q}\right)\sin^2(v\tau + b)}\,.
\ee
Imposing $g_+(1-\tau) = g_+(\tau)$ implies that $v + 2b = \pi n$, where $n$ is an odd integer since $b$ is defined modulo $\pi$. Eliminating $b = \frac{\pi n}{2}-\frac{v}{2}$ allows us to rewrite $\sin^2(v\tau + b) = \cos^2(v\tau -v/2)$. Subsequently, imposing $g_+(0^+) = g_+(1^-)=\pi i q(n_+ + n_-)$ in \eqref{large q bc} implies
\be\label{v det cond}
\frac{2v^2}{(\beta\cJ)^2\left(1-\frac{2}{q}\right)\cos^2\left(\frac{v}{2}\right)} = 1\,,\quad v\in (0,\pi)\,,
\ee
since $q$ is even. This allows us to determine $v$ given $\beta\cJ$. We assume that this equation has no solutions for $v\notin \bR$. Note that $v$ and $-v$ give the same solution for $g_+$, so we can restrict to $v > 0$ without loss of generality. However, $v > \pi$ must be excluded, since otherwise $G(\tau)$ will have singularities in the range $\tau\in(0,1)$. This gives the range $v\in (0,\pi)$. Since $q - 2$ is even, there is a choice of positive or negative roots in 
\be\label{exp G_+ sol}
e^{\frac{1}{q}g_+(\tau)}=e^{\pi i(n_+ + n_-)}\left[\frac{\cos\left(\frac{v}{2}\right)}{\cos\left(v\tau-\frac{v}{2}\right)}\right]^{\frac{2}{q-2}}\,,
\ee
which is determined by $g_+(0^+) = g_+(1^-)=\pi i q(n_+ + n_-)$.

We now turn to the solution for $g_-$. Plugging the solution for $g_+$ into \eqref{deriv Gm} gives
\be
\partial_\tau g_-(\tau) = -\frac{4\cQ v\tan\frac{v}{2}}{1-\frac{2}{q}}\,,
\ee
and therefore
\be\label{G_- sol}
g_-(\tau) = -\frac{4\cQ v\tan\frac{v}{2}}{1-\frac{2}{q}}\tau + b\,,\quad b\in\bC\,.
\ee
The boundary condition $g_-(0^+) = \pi i q(n_+ - n_-)$ in \eqref{large q bc} then sets $b = \pi i q(n_+ - n_-)$, while the boundary condition for $g_-(1^-)$ implies
\be\label{eos u}
iu = \log\left(\frac{1+2\cQ}{1-2\cQ}\right) + \frac{4\cQ v\tan\frac{v}{2}}{q-2} + 2\pi i n\,,\quad n = n_- - n_+ + l\,.
\ee
The $2\pi\bZ$ ambiguity in $u$ arises from the invariance of the Schwinger-Dyson equations \eqref{SD eqn unit beta} under large gauge transformations. If we fix the gauge by restricting $u\in (0,2\pi)$, the integer $n$ in \eqref{eos u} is fixed for a given $\cQ$ such that $u$ is within $(0,2\pi)$.  
\begin{figure}[h]
    \centering
    \includegraphics[width=0.8\textwidth]{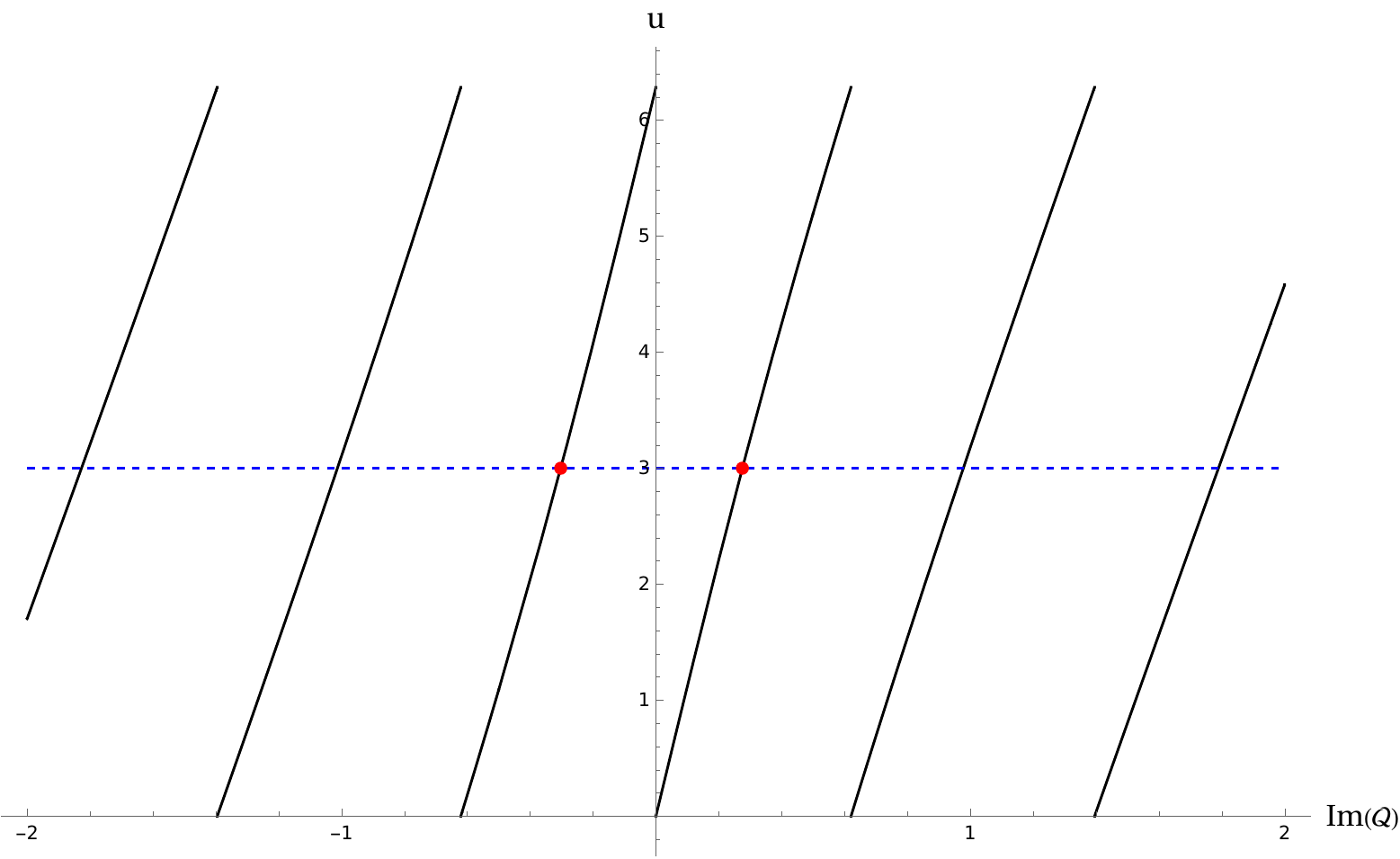}
    \caption{\small The black curve plots $u$ against $\im\cQ$ according to \eqref{eos u}, where $\cQ = i\im\cQ$. The integer $n$ in \eqref{eos u} is chosen such that $u\in (0,2\pi)$. The parameters of the model are set to $q = 40$ and $\beta\cJ = 100$, for which $v\approx 3.05$. At each value of $u$, say $u = 3$ which is traced by the blue dotted line, there seem to be infinitely many solutions corresponding to the intersections between the blue dotted line and the black curve. However, only the two coloured red are consistent.}
    \label{fig:u vs imQ at large q}
\end{figure}
Figure \ref{fig:u vs imQ at large q} is a plot of $u\in (0,2\pi)$ where $n$ has been determined in this way. Note that at each value of $u$ there seem to be infinitely many solutions with different values of $\cQ$ and $n$. As expected, this is consistent with the situation at $q = 4$ illustrated by Figure \ref{fig:uIR vs uUV}, where for each $u$ there are infinitely many solutions with different $u_\text{IR}$ and therefore different $\cQ$. However, not all values of $\cQ$ are expected to give consistent solutions at large $q$ because we need $\cQ\sim\cO(1)$ in order to be compatible with the derivation thus far. In particular, values of $\cQ$ which satisfy \eqref{eos u} for $n \leq -1$ or $n\geq 2$ necessarily have $\cQ\sim\cO(q)$ and should be neglected. To show this, we first note that
\be
-i\log\left(\frac{1+2\cQ}{1-2\cQ}\right)+\frac{4\im\cQ\, v\tan\frac{v}{2}}{q-2} \begin{cases}> 2\pi\,,\quad \;\;\;\text{if}\;\; n\leq -1 \\ <-2\pi\,,\quad\text{if}\;\; n\geq 2\end{cases}\,,
\ee
since otherwise the shifts by $2\pi n$ would not have been necessary for $u\in(0,2\pi)$. Since $-i\log\left(\frac{1+2\cQ}{1-2\cQ}\right)\in(-\pi,\pi)$, we must have $\im\cQ > 0$ if $n\leq -1$ and $\im\cQ < 0$ if $n\geq -2$. The two cases above can then be written uniformly as
\be
2\pi < -i\sgn(\im\cQ)\log\left(\frac{1+2\cQ}{1-2\cQ}\right)+\frac{4|\im\cQ|\, v\tan\frac{v}{2}}{q-2} < \pi + \frac{4|\im\cQ|\, v\tan\frac{v}{2}}{q-2}\,, 
\ee
where the last inequality is again due to $\mp i\log\left(\frac{1+2\cQ}{1-2\cQ}\right)\in(-\pi,\pi)$. This then implies
\be
|\im\cQ| > \frac{\pi(q-2)}{4v\tan\frac{v}{2}}\,,
\ee
which contradicts the assumption $\cQ\sim\cO(1)$. Therefore, we expect there to be at most two solutions at large $q$ for each value of $u$, corresponding to $n = 0,1$. For example, at $u = 3$, which is traced by the blue dotted line in Figure \ref{fig:u vs imQ at large q}, the valid solutions and their values of $\cQ$ are indicated by the intersections in red closest to the vertical axis. The two connected components of the black curve which are to the immediate right and left of the vertical axis correspond to $n = 0$ and $n = 1$ respectively.  

We now take the two parts $g_\pm$ together, plugging \eqref{exp G_+ sol} and \eqref{G_- sol} back into the ansatz \eqref{large q ansatz}. This gives
\be
G(\tau) = \left(\cQ-\frac{1}{2}\right)e^{-\frac{4\cQ v\tan\left(\frac{v}{2}\right)}{q-2}\tau}\left[\frac{\cos\left(\frac{v}{2}\right)}{\cos\left(v\tau-\frac{v}{2}\right)}\right]^{\frac{2}{q-2}}\,,\quad \tau\in(0,1)\,,
\ee
where $v$ and $u$ are determined by \eqref{v det cond} and \eqref{eos u} respectively. If we undo the initial redefinition \eqref{redef rem mu}, the solution for the original 2-point function is
\be\label{large q sol}
G(\tau) = \left(\cQ - \frac{1}{2}\right)e^{\left[\log\left(\frac{1+2\cQ}{1-2\cQ}\right) + 2\pi i n\right]\tau}\left[\frac{\cos\left(\frac{v}{2}\right)}{\cos\left(v\tau-\frac{v}{2}\right)}\right]^{\frac{2}{q-2}}\,.
\ee

\begin{figure}[H]
    \centering
    \includegraphics[width=0.9\textwidth]{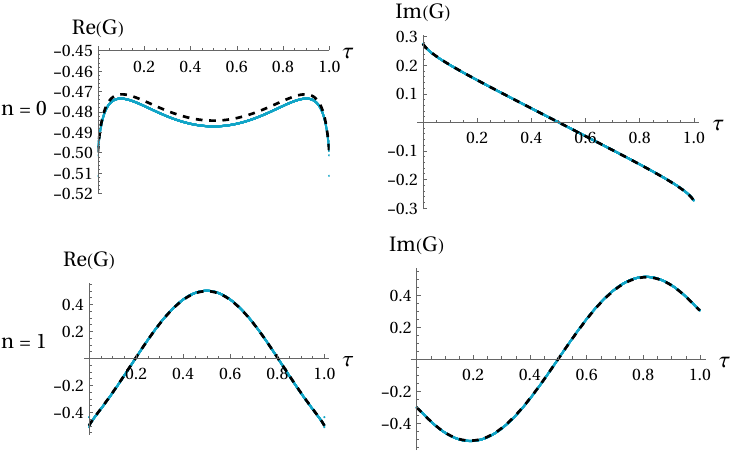}
    \caption{\small Plots of $\re G(\tau)$ and $\im G(\tau)$ against $\tau$ at $q=40$, $\beta\cJ = 100$, and $u = 3$. The black dotted curves are the analytic predictions according to \eqref{large q sol}, while the blue solid curves are the numerical solutions obtained via the method of Section \ref{sec:num}. In the first row, we pick the case where $n = 0$, and $\cQ\approx 0.275$ is the horizontal position of the red point to the right of the vertical axis in Figure \ref{fig:u vs imQ at large q}. In the second row, we took $n = -1$, and $\cQ\approx -0.303$ is given by the horizontal position of the red point to the left of the vertical axis in Figure \ref{fig:u vs imQ at large q}. In the two cases, the original parameter $\beta J$ which appears in the Schwinger-Dyson equations \eqref{SD eqn unit beta} is determined by \eqref{first deriv pos}, and it evaluates to $\beta J \approx 671943, 426399$ for $n=0,1$ respectively.}
    \label{fig:large q num}
\end{figure}

\paragraph{Numerical checks} We now check that the analytic large $q$ solutions that were found are a good approximation to the exact numerical solutions. In particular, we want to examine the two solutions which exist simultaneously for the same value of $u$. As such, we fix $q=40$ and $\beta\cJ=100$ so that $v$ is fixed, and consider the two solutions at $u = 3$. Their predicted values of $\cQ$ are given by the horizontal positions of the red points to the left and right of the vertical axis in Figure \ref{fig:u vs imQ at large q}. We then compare the numerical solutions obtained via the standard method in Section \ref{sec:num} with the predicted solution in \eqref{large q sol}. The result is given in Figure \ref{fig:large q num}, where the blue curves plot the numerical solutions while the dotted black curves plot the analytic prediction. We observe that there is very good agreement between the two sets of curves.

\section{The partition function}\label{sec: partition fn}

The first observable that one can think of computing is the partition function \eqref{ferm path integral}
\be\label{Zk saddle pt}
Z_k = \int_{-\pi}^{\pi} du\,e^{-iku}\wt Z(u)\approx\sum_{n\,\in\,\bZ}\int_{-\pi}^{\pi}du\,\frac{1}{\sqrt{\Delta(G_0(u,n))}}e^{-iku-S[u,G_0(u,n),\Sigma_0(u,n)]}\,,
\ee
where we have used the large $N$ saddle point approximation for $\wt Z(u)$ in \eqref{SD sum Z G}. Since there are multiple solutions $G_0(\tau;u,n)$ to the Schwinger-Dyson equations for each $u$, as introduced in Section \ref{sec:num}, we need to sum over all of these saddles. When it is not necessary to consider the time dependence of the solutions $G_0(\tau;u,n)$, we will sometimes denote them by $G_0(u,n)$, as we have done in \eqref{Zk saddle pt}.

We first examine the on-shell action $S[u, G_0,\Sigma_0]$. Using the second equation of \eqref{full SD eqn} to simplify the integral term in \eqref{wt S def}, and passing to frequency space, the expression for the on-shell action becomes
\be\label{on-shell S before reg}
-\wt S[u, G_0,\Sigma_0]=\sum_{n\,\in\,\bZ}\log\left[-2\pi i \left(n+\frac{1}{2}\right)+iu-\Sigma_0(\omega_n)\right]+\frac{q-1}{q}\sum_{n\,\in\,\bZ}\Sigma_0(\omega_n)G_0(\omega_n)\,.
\ee
In the above, $\Sigma_0$ is determined by $G_0$ according to the second equation of \eqref{full SD eqn}. The first term is a divergent sum, and different regularization schemes are possible. We choose the scheme used in \cite{Davison:2016ngz,Gu:2019jub}, which replaces $\log\det(-\partial_\tau +iu)$ with $\log Z_\text{free}$, where $Z_\text{free}$ is the partition function of a free fermion. We choose the charge conjugation invariant answer for $Z_\text{free}$ in order to be consistent with the choice of operator ordering made in \eqref{H op def}, as we shall see shortly from how the regularised on-shell action transforms under large gauge transformations. The regularized on-shell action is\footnote{For special values $u=(2m+1)\pi$, $m\in\bZ$, the $\log$ term in \eqref{on-shell S before reg} is $\sum_{n\in\bZ}\log\left[-i\pi(2n+1)-\Sigma_0(\omega_{m+n})\right]$, which is regularised to $\log 2+\sum_{n\in\bZ}\log\left[1-\frac{i\Sigma_0(\omega_{m+n})}{\pi(2n+1)}\right]$.}
\be\label{on-shell S after reg}
-\wt S[u, G_0,\Sigma_0]=\log\bigg(2\cos\frac{u}{2}\bigg)+\sum_{n\,\in\,\bZ}\log\left[1-\frac{i\Sigma_0(\omega_n)}{2\pi\left(n+\frac{1}{2}\right)-u}\right]+\frac{q-1}{q}\sum_{n\,\in\,\bZ}\Sigma_0(\omega_n)G_0(\omega_n)\,.
\ee
Recall from \eqref{G FT imag} that $G_0(\omega_n)$ and $\Sigma_0(\omega_n)$ are imaginary, which implies that the last two terms of \eqref{on-shell S after reg} are real, and only contribute to the magnitude of $e^{-S}$.  

We want to first examine the behaviour of $S=N\wt S$ and therefore $\wt Z(u)$ under a large gauge transformation. As discussed below \eqref{full SD eqn}, a large gauge transformation maps $G_0(\omega_m)\mapsto G_0(\omega_{m-n})$, $\Sigma_0(\omega_m)\mapsto \Sigma_0(\omega_{m-n})$, and $u\mapsto u+2\pi n$, under the last two terms of \eqref{on-shell S after reg} are invariant since
\bea
\sum_{m\,\in\,\bZ}\log\left[1-\frac{i\Sigma_0(\omega_{m-n})}{2\pi\left(m+\frac{1}{2}\right)-u-2\pi n}\right]+\frac{q-1}{q}\sum_{m\,\in\,\bZ}\Sigma_0(\omega_{m-n})G_0(\omega_{m-n})\\
=\sum_{m'\,\in\,\bZ}\log\left[1-\frac{i\Sigma_0(\omega_{m'})}{2\pi\left(m'+\frac{1}{2}\right)-u}\right]+\frac{q-1}{q}\sum_{m'\,\in\,\bZ}\Sigma_0(\omega_{m'})G_0(\omega_{m'})\,,
\eea
by a simple relabelling $m'=m-n$. The remaining part of $e^{-S}$ coming from the first term in \eqref{on-shell S after reg} transforms to 
\be
\left(2\cos\frac{u+2\pi n}{2}\right)^N=(-1)^{Nn}\left(2\cos\frac{u}{2}\right)^N\,,
\ee
which is not invariant if $N$ is odd. Lastly, since we are expanding a gauge invariant action around $G_0$, the term at quadratic order in the fluctuations must be gauge invariant. This can be shown from \eqref{S quad expansion}. As $\Delta(G_0)$ is the determinant of the operator appearing at quadratic order, it will be gauge invariant provided that a gauge invariant regularization scheme can be found. We shall assume this. Then one has
\be\label{on-shell S gauge inv}
\frac{1}{\sqrt{\Delta(e^{2\pi in\tau}G_0)}}e^{-S[u+2\pi n,e^{2\pi in\tau}G_0,e^{2\pi in\tau}\Sigma_0]}=(-1)^{Nn}\frac{1}{\sqrt{\Delta(G_0)}}e^{-S[u,G_0,\Sigma_0]}\,.
\ee

It is now apparent why we need infinitely many solutions $G_0(u,n)$ labelled by $n\in\bZ$ in order for $\wt Z$ to satisfy the periodicity $\wt Z(u-2\pi n)=(-1)^{Nn}\wt Z(u)$, thereby ensuring $\U(1)$ charge quantization. Suppose we compute $\wt Z(u)$ for $u\in\bR$ by including only the contribution of the single solution $G_0(u,0)$ for each value of $u$. In other words, for $u\in (-\pi,\pi)$, we only include the solution on the branch $n=0$, while for $u\notin (-\pi,\pi)$ we only include the gauge transformed solution defined in \eqref{gt sols branch 0}. Visually, we are taking the solutions whose values of $u_\text{IR}$ are plotted in the right panel of Figure \ref{fig:uIR vs uUV}. This gives for $u\in (-\pi,\pi)$ and $n\neq 0$,
\bea
\wt Z(u) &= \frac{1}{\sqrt{\Delta(G_0(u,0))}}e^{-S[u,G_0(u,0),\Sigma_0(u,0)]}\\
\wt Z(u - 2\pi n) &= \frac{1}{\sqrt{\Delta(e^{-2\pi in\tau}G_0(u,n))}}e^{-S[u-2\pi n,e^{-2\pi in\tau}G_0(u,n),e^{-2\pi in\tau}\Sigma_0(u,n)]} \\
&= (-1)^{Nn}\frac{1}{\sqrt{\Delta(G_0(u,n))}}e^{-S[u,G_0(u,n),\Sigma_0(u,n)]}\neq (-1)^{Nn}\wt Z(u)\,.
\eea
We have used \eqref{on-shell S gauge inv} to write $\wt Z(u-2\pi n)$ as $(-1)^{Nn}$ times the contribution of the solution $G_0(u,n)$ on the branch $n\neq 0$, which is distinct from the solution $G_0(u,0)$ on the branch $n = 0$. Therefore, if only a single solution is included at every $u$, the periodicity $\wt Z(u-2\pi n)=(-1)^{Nn}\wt Z(u)$ cannot be satisfied. Visually, say for $n=-3$, we have written the contribution of a solution in the labelled segment in the right panel of Figure \ref{fig:uIR vs uUV} as $(-1)^N$ times the contribution of the corresponding gauge equivalent solution on the labelled $n=-3$ branch in the left panel. This is distinct from the $n = 0$ branch, whose points are very close to the horizontal $u$ axis in the left panel. It should now be clear that the appropriate remedy to restore the periodicity of $\wt Z$ is to include the contributions of all the solutions $G_0(u,n)$ on every branch $n\in\bZ$, as we have done in \eqref{Zk saddle pt} for $u\in (-\pi,\pi)$. For $u-2\pi n\in (-2\pi n-\pi,-2\pi n+\pi)$, we include the contributions of all the gauge transformed solutions
\be
G_0(u-2\pi n,m-n)\equiv e^{-2\pi in\tau}G_0(u,m)\,,\quad m\in\bZ\,.
\ee
Note that unlike before, the gauge transformations of $G_0(u,m)$ with $m\neq n$ are also included. In this way, \eqref{on-shell S gauge inv} implies that the periodicity of $\wt Z$ is indeed restored:
\bea\label{eq: wt Z periodicity}
\wt Z(u-2\pi n) &=\sum_{m\in\bZ}\frac{1}{\sqrt{\Delta(G_0(u-2\pi n,m-n))}}e^{-S[u-2\pi n,G_0(u-2\pi n,m-n),\Sigma(u-2\pi n,m-n)]}\\
&= \sum_{m\in\bZ}\frac{1}{\sqrt{\Delta(e^{-2\pi in\tau}G_0(u,m))}}e^{-S[u-2\pi n,e^{-2\pi in\tau}G_0(u,m),e^{-2\pi in\tau}\Sigma_0(u,m)]} \\
&= (-1)^{Nn}\sum_{m\in\bZ}\frac{1}{\sqrt{\Delta(G_0(u,m))}}e^{-S[u,G_0(u,m),\Sigma_0(u,m)]} = (-1)^{Nn}\wt Z(u)\,.
\eea

Notice from \eqref{gt sols branch 0} that the sum over $n\in\bZ$ with corresponding solutions $G_0(u,n)$ can be seen as a sum over the winding numbers of the gauge transformation $e^{2\pi i n\tau}$ acting on the solution $G_0(u-2\pi n,0)$. On the other hand, it is known from \cite{Gu:2019jub,Davison:2016ngz} that the effective low energy degrees of freedom in the complex SYK model are the Schwarzian mode $\varphi(\tau)$, associated with reparametrizations of $S^1$ taking $\tau\mapsto\varphi(\tau)$, and the axion $\lambda(\tau)$, which comes from gauge transformations $G_0(\tau)\mapsto e^{i(\lambda(\tau)-\lambda(0))}G_0(\tau)$. When computing the grand canonical partition function using the low temperature effective action, one has to sum over the winding numbers of $\lambda(\tau)$ (see (2.86) of \cite{Gu:2019jub} and \cite{Heydeman:2024ezi}). Due to their exact same origin as the winding numbers of $\U(1)$ gauge transformations acting conformal solutions $G_0$, we can therefore identify the sum over solutions in \eqref{Zk saddle pt} with the sum over winding numbers in the low temperature approximation to the grand canonical partition function. In the latter context, the sum over winding numbers serves the same purpose of ensuring that the partition function has the required periodicity under $iu=\beta\mu\mapsto \beta\mu + 2\pi i n$. From the perspective of the dual bulk AdS$_2$ dilaton gravity theory with a $\U(1)$ gauge field, we expect from \cite{Sachdev:2019bjn} that the sum over winding numbers can be identified with a sum over different bulk gauge field configurations where the boundary value of the time component $A_\tau$ is shifted by $2\pi n$. They contribute to the same boundary partition function due to gauge invariance. This sum over gravitational saddles was also found to be responsible for the periodicity of the supersymmetric indicies of higher-dimensional AdS/CFT duals such as \cite{Aharony:2021zkr,Iliesiu:2021are,Cabo-Bizet:2018ehj} under shifts of the chemical potentials.  

The sign in \eqref{eq: wt Z periodicity} is compensated by the transformation of the Wilson line
\be
e^{-ik(u+2\pi n)}=(-1)^{2kn}e^{-iku}\,,
\ee
since $(-1)^{2n(k+N/2)}=1$ due to \eqref{anomaly free condition}. Note that this explicitly realises the discussion in Section \ref{subsec: k quantisation} regarding the gauge anomaly in the path integral formalism. 

Now note that \eqref{Zk saddle pt} can be rewritten as
\bea\label{Zk u int}
Z_k &\approx\sum_{n\,\in\,\bZ}\int_{-\pi}^{\pi}du\,\frac{1}{\sqrt{\Delta(G_0(u,n))}}e^{-iku- S[u,G_0(u,n),\Sigma_0(u,n)]}\\
&=\sum_{n\,\in\,\bZ}\int_{-\pi}^{\pi}du\,\frac{1}{\sqrt{\Delta(G_0(u-2\pi n,0))}}e^{-ik(u-2\pi n)-S[u-2\pi n,G_0(u-2\pi n,0),\Sigma_0(u-2\pi n,0)]}\\
&=\sum_{n\,\in\,\bZ}\int_{-(2n+1)\pi}^{-(2n-1)\pi}du\,\frac{1}{\sqrt{\Delta(G_0(u,0))}}e^{-iku-S[u,G_0(u,0),\Sigma_0(u,0)]}\\
&=\int_{-\infty}^{\infty}du\,\frac{1}{\sqrt{\Delta(G_0(u,0))}}e^{-iku- S[u,G_0(u,0),\Sigma_0(u,0)]}\,.
\eea
In the second line, we have used the property \eqref{on-shell S gauge inv}, where $G_0(u-2\pi n,0)$ is the gauge transformed solution as defined in \eqref{gt sols branch 0}. In addition, recall that $\Delta(G_0)$ is gauge invariant. In the third line, we performed the change of variables $u'=u-2\pi n$, after which the integrand becomes $n$-independent, and the integrals add up to an integral over the full range $(-\infty,\infty)$. The resulting expression resembles what one would get for a theory with gauge group $\bR$ instead of $\U(1)$. Physically, this is reasonable because the only difference between the gauge groups is the presence of charge quantization for $\U(1)$, which makes large gauge transformations with nontrivial winding on $S^1$ possible, which in turn makes $u$ periodic modulo $2\pi$. Therefore, it makes sense that the crucial point which makes \eqref{Zk u int} different from the partition function of a $\bR$ gauge theory is the fact that the integrand is invariant under large gauge transformations, which allows us to rewrite it as an integral over $u\in (-\pi,\pi)$. 

\subsection{Naive saddle point analysis}\label{subsec: naive saddle}

Since the allowed charges $k$ of the Wilson line scale linearly with $N$, and so does the on-shell action $S$, one might try to compute the integral in \eqref{Zk u int} via the saddle point method. Defining $\kappa = k/N \in [-\frac{1}{2},\frac{1}{2}]$, \eqref{Zk u int} can be written in the form
\be\label{Zk u int overall N}
Z_k\approx\int_{-\infty}^{\infty}du\,\exp \left(N\left\{-i\kappa u- \wt S[u,G_0(u,0),\Sigma_0(u,0)]\right\}-\frac{1}{2} \log({\Delta(G_0(u,0))})\right)\,.
\ee
Deriving the exponent with respect to $u$, one finds that the saddle point equation is\footnote{Note that \eqref{on-shell S before reg} and \eqref{on-shell S after reg} differ by a $u$-independent infinite constant $\sum_{n\in\bZ}\log\left[-2\pi i\left(n+\frac{1}{2}\right)\right]$, which should not affect their $u$ derivatives. This can be checked explicitly using the Schwinger-Dyson equations \eqref{full SD eqn} and the identity $\frac{1}{2}\tan\frac{u}{2}=\sum_{n\in\bZ}\frac{1}{2\pi\left(n+\frac{1}{2}\right)-u}$.}
\be\label{u saddle eqn}
\kappa = i\partial_u\wt S[u,G_0,\Sigma_0] =\sum_{n\in\bZ}G_0(\omega_n) = \frac{G_0(0^+)+G_0(0^-)}{2} = i\im G_0(0^+)\,.
\ee
The second equality uses the first equation in \eqref{full SD eqn}, and the fact that the implicit $u$-dependence of $\wt S$ via $G_0$ does not matter, since $\wt S$ is stationary with respect to variations in $G_0$. In the third equality, we recognise that $G_0$ is discontinuous at $\tau=0$ and $G_0(0)$ is ambiguous. However, numerically, it turns out that $\sum_{n\in\bZ}G_0(\omega_n)$ coincides with the average of $G_0(0^+)$ and $G_0(0^-)$ to within working precision. The last equality follows from \eqref{G cc relation}, which holds when $u$ is real. Note that we have explicitly verified the relation \eqref{imG S deriv}, which was derived from general considerations. 

Therefore, \eqref{u saddle eqn} cannot be satisfied, unless $\kappa = 0$ and $u = 0$, in which case $G_0(\tau)$ is real and $\im G_0(0^+) = 0$. For $\kappa \neq 0$, it would naively seem that the saddle point method cannot be applied to the integral \eqref{Zk u int overall N} due to the absence of saddle points for real $u$. Later, in Section \ref{subsec: analy cont}, we shall see that this is simply an indication that one should deform the contour of integration from real $u$ to imaginary $u$, where the contributing saddles lie.     

\subsection{Direct integration via fitting}\label{subsec: Zk direct int}

For now, we temporarily put aside the saddle point approximation and attempt to evaluate the integral \eqref{Zk u int overall N} directly. Firstly, the determinant factor $\Delta(G_0)^{-\frac{1}{2}}$ is subleading in $N$, so we neglect this factor in our computation\footnote{If we were about to compute this factor, we need to find the eigenvalues of the operator appearing at second order in the expansion of $\wt S$ around $G_0$, and taking the infinite product of eigenvalues, which might require a carefully chosen regularisation.}, and compute
\be\label{det negl}
Z_k\approx\int_{-\infty}^{\infty}du\,\exp \left(N\left\{-i\kappa u- \wt S[u,G_0(u),\Sigma_0(u)]\right\}\right)\,.
\ee
As we will see in Section \ref{subsec: Zk ED}, this is justified a posteriori when comparing $\log(Z_k)/N$ computed in this way with the results coming from exact diagonalization. In this subsection and the next, we will always be working with the gauge transformed solutions $G_0(\tau;u,0)$. The argument specifying the branch $n = 0$ is redundant and we will denote these solutions by $G_0(\tau;u)$ for simplicity, or sometimes $G_0(u)$ if the time dependence can be left implicit, as is the case above.

Next, notice that the on-shell action is an even function of $u$ since
\bea
&-\wt S[-u,G_0(-u),\Sigma_0(-u)]\\
\nn &=\log\bigg(2\cos\frac{u}{2}\bigg)+\sum_{n\,\in\,\bZ}\log\left[1-\frac{i\Sigma_0(\omega_n;-u)}{2\pi\left(n+\frac{1}{2}\right)-u}\right]+\frac{q-1}{q}\sum_{n\,\in\,\bZ}\Sigma_0(\omega_n;-u)G_0(\omega_n;-u)\\
\nn &=\log\bigg(2\cos\frac{u}{2}\bigg)+\sum_{n\,\in\,\bZ}\log\left[1+\frac{i\Sigma_0(-\omega_n;u)}{2\pi\left(n+\frac{1}{2}\right)+u}\right]+\frac{q-1}{q}\sum_{n\,\in\,\bZ}\Sigma_0(-\omega_n;u)G_0(-\omega_n;u)\\
&=\log\bigg(2\cos\frac{u}{2}\bigg)+\sum_{n\,\in\,\bZ}\log\left[1-\frac{i\Sigma_0(\omega_n)}{2\pi\left(n+\frac{1}{2}\right)-u}\right]+\frac{q-1}{q}\sum_{n\,\in\,\bZ}\Sigma_0(\omega_n)G_0(\omega_n)\\
&=-\wt S[u,G_0(u),\Sigma_0(u)]\,.
\eea
The second equality uses \eqref{G FT imag}, while the third equality follows from a relabelling $n' = -n-1$. Consequently, \eqref{Zk u int overall N} can be written as
\be\label{Zk cos int}
Z_k\approx\int_{-\infty}^{\infty}du\,\cos(ku) \exp\left\{-N\wt S\left[u,G_0,\Sigma_0\right]\right\}\,.
\ee
We then generate solutions $G_0$ for a large range $u\in (-200\pi,200\pi)$ and compute $\wt S[u,G_0,\Sigma_0]$. The data for $\wt S$ is subsequently fitted to an even polynomial of degree $2\Lambda$, where $\Lambda$ is some positive integer cutoff. Namely,
\bea\label{wt S poly fit}
\exp\left[-N\wt S(u)\right] \approx \exp\left[ N\left(s_0 - s_2 u^2 + \sum_{i=2}^\Lambda s_{2i} u^{2i}\right)\right]\approx e^{N(s_0 - s_2 u^2)}\left(1+\sum_{i=2}^\Lambda z_{2i}(s)\, u^{2i}\right)\,.
\eea
In the last approximation, we have expanded the exponential $\exp N\sum_{i=2}^\Lambda s_{2i} u^{2i}$, also up to order $2\Lambda$. The coefficients $z_{2i}$ are polynomial functions of $s_{2i}$. Numerically, it turns out that $s_2 > 0$, and therefore we can apply 
\be
\int_{-\infty}^\infty du\,\cos(ku)\exp\left[N(s_0-s_2 u^2)\right]=\sqrt{\frac{\pi}{Ns_2}}\exp\left(Ns_0-\frac{k^2}{4Ns_2}\right)\,.
\ee
Using derivatives of the above with respect to $k$, we also have
\bea
&\int_{-\infty}^\infty du\, u^{2i}\cos(ku)\exp\left[N(s_0-s_2 u^2)\right]=(-1)^i\frac{d^{2i}}{dk^{2i}}\sqrt{\frac{\pi}{Ns_2}}\exp\left(Ns_0-\frac{k^2}{4Ns_2}\right)\\
&=\sqrt{\frac{\pi}{Ns_2}}\exp\left(Ns_0-\frac{k^2}{4Ns_2}\right)(-4Ns_2)^{-i}H_{2i}\left(\frac{k}{2\sqrt{Ns_2}}\right)\,,
\eea
where $H_n$ are the Hermite polynomials defined as
\be\label{hermite def}
H_n(x)\equiv (-1)^ne^{x^2}\frac{d^n}{dx^n}e^{-x^2}\,.
\ee
The integral \eqref{Zk cos int} can now be evaluated as 
\bea
Z_k&\approx \int_{-\infty}^\infty du\, \cos(ku)\,e^{N(s_0 - s_2 u^2)}\left(1+\sum_{i=2}^\Lambda z_{2i}(s)\, u^{2i}\right)\\
&=\sqrt{\frac{\pi}{Ns_2}}\exp\left(Ns_0-\frac{k^2}{4Ns_2}\right)\left[1+\sum_{i=2}^\Lambda z_{2i}(s)(-4Ns_2)^{-i}H_{2i}\left(\frac{k}{2\sqrt{Ns_2}}\right)\right]\,.
\eea
Numerically, it turns out that the series above converges rather quickly and further corrections are negligible once $\Lambda$ is sufficiently large. The result is presented in Section \ref{subsec: Zk ED}. Let us emphasize that as is evident from the derivation, in this computation of direct integration no contour deformation is involved; we simply perform the original integral over $u$ explicitly. 

\subsection{Saddle point method}\label{subsec: analy cont}

As we saw in Section \ref{subsec: naive saddle}, the naive saddle point equation in $u$ cannot be solved for real $u$ and one is compelled to search for complex saddles. We return to~\eqref{det negl}
\be\label{Zk analy cont}
Z_k\approx\int_{-\infty}^{\infty}du\,\exp \left(N\left\{-i\kappa u- \wt S[u,G_0(u),\Sigma_0(u)]\right\}\right)\,,
\ee
and attempt to look for solutions to the saddle point equation at leading order in $N$
\be\label{u saddle cmplx u}
\kappa = \frac{G_0(0^+;u)+G_0(0^-;u)}{2}=\frac{G_0(0^+;u)-G_0(1^-;u)}{2}\,.
\ee
The second equality uses the KMS condition \eqref{KMS condition sd}, which still holds for complex $u$. Since $u$ is generically complex, we do not run into an immediate contradiction with the properties of $G_0$ for real $u$ like in \eqref{u saddle eqn}. For consistency, we need to find values of $u$ such that the right hand side of \eqref{u saddle cmplx u} is real. 

One obvious possibility is that $u = -i\beta\mu$ is purely imaginary, and we are back in the case where $G_0$ solves the Schwinger-Dyson equations in the presence of a chemical potential. In this case, $G_0(\tau;-i\beta\mu)$ is real and~\eqref{u saddle cmplx u} can be solved consistently, so we indeed have a saddle point. One can additionally observe that for a given $\kappa$, we find only one numerical solution to the Schwinger-Dyson equations $G_0(\tau;\kappa)$ at a corresponding value of $\beta\mu(\kappa)$. This solution corresponds to a single saddle point on the imaginary axis, and we shall see in Section \ref{subsec: Zk ED} that the contribution from this saddle agrees well with the results from ED.

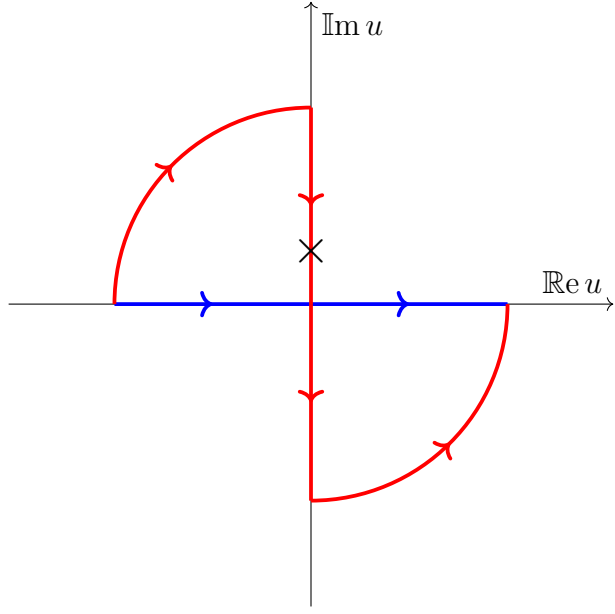
\begin{figure}[h]
\centering
\begin{tikzpicture}
\begin{scope}[decoration={
    markings,
    mark=at position 0.5 with {\arrow{>}}}]
    \draw [->] (-4,0) -- (4,0) node [above left]  {$\re u$};
    \draw [->] (0,-4) -- (0,4) node [below right] {$\im u$};
    \draw[line width=0.5mm, blue, postaction={decorate}] (-2.6,0) -- (0,0);
    \draw[line width=0.5mm, blue, postaction={decorate}] (0,0) -- (2.6,0);
    \draw[line width=0.5mm, red, postaction={decorate}](180:2.6) arc[start angle = 180, end angle = 90,radius = 2.6 cm];
    \draw[line width=0.5mm, red, postaction={decorate}] (0,2.6) -- (0,0);
    \draw[line width=0.5mm, red, postaction={decorate}] (0,0) -- (0,-2.6);
    \draw[line width=0.5mm, red, postaction={decorate}](270:2.6) arc[start angle = 270, end angle = 360,radius = 2.6 cm];
    \draw (0,0.7) node {\resizebox{0.5cm}{!}{$\times$}};
\end{scope}
\end{tikzpicture}
\caption{Contours of integration in the complex $u$ plane. The blue contour is the original contour along the real axis from $-\infty$ to $\infty$, while the red contour is the target contour that is reached via a continuous deformation. The red contour passes through the saddle point for some fixed $\kappa$ on the imaginary axis, indicated by a black cross, and is made up of the imaginary axis and two circular arcs. It should be understood that the arcs are actually infinitely far away from the origin, and are only schematically drawn at a finite distance.}
\label{fig: contour def}
\end{figure}

Importantly, as noted below \eqref{u saddle eqn}, $\wt S$ only depends on $u$ through its \emph{explicit} appearance in the first two terms of \eqref{on-shell S after reg}, and not through how it enters \emph{implicitly} in $G_0$ and $\Sigma_0$. Consequently, $\wt S$ is holomorphic in $u$. Next, we argue that the determinant factor $\Delta(G_0)^{-\frac{1}{2}}$ is also a holomorphic function of $u$. Recall that $\Delta(G_0)$ is the determinant of the operator in the quadratic action for $\delta G$ in \eqref{S quad expansion}. This operator depends holomorphically on $G_0$, and we shall show that $G_0$ is holomorphic in $u$. Firstly, the holomorphicity of $G_0$ would be consistent with the Schwinger-Dyson equations. Supposing that $G_0(\tau)$ is holomorphic in $u$ for all $\tau$, the second equation of \eqref{full SD eqn} implies that the same is true of $\Sigma_0(\tau)$ and $\Sigma_0(\omega_n)$. The first equation of \eqref{full SD eqn} is then consistent with the assumption that $G_0(\tau)$ and $G_0(\omega_n)$ are holomorphic. In addition, we can imagine tuning $\beta J$ from $0$ to some finite value. At $\beta J=0$, the free solution for $G_0$ is $-e^{iu\tau-\frac{iu}{2}}$, which is manifestly holomorphic in $u$. As $\beta J$ is increased, we expect this holomorphicity to persist. This completes the argument. Consequently, the integral \eqref{Zk analy cont} can be analysed using the conventional saddle point method. Namely, we can deform the contour of integration such that it passes through the saddle point on the imaginary axis. An additional requirement, coming from the applicability of the saddle point analysis, is that the exponent of the integrand in \eqref{Zk analy cont} should have a constant imaginary part on the deformed contour. As such, the imaginary axis $u\in (-i\infty, i\infty)$ would work, because the exponent at any point $u = -i\beta\mu$ is      
\bea
N\left\{-\kappa\beta\mu+\log\left(2\cosh\frac{\beta\mu}{2}\right)+\sum_{n\in\bZ}\log\left[1+\frac{\Sigma_0(\omega_n)}{2\pi i\left(n+\frac{1}{2}\right)-\beta\mu}\right]+\frac{q-1}{q}\sum_{n\in\bZ}\Sigma_0(\omega_n)G_0(\omega_n)\right\}\,.
\eea
This is real because the first two terms are manifestly real and the complex conjugate of the remaining two terms are
\bea
&\sum_{n\in\bZ}\log\left[1+\frac{\Sigma_0(-\omega_n)}{-2\pi i\left(n+\frac{1}{2}\right)-\beta\mu}\right]+\frac{q-1}{q}\sum_{n\in\bZ}\Sigma_0(-\omega_n)G_0(-\omega_n)\\
&=\sum_{n\in\bZ}\log\left[1+\frac{\Sigma_0(\omega_n)}{2\pi i\left(n+\frac{1}{2}\right)-\beta\mu}\right]+\frac{q-1}{q}\sum_{n\in\bZ}\Sigma_0(\omega_n)G_0(\omega_n)\,.
\eea
Since $G_0(\tau)$ and $\Sigma_0(\tau)$ are real for imaginary $u$, we have $G_0(\omega_n)^*=G_0(-\omega_n)$ and $\Sigma_0(\omega_n)^*=\Sigma_0(-\omega_n)$. The equality above follows from a relabelling $n' = -n-1$. Therefore, we can perform the deformation shown in Figure \ref{fig: contour def}, where the original blue contour along the real axis is reshaped into the red contour, consisting of two arcs at infinity and the entire imaginary axis. Numerically, it can be checked that the real part of the exponent in \eqref{Zk analy cont} decays to $-\infty$ for large $|u|$. The contribution from the arcs at infinity can therefore be neglected, and one is left with an integral along the imaginary axis.      

The remaining integral has a real integrand and can be analysed using Laplace's method. To leading order in $N$, the contribution of the a formentioned saddle point $G_0(\tau)=G_0(\tau;\kappa)$ at $u=-i \beta\mu(\kappa)$ is
\bea\label{Zk laplace}
Z_k &\approx\int_{-\infty}^{\infty}d(\beta\mu)\,\frac{1}{\sqrt{\Delta(G_0(-i\beta\mu))}}\exp \left(N\left\{-\kappa \beta\mu- \wt S[-i\beta\mu,G_0(-i\beta\mu),\Sigma_0(-i\beta\mu)]\right\}\right)\\
&\approx\frac{1}{\sqrt{-\partial_u^2\wt S(\kappa)}}\frac{1}{\sqrt{\Delta(G_0(\kappa))}}\exp \left(N\left\{-\kappa \beta\mu(\kappa)- \wt S[-i\beta\mu(\kappa),G_0(\kappa),\Sigma_0(\kappa)]\right\}\right)\,,\\
&\approx\exp \left(N\left\{-\kappa \beta\mu(\kappa)- \wt S[-i\beta\mu(\kappa),G_0(\kappa),\Sigma_0(\kappa)]+\cO(N^{-1})\right\}\right)\,,\\
&\hspace{3cm} \kappa =\frac{G_0(0^+;\kappa)-G_0(1^-;\kappa)}{2}\,.
\eea
In the second line, $\partial_u^2\wt S(\kappa)$ denotes the coefficient at second order in $u$ when expanding $\wt S$ around $u = -i\beta\mu(\kappa)$. In the third line, we used the fact that to leading order in $N$, the contribution of $(\partial_u^2\wt S)^{-\frac{1}{2}}(\Delta(G_0))^{-\frac{1}{2}}$ to $\log Z_k/N$ can be neglected. In practice, we generate data for $\log(Z_k)/N$ as a function of $\kappa$ by numerically solving the Schwinger-Dyson equations for a large range of values of $u=-i\beta\mu$. For each value of $\beta\mu$, the associated value of $\kappa$ is computed using $(G_0(0^+)-G_0(1^-))/2$ and $\log(Z_k)/N$ is computed via \eqref{Zk laplace}. The result of this computation will be exhibited in the following subsection and compared with the data from exact diagonalization.

\subsection{Comparison with exact diagonalization}\label{subsec: Zk ED}

\begin{figure}[htbp]
\centering
\includegraphics[width=0.8\textwidth]{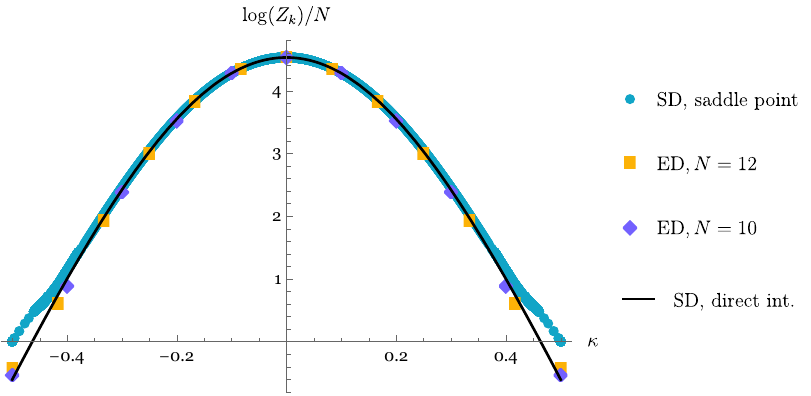}
\caption{Plot of $\log(Z_k)/N$ against $\kappa = k/N$, computed in $3$ ways. The diamond and square shaped points are computed using exact diagonalization for $N = 10$ and $12$ fermions respectively and $\beta J = 50$. We used $4000$ samples for $N = 10$ and $200$ samples for $N =12$, which turned out to be sufficient for convergence. The black line is obtained via the direct integration method in Section \ref{subsec: Zk direct int} with cutoff $\Lambda = 50$, while the blue points are obtained via the saddle point method of Section \eqref{subsec: analy cont}. In both methods, we used a discretisation of $10^{12}$ points when solving the Schwinger-Dyson equations.}
\label{fig: Zk result}
\end{figure}

As seen from \eqref{Zk sum Q}, the integral over $u$ restricts the partition function to the charge $k$ sector, namely
\be
Z_k = \sum_{Q=-\frac{N}{2}}^{\frac{N}{2}}\int_{-\pi}^\pi du\, e^{iu(Q-k)}\Tr_{\cH_Q}e^{-\hat{H}_0}=\Tr_{\cH_k}e^{-\hat{H}_0}\,.
\ee
To compute $Z_k$, we generate a particular set of couplings from the distribution \eqref{variance couplings} and construct $\hat{H}_0$ as a matrix acting on the basis of the charge $k$ Hilbert space $\cH_k$ shown in \eqref{Q subspace}. After diagonalising this matrix representing the action of $\hat{H}_0$, we sum the exponentials of minus the eigenvalues to obtain $Z_k$ for a particular instance. The result is then averaged over a sufficiently large number of instances such that including additional samples in the average does not change its value significantly. Unless stated otherwise, we will focus on the $q = 4$ model with quartic interactions. The result is shown in Figure \ref{fig: Zk result} for $N=10,12$ and $\beta J=50$.  

The results of the direct integration method in Section \ref{subsec: Zk direct int}, and the saddle point method in Section \ref{subsec: analy cont} are also displayed in the same figure. The data have been shifted for all $\kappa$ by an overall constant such that they agree at $\kappa = 0$. This is allowed because the normalisation of $Z_k$ is an ambiguity. One observes that the results agree with each other rather well, except for values of $\kappa$ near $\pm\frac{1}{2}$. We expect that this difference is due to the limited number of fermions $N=10,12$ used during exact diagonalization. It might also be due to finite size effects coming from the limited number of discretisation points used in solving the Schwinger-Dyson equations.  It might also because of the saddle point results becoming less accurate near the edge of the allowed region of $\kappa$. 

\section{The 2-point function}\label{sec: 2 pt fn}

In general, we want to consider observables that are amenable to computations in the annealed approximation and the large $N$ limit. At the very least, they must be expressed in terms of the bilocal field $G$ as defined in \eqref{id bilocal}. Clearly, other than the partition function, the next simplest observable would naively be the expectation value of $\frac{1}{N}\psi_i^\dagger(0)\psi_i(\tau)$. However, this operator transforms as 
\be\label{2 pt op gauge transf}
\frac{1}{N}\psi_i^\dagger(0)\psi_i(\tau)\mapsto \frac{1}{N}e^{i(\alpha(\tau)-\alpha(0))}\psi_i^\dagger(0)\psi_i(\tau)
\ee
under a gauge transformation with parameter $\alpha$. Since it is not gauge invariant, it is not a consistent operator insertion in the path integral of the gauged theory. However, one can attach a Wilson line between $\psi_i$ and $\psi_i^\dagger$ to form
\be\label{full gauge inv 2pt op}
\frac{1}{N}\psi_i^\dagger(0)\exp \left[i\int_\tau^0 d\tau' A_{\tau'}(\tau')\right]\psi_i(\tau)\,,
\ee
which is now gauge invariant. After the gauge fixing \eqref{gauge fixing}, this reduces to\footnote{In principle, we could have used a Wilson line which winds $n$ times around $S^1$, i.e. in \eqref{full gauge inv 2pt op} we attach $\exp\left[i\int_{\tau+n}^0 d\tau' A_{\tau'}(\tau')\right]$ instead, which reduces to $e^{-iu\tau - inu}$ after gauge fixing. Comparing with the Wilson line contribution $e^{-iku}$ in \eqref{ferm path integral}, we observe that the extra term $e^{-inu}$ can be absorbed as a shift of the charge $k\mapsto k+n$, and thus constructing the two-point function using a Wilson line with winding is not distinct from \eqref{full gauge inv 2pt op}. }
\be\label{gauge fixed 2pt insertion}
\frac{1}{N}e^{-iu\tau}\psi_i^\dagger(0)\psi_i(\tau)\,.
\ee
The normalised expectation value of \eqref{full gauge inv 2pt op} is therefore computed as
\bea
G_k(\tau)&\equiv\frac{\int_{-\pi}^\pi du\,e^{-iu(\tau+k)}\int\cD[\psi^\dagger]\cD[\psi]\frac{1}{N}\psi_i^\dagger(0)\psi_i(\tau)e^{- S[u,\psi,\psi^\dagger]}}{\int_{-\pi}^\pi du\,e^{-iku}\int\cD[\psi^\dagger]\cD[\psi]e^{- S[u,\psi,\psi^\dagger]}}\\
&=\frac{1}{Z_k}\int_{-\pi}^\pi du\,e^{-iu(\tau+k)}\wt G(\tau)\,,
\eea
where $\wt G$ was defined in \eqref{2pt fn def}. Extracting the $u$ dependence in \eqref{2pt fn def} by expanding in charge sectors, we observe that for $\tau > 0$,
\bea\label{Gk hamiltonian expr}
G_k(\tau)&=-\frac{1}{NZ_k}\int_{-\pi}^\pi du\,e^{-iu(\tau+k)}\Tr_{\cH}\left[e^{-\hat{H}}\hat{\psi}_i(\tau)\hat{\psi}_i^\dagger(0)\right]\\
&=-\frac{1}{NZ_k}\int_{-\pi}^\pi du\,e^{-iu(\tau+k)}\Tr_{\cH}\left[e^{(\tau-1)\hat{H}}\hat{\psi}_i(0)e^{-\tau\hat{H}}\hat{\psi}_i^\dagger(0)\right]\\
&=-\frac{1}{NZ_k}\int_{-\pi}^\pi du\,e^{-iu(\tau+k)}\sum_{Q=-\frac{N}{2}}^{\frac{N}{2}}e^{iu(Q+\tau)}\Tr_{\cH_Q}\left[e^{(\tau-1)\hat{H}_0}\hat{\psi}_i(0)e^{-\tau\hat{H}_0}\hat{\psi}_i^\dagger(0)\right]\\
&=-\frac{1}{NZ_k}\Tr_{\cH_k}\left[e^{(\tau-1)\hat{H}_0}\hat{\psi}_i(0)e^{-\tau\hat{H}_0}\hat{\psi}_i^\dagger(0)\right]\,.
\eea
This is nothing but the 2-point function of the \emph{ungauged} theory, restricted to states of charge $k$. From \eqref{Gk hamiltonian expr}, it also follows that the boundary values of $G_k$ are  
\bea\label{Gk bdy vals}
G_k(0^+)&=-\frac{1}{NZ_k}\Tr_{\cH_k}\left[e^{-\hat{H}_0}\hat{\psi}_i(0)\hat{\psi}_i^\dagger(0)\right]=-\frac{1}{NZ_k}\Tr_{\cH_k}\left[e^{-\hat{H}_0}\left(\frac{N}{2}-\hat{Q}\right)\right]=\kappa-\frac{1}{2}\,,\\
G_k(1^-)&=-\frac{1}{NZ_k}\Tr_{\cH_k}\left[\hat{\psi}_i(0)e^{-\hat{H}_0}\hat{\psi}_i^\dagger(0)\right]=-\frac{1}{NZ_k}\Tr_{\cH_{k+1}}\left[e^{-\hat{H}_0}\hat{\psi}_i^\dagger(0)\hat{\psi}_i(0)\right]\\
&=-\frac{1}{NZ_k}\Tr_{\cH_{k+1}}\left[e^{-\hat{H}_0}\left(\hat{Q}+\frac{N}{2}\right)\right]=-\frac{Z_{k+1}}{Z_k}\left(\frac{k+1}{N}+\frac{1}{2}\right)\,.
\eea
To obtain the second equality of the second line, we have inserted a complete basis of states between $\hat{\psi}_i(0)$ and $e^{-\hat{H}_0}$, and only those with charge $k + 1$ give non-vanishing contributions.

In the large $N$ saddle point approximation for the integrals over the bilocal fields, $\wt G$ can be computed like in \eqref{SD sum Z G}, which implies that
\bea
G_k(\tau)\approx\frac{1}{Z_k}\int_{-\pi}^\pi du\,\sum_{n\in\bZ}\frac{1}{\sqrt{\Delta(G_0(u,n))}}e^{-iku-S[u,G_0(u,n),\Sigma_0(u,n)]}e^{-iu\tau}G_0(\tau;u,n)\,.
\eea
Here $G_0(\tau;u,n)$ are the infinite collection of solutions labelled by $n\in\bZ$, which have been introduced in Section \ref{sec:num}. Following the same steps as \eqref{Zk u int}, we work with the gauge transformed solutions \eqref{gt sols branch 0} when $n\neq 0$ and rewrite  
\bea\label{Gk int u inf}
G_k(\tau)&\approx \frac{1}{Z_k}\int_{-\pi}^\pi du\sum_{n\in\bZ}\frac{1}{\sqrt{\Delta(G_0(u-2\pi n,0))}}e^{-ik(u-2\pi n)-S[u-2\pi n,G_0(u-2\pi n,0),\Sigma_0(u-2\pi n,0)]}\\
&\hspace{6.5cm}\times e^{-i(u-2\pi n)\tau}G_0(\tau;u-2\pi n,0)\\
&=\frac{1}{Z_k}\sum_{n\in\bZ}\int_{-(2n+1)\pi}^{-(2n-1)\pi}du\,\frac{1}{\sqrt{\Delta(G_0(u,0))}}e^{-iku-S[u,G_0(u,0),\Sigma_0(u,0)]}e^{-iu\tau}G_0(\tau;u,0)\\
&=\frac{1}{Z_k}\int_{-\infty}^{\infty}du\,\frac{1}{\sqrt{\Delta(G_0(u,0))}}e^{-iku-S[u,G_0(u,0),\Sigma_0(u,0)]}e^{-iu\tau}G_0(\tau;u,0)
\eea
As in \eqref{Zk u int}, we used the fact that the integrand is gauge invariant in the first line. 

Similarly to the partition function, the integral in \eqref{Gk int u inf} is computed either by approximating the integrand with a simpler fitted function, or by using the saddle point approximation in $u$. We will describe the methods in turn in the next subsections.

\subsection{Direct integration via fitting}\label{subsec: Gk direct int}

Like in~\eqref{det negl}, we do not compute the determinant factor $\Delta(G_0)^{-\frac{1}{2}}$ that is subleading in $1/N$
\be\label{Gk direct int neg det}
G_k(\tau)\approx \frac{1}{Z_k}\int_{-\infty}^{\infty}du\,e^{-i(k+\tau)u-S[u,G_0(u),\Sigma_0(u)]}G_0(\tau;u)\,.
\ee
Next, we split $G_0$ into its real and imaginary parts, and use the fact that $\re G_0$ and $\im G_0$ are even and odd functions of $u$ respectively, as proven in \eqref{G cc re and im}:
\bea\label{Gk re im split}
G_k(\tau)\approx &\frac{1}{Z_k}\int_{-\infty}^\infty du\,\cos[(k+\tau)u]\re G_0(\tau; u)\,e^{-S[u,G_0(u),\Sigma_0(u)]}\\
&~+\frac{1}{Z_k}\int_{-\infty}^\infty du\,\sin[(k+\tau)u]\im G_0(\tau; u)\,e^{-S[u,G_0(u),\Sigma_0(u)]}\,.
\eea
For consistency, the normalization factors of $Z_k$ above will also be computed by approximating the integrand and integrating directly as in Section \ref{subsec: Zk direct int}. At this point, it is possible to perform a check that the expression in \eqref{Gk re im split} is consistent with the boundary values of $G_k$ derived in \eqref{Gk bdy vals}. Using the boundary values of $\re G_0$ and $\im G_0$ in \eqref{reG bdy} and \eqref{imG S deriv} respectively,
\bea\label{Gk direct int 0 bdy}
G_k(0^+) &\approx -\frac{1}{2}+\frac{1}{NZ_k}\int_{-\infty}^{\infty}du\,\sin(ku)\,\partial_uS\,e^{-S[u,G_0,\Sigma_0]}\\
&= -\frac{1}{2}-\frac{1}{NZ_k}\sin(ku)e^{-S[u,G_0,\Sigma_0]}\bigg|_{u=-\infty}^{u=\infty}+\frac{k}{NZ_k}\int_{-\infty}^\infty du\,\cos(ku)\,e^{-S[u,G_0,\Sigma_0]}\\
&=\kappa-\frac{1}{2}\,.
\eea
After integrating by parts in the second line, the boundary time vanishes since $S\rightarrow\infty$ as $u\rightarrow\pm\infty$, which can be checked numerically. Similarly,
\bea\label{Gk direct int 1 bdy}
G_k(1^-) &\approx -\frac{1}{2Z_k}\int_{-\infty}^\infty du\,\cos[(k+1)u]\,e^{-S[u,G_0,\Sigma_0]}-\frac{1}{NZ_k}\int_{-\infty}^\infty du\,\sin[(k+1)u]\partial_uSe^{-S[u,G_0,\Sigma_0]}\\
&=-\frac{Z_{k+1}}{2Z_k}+\frac{1}{NZ_k}\sin[(k+1)u]e^{-S[u,G_0,\Sigma_0]}\bigg|_{u=-\infty}^{u=\infty}-\frac{k+1}{NZ_k}\int_{-\infty}^\infty du\,\cos[(k+1)u]\,e^{-S[u,G_0,\Sigma_0]}\\
&=-\frac{Z_{k+1}}{Z_k}\left(\frac{k+1}{N}+\frac{1}{2}\right)\,.
\eea
Similarly to how $Z_k$ was computed in Section \ref{subsec: Zk direct int}, we generate numerical solutions $G_0$ for a large range $u\in(-200\pi,200\pi)$. For each fixed value of $\tau$ in \eqref{discrete tau}, discretised over $M=10^{12}$ points, we fit $\re G_0(\tau;u)$ and $\im G_0(\tau; u)$ to even and odd polynomials in $u$ respectively, with degrees $2\Lambda$, $2\Lambda +1$, $\Lambda\in\bN$. We also fit the numerical data for $\wt S = S/N$ to an even polynomial of degree $2\Lambda$, like in \eqref{wt S poly fit}. The integrands in \eqref{Gk re im split} can then be approximated by series of the form
\bea\label{re im G poly fit}
&\cos[(k+\tau)u]\re G_0(\tau; u)\,e^{-S[u,G_0(u),\Sigma_0(u)]}\\
&\approx \cos[(k+\tau)u]\left(\sum_{j=0}^{\Lambda}a_{2j}(\tau)u^{2j}\right) e^{N(s_0 - s_2 u^2)}\left(1+\sum_{i=2}^\Lambda z_{2i}\, u^{2i}\right)\\
&\approx \cos[(k+\tau)u]e^{N(s_0 - s_2 u^2)}\sum_{i=0}^\Lambda f_{2i}(\tau)u^{2i}\,,\\
&\sin[(k+\tau)u]\im G_0(\tau; u)\,e^{-S[u,G_0(u),\Sigma_0(u)]}\\
&\approx \sin[(k+\tau)u]\left(\sum_{j=0}^{\Lambda}b_{2j}(\tau)u^{2j+1}\right) e^{N(s_0 - s_2 u^2)}\left(1+\sum_{i=2}^\Lambda z_{2i}\, u^{2i}\right)\\
&\approx \sin[(k+\tau)u]e^{N(s_0 - s_2 u^2)}\sum_{i=0}^\Lambda g_{2i}(\tau)u^{2i+1}\,.
\eea
We emphasize that the coefficients $a_{2i}$ and $b_{2j}$ of the polynomial fits to $\re G_0$ and $\im G_0$ depend on $\tau$. Using the identities
\bea
&\int_{-\infty}^\infty du\, u^{2i}\cos[(k+\tau)u]\exp\left[N(s_0-s_2 u^2)\right]=(-1)^i\frac{d^{2i}}{dk^{2i}}\sqrt{\frac{\pi}{Ns_2}}\exp\left[Ns_0-\frac{(k+\tau)^2}{4Ns_2}\right]\\
&=\sqrt{\frac{\pi}{Ns_2}}\exp\left[Ns_0-\frac{(k+\tau)^2}{4Ns_2}\right](-1)^i(2\sqrt{Ns_2})^{-2i}H_{2i}\left(\frac{k+\tau}{2\sqrt{Ns_2}}\right)\,,\\
&\int_{-\infty}^\infty du\, u^{2i+1}\sin[(k+\tau)u]\exp\left[N(s_0-s_2 u^2)\right]=(-1)^{i+1}\frac{d^{2i+1}}{dk^{2i+1}}\sqrt{\frac{\pi}{Ns_2}}\exp\left[Ns_0-\frac{(k+\tau)^2}{4Ns_2}\right]\\
&=\sqrt{\frac{\pi}{Ns_2}}\exp\left[Ns_0-\frac{(k+\tau)^2}{4Ns_2}\right](-1)^i(2\sqrt{Ns_2})^{-2i-1}H_{2i+1}\left(\frac{k+\tau}{2\sqrt{Ns_2}}\right)\,,
\eea
where $H_n$ are the Hermite polynomials in \eqref{hermite def}, we can finally compute \eqref{Gk re im split} as
\bea\label{Gk direct int result}
G_k(\tau)\approx\frac{1}{Z_k}\sqrt{\frac{\pi}{Ns_2}}\exp\left[Ns_0-\frac{(k+\tau)^2}{4Ns_2}\right]\sum_{i=0}^\Lambda&(-1)^i(2\sqrt{Ns_2})^{-2i}\bigg[f_{2i}(\tau)H_{2i}\left(\frac{k+\tau}{2\sqrt{Ns_2}}\right)\\
&+\frac{1}{2\sqrt{Ns_2}}g_{2i}(\tau)H_{2i+1}\left(\frac{k+\tau}{2\sqrt{Ns_2}}\right)\bigg]\,.
\eea
The result will be presented in Section \ref{subsec: 2pt ED}.

\subsection{Saddle point method}\label{subsec: Gk saddle pt}

We return to the integral
\bea\label{Gk u int}
G_k(\tau)&\approx\frac{1}{Z_k}\int_{-\infty}^{\infty}du\,
\exp\left(-i(k+\tau)u-S[u,G_0(u),\Sigma_0(u)]\right)G_0(\tau;u)\\
&=\frac{1}{Z_k}\int_{-\infty}^{\infty}du\,
\exp \left(N\left\{-i\left(\kappa+\frac{\tau}{N}\right)u-\wt S[u,G_0(u),\Sigma_0(u)]\right\}\right)G_0(\tau;u)\,,
\eea
and attempt to approximate it using the saddle point method. Deriving the exponent with respect to $u$, the saddle point equations are
\be\label{u saddle pt eqn Gk}
\kappa + \frac{\tau}{N} = \frac{G_0(0^+;u)-G_0(1^-;u)}{2}\,.
\ee
If the large $N$ limit is taken at a fixed value of $\kappa=\frac{k}{N}$, the shift by $\tau/N$ in the left hand side as compared to \eqref{u saddle cmplx u} will become negligible for sufficiently large $N$, because $\tau\in(0,1)$. However, since we will also be interested in comparing the saddle point results with finite $N$ exact diagonalization results at \emph{all} possible values of $k=-\frac{N}{2},\ldots,\frac{N}{2}$, there will always be values (eg. $k = \pm\frac{1}{2}$ for $N$ odd and $k = 0$ for $N$ even) where $\tau/N$ is not negligible. For this reason, it is kept in \eqref{u saddle pt eqn Gk}.  

As discussed in Section \ref{subsec: analy cont}, at a fixed value of $\kappa$ and $\tau$, \eqref{u saddle pt eqn Gk} has a single solution, denoted $G_0(\tau;\kappa+\frac{\tau}{N})$, at a corresponding value of $u = -i\beta\mu(\kappa+\frac{\tau}{N})$. To deform the contour to the imaginary axis, we need to make sure that the integrand of \eqref{Gk u int} is holomorphic in $u$. Recall from the discussion below \eqref{u saddle eqn} that only the explicit $u$ dependence of $\wt S$ matters, which is holomorphic. In Section \ref{subsec: analy cont}, we also argued that $G_0$ is holomorphic, from which it follows that the determinant $\Delta(G_0)$ is holomorphic. Consequently, the entire integrand of \eqref{Gk u int} holomorphic. The contour can then be deformed to the red contour in Figure \ref{fig: contour def}, where the arcs at infinity are negligible due to numerical evidence suggesting that $\wt S\rightarrow \infty$ as $|u|\rightarrow\infty$. This leaves an integral over the imaginary $u$ axis.  

Using the Laplace method for real integrals, the remaining integral can be approximated as 
\bea
&\frac{1}{Z_k}\int_{-\infty}^\infty \frac{d(\beta\mu)}{\sqrt{\Delta(G_0(-i\beta\mu))}}\exp \left(N\left\{-\left(\kappa+\frac{\tau}{N}\right)\beta\mu-\wt S[-i\beta\mu,G_0(-i\beta\mu),\Sigma_0(-i\beta\mu)]\right\}\right)G_0(\tau;-i\beta\mu)\\
&\approx\frac{1}{Z_k}\frac{1}{\sqrt{-\partial_u^2\wt S(\kappa+\frac{\tau}{N})}\sqrt{\Delta(G_0(\kappa+\frac{\tau}{N}))}}\exp \left(N\bigg\{-\left(\kappa+\frac{\tau}{N}\right)\beta\mu\left(\kappa+\frac{\tau}{N}\right)\right.\\
&\hspace{3cm}\left.-\wt S\left[-i\beta\mu\left(\kappa+\frac{\tau}{N}\right),G_0\left(\kappa+\frac{\tau}{N}\right),\Sigma_0\left(\kappa+\frac{\tau}{N}\right)\right]\bigg\}\right)G_0\left(\tau;\kappa+\frac{\tau}{N}\right)\,.
\eea
Taking the expression for $Z_k$ from the second line of \eqref{Zk laplace}, this is
\bea
&\frac{\sqrt{-\partial_u^2\wt S(\kappa)}\sqrt{\Delta(G_0(\kappa))}}{\sqrt{-\partial_u^2\wt S(\kappa+\frac{\tau}{N})}\sqrt{\Delta(G_0(\kappa+\frac{\tau}{N}))}}\exp \left(N\bigg\{-\left(\kappa+\frac{\tau}{N}\right)\beta\mu\left(\kappa+\frac{\tau}{N}\right)\right.\\
&\hspace{2cm}\left.-\wt S\left[-i\beta\mu\left(\kappa+\frac{\tau}{N}\right),G_0\left(\kappa+\frac{\tau}{N}\right),\Sigma_0\left(\kappa+\frac{\tau}{N}\right)\right]\bigg\}\right)G_0\left(\tau;\kappa+\frac{\tau}{N}\right)\bigg/\\
&\hspace{3cm}\exp \left(N\left\{-\kappa\beta\mu(\kappa)-\wt S[-i\beta\mu(\kappa),G_0(\kappa),\Sigma_0(\kappa)]\right\}\right)\,.
\eea
The factor $\frac{\sqrt{-\partial_u^2\wt S(\kappa)}\sqrt{\Delta(G_0(\kappa))}}{\sqrt{-\partial_u^2\wt S(\kappa+\frac{\tau}{N})}\sqrt{\Delta(G_0(\kappa+\frac{\tau}{N}))}}$ in this expression is unity to eading order in $1/N$  (while $\kappa$ is kept fixed) due to the following:
\bea\label{ratio det neg}
&\frac{\sqrt{-\partial_u^2\wt S(\kappa)}\sqrt{\Delta(G_0(\kappa))}}{\sqrt{-\partial_u^2\wt S(\kappa+\frac{\tau}{N})}\sqrt{\Delta(G_0(\kappa+\frac{\tau}{N}))}}=\exp\bigg[\log\left(\sqrt{-\partial_u^2\wt S(\kappa)}\sqrt{\Delta(G_0(\kappa))}\right)\\
&-\log\left(\sqrt{-\partial_u^2\wt S(\kappa+\frac{\tau}{N})}\sqrt{\Delta(G_0(\kappa+\frac{\tau}{N}))}\right)\bigg]=\exp\bigg[ -\partial_\kappa\log\left(\sqrt{-\partial_u^2\wt S(\kappa)}\sqrt{\Delta(G_0(\kappa))}\right)\frac{\tau}{N}+\cO(N^{-2})\bigg]\\
&=1+\cO({\tau}/{N})\,.
\eea
It can then be safely neglected. We thus have
\bea\label{Gk fin saddle}
&G_k(\tau)\approx \frac{e^ {N\left\{-\left(\kappa+\frac{\tau}{N}\right)\beta\mu\left(\kappa+\frac{\tau}{N}\right)-\wt S\left[-i\beta\mu\left(\kappa+\frac{\tau}{N}\right),G_0\left(\kappa+\frac{\tau}{N}\right),\Sigma_0\left(\kappa+\frac{\tau}{N}\right)\right]\right\}}}{e^{N\left\{-\kappa\beta\mu(\kappa)-\wt S[-i\beta\mu(\kappa),G_0(\kappa),\Sigma_0(\kappa)]\right\}}}G_0\left(\tau;\kappa+\frac{\tau}{N}\right)\,.
\eea
The result of this computation at generic $k$ will be presented and compared with those coming from exact diagonalization in Section \ref{subsec: 2pt ED}. It turns out a posteriori that \eqref{Gk fin saddle} reproduces the exact diagonalization result extremely accurately in these cases.~\footnote{In principle, it is legitimate to worry if this expansion is fully generic when $\kappa$ is also small, i.e. $\kappa\sim\frac{1}{N}$. But this is not a problem since the expansion in~\eqref{ratio det neg} can be regarded in terms of a small variable $\frac{\tau}{N}$. The variable $\kappa$, when small, should be considered as another independent variable with respect to which we have the freedom to expand or not. So~\eqref{ratio det neg} is a valid expansion in $\frac{\tau}{N}$ and we choose to not expand in $\kappa$. Since this is further confirmed by a comparison with the ED results, the expansion~\eqref{ratio det neg} is well-grounded. } Note that \eqref{Gk fin saddle} is consistent with the boundary values of $G_k$ in \eqref{Gk bdy vals}, since evaluating \eqref{Gk fin saddle} at $\tau = 0^+, 1^-$ gives
\bea\label{Gk saddle pt bdy}
G_k(0^+)&=G_0(0,\kappa)=\kappa-\frac{1}{2}\,,\\
G_k(1^-)&=\frac{Z_{k+1}}{Z_k}G_0(1^-;\kappa+\frac{1}{N})=-\frac{Z_{k+1}}{Z_k}\left(\frac{k+1}{N}+\frac{1}{2}\right)\,,
\eea
where we have used the boundary values of $G_0(0^+;\kappa)=\kappa-\frac{1}{2}$ and $G_0(1^-;\kappa)=-\frac{1}{2}-\kappa$ given in \cite{Gu:2019jub}.

In the limit of large $N$ and fixed $\kappa$, further simplifications are possible. For the ratio of exponential factors, we have
\bea
&\frac{e^{N\left\{-\left(\kappa+\frac{\tau}{N}\right)\beta\mu\left(\kappa+\frac{\tau}{N}\right)-\wt S\left[-i\beta\mu\left(\kappa+\frac{\tau}{N}\right),G_0\left(\kappa+\frac{\tau}{N}\right),\Sigma_0\left(\kappa+\frac{\tau}{N}\right)\right]\right\}}}{e^{N\left\{-\kappa\beta\mu(\kappa)-\wt S[-i\beta\mu(\kappa),G_0(\kappa),\Sigma_0(\kappa)]\right\}}}\\
&=\exp \left(N\bigg\{\left(\partial_\kappa + \frac{d\beta\mu(\kappa)}{d\kappa}\partial_{\beta\mu}\right)\left(-\kappa\beta\mu(\kappa)-\wt S[-i\beta\mu(\kappa),G_0(\kappa),\Sigma_0(\kappa)]\right)\frac{\tau}{N}+\cO(N^{-2})\bigg\}\right)\\
&=\exp(-\beta\mu(\kappa)\tau)(1+\cO(N^{-1}))\,.
\eea
In the first equality, we noticed that the exponent in the second line of \eqref{Zk laplace} depends on $\kappa$ both implicitly through $\beta\mu(\kappa)$ and explicitly in the term $-\kappa\beta\mu$. Since $\beta\mu(\kappa)$ was chosen precisely to solve the saddle point equation and extremise the exponent, the implicit dependence via $\beta\mu$ does not matter. The contribution of the explicit $\kappa$ dependence gives the remaining term in the second equality. We can also approximate the last factor in~\eqref{Gk fin saddle}
\be
G_0\left(\tau;\kappa+\frac{\tau}{N}\right)=G_0(\tau;\kappa)+\cO(N^{-1})\,.
\ee
Therefore, if $\kappa$ is kept fixed while taking the large $N$ limit,
\be\label{Gk saddle pt fixed kappa}
\lim_{\substack{N\rightarrow\infty }}G_{N\kappa}(\tau) = e^{-\beta\mu(\kappa)\tau}G_0(\tau;\kappa)\,.
\ee
The exponential factor can be traced back to the connecting Wilson line in \eqref{gauge fixed 2pt insertion}, which is required to make the operator insertion gauge invariant.

\subsection{Comparison with exact diagonalization}\label{subsec: 2pt ED}

In a fixed instance of $\hat{H}_0$, suppose that $\ket{n;k}$ is the $n^\text{th}$ energy eigenstate with charge $k$ and energy $E_{n;k}$, where $n=1,\ldots,\dim\cH_k$. Starting from the expression of the gauge invariant two-point function in the operator formalism in \eqref{Gk hamiltonian expr}, we can write it more explicitly as
\bea\label{Gk exact diag exp}
G_k(\tau)&=-\frac{1}{NZ_k}\sum_{m=1}^{\dim\cH_k}e^{(\tau-1)E_{m;k}}\bra{m;k}\hat{\psi}_i(0)e^{-\tau\hat{H}_0}\hat{\psi}_i^\dagger(0)\ket{m;k}\\
&=-\frac{1}{NZ_k}\sum_{m=1}^{\dim\cH_k}\sum_{n=1}^{\dim\cH_{k+1}}e^{(\tau-1)E_{m;k}-\tau E_{n;k+1}}\bra{m;k}\hat{\psi}_i(0)\ket{n;k+1}\bra{n;k+1}\hat{\psi}_i^\dagger(0)\ket{m;k}\\
&=-\frac{1}{N}\sum_{m=1}^{\dim\cH_k}\sum_{n=1}^{\dim\cH_{k+1}}e^{(\tau-1)E_{m;k}-\tau E_{n;k+1}}\sum_{i=1}^N|\bra{n;k+1}\hat{\psi}_i^\dagger(0)\ket{m;k}|^2\bigg/\sum_{m=1}^{\dim\cH_k}e^{-E_{m;k}}\,.
\eea
The energy eigenvalues $E_{n;k}$ and their corresponding eigenstates $\ket{n;k}$ can be computed via exact diagonalization, which then allows us to assemble $G_k$ using \eqref{Gk exact diag exp}. This result is then averaged over multiple instances until we reach a point where including further instances in the average does not change the result significantly. For $N = 12$ and $\beta J = 50$, it turns out that $200$ instances are sufficient.

\begin{figure}[H]
\centering
\includegraphics[width=0.9\textwidth]{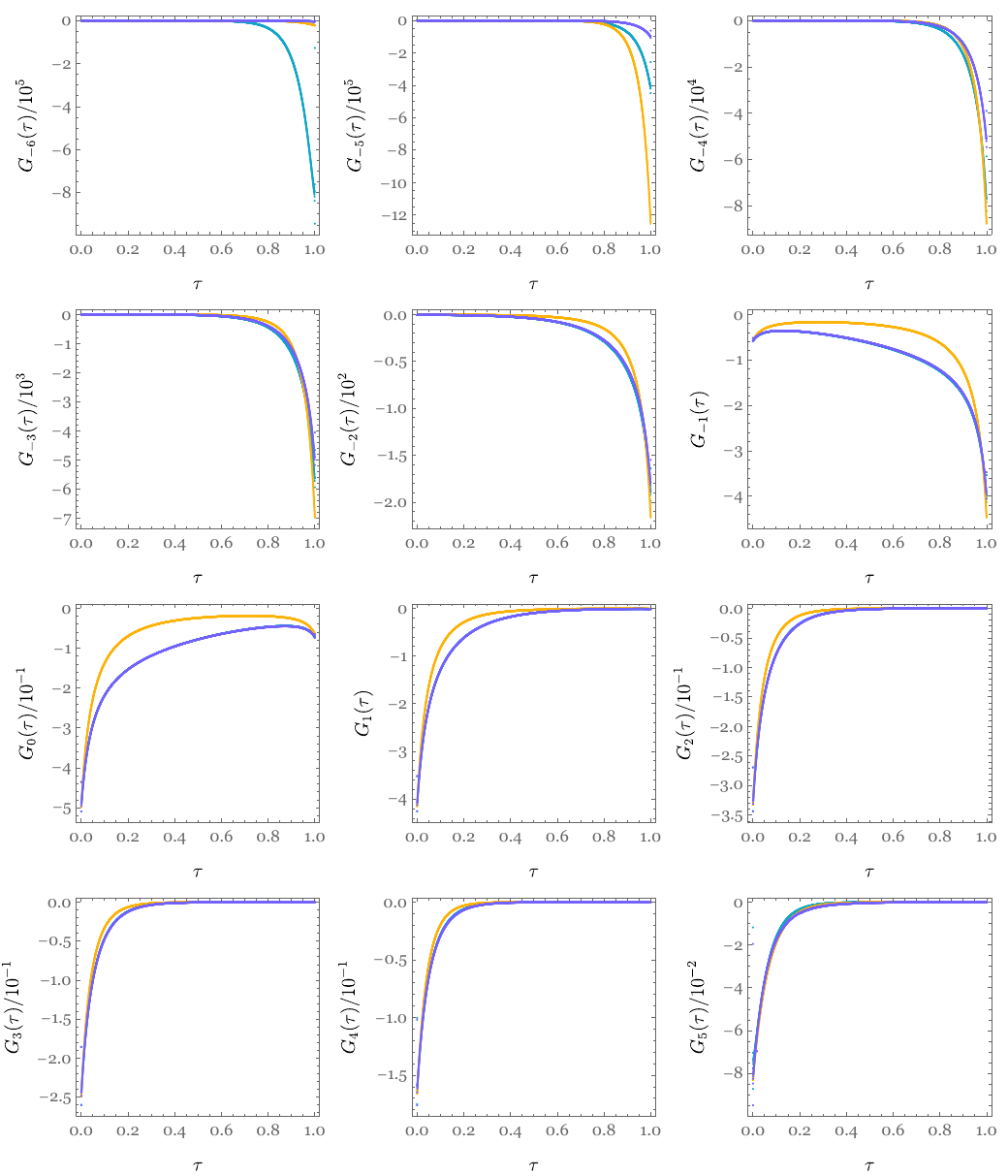}
\caption{Plots of the gauge invariant two-point function $G_k(\tau)$ at $N = 12$, and various $k=-6,\ldots,5$, computed in $3$ ways: exact diagonalization in blue, direct integration \eqref{Gk direct int result} in yellow, and the saddle point method \eqref{Gk fin saddle} in purple. The results from the saddle point analysis and exact diagonalization are almost identical for all $k$ except for $k=-4,-5,-6$. The data are generated at $M = 10^{12}$ discrete values of $\tau$, and we chose the coupling $\beta J = 50$.}
\label{fig: Gk result}
\end{figure}

In Figure \ref{fig: Gk result}, we display the averaged exact diagonalization result in blue for all possible values of the charge $k$ when $N = 12$. Note that $G_k$ is trivially zero at the maximum charge $k = N/2=6$ because the unique state with $k=N/2$ is annihilated by $\psi_i^\dagger$ in \eqref{Gk exact diag exp}. The plot for $G_6(\tau)$ is therefore omitted in the figure. In the same panels, we also exhibit the results for $G_k(\tau)$ computed using \eqref{Gk direct int result} via the direct integration method described in Section \ref{subsec: Gk direct int} (yellow), and also the result according to \eqref{Gk fin saddle}, which uses the saddle point method described in Section \ref{subsec: Gk saddle pt} (purple). Since we are considering all possible values of $k$ or $\kappa$, including $\kappa\sim\cO(N^{-1})$, we have chosen to compare the exact diagonalization results with \eqref{Gk fin saddle} instead of \eqref{Gk saddle pt fixed kappa} since the simplifications leading up to \eqref{Gk saddle pt fixed kappa} are not applicable when $\kappa\sim\cO(N^{-1})$. Note that the saddle point result agrees extremely well and almost coincides with the exact diagonalization result for most charges except $k = -6,-5,-4$. In particular, they agree even in the cases $\kappa\sim\cO(N^{-1})$.
This result verifies a posteriori our treatment that the determinant factors can be neglected. 
We expect that the discrepancies between the saddle point and exact diagonalization results near $\tau = 1$ at $k = -6,-5,-4$ are due to numerical error when computing $Z_k$, see e.g.~\eqref{Gk bdy vals}, which originates from the fact that $N$ is not sufficiently large. But we were not able to verify this due to insufficient computational resources. On the other hand, the discrepancies between the direct integration results in yellow and the exact diagonalization results are consistently larger. This is most probably due to the fact that there are many more sources of error in the direct integration method, such as those introduced when fitting $\re G_0$ and $\im G_0$ to polynomials in~\eqref{re im G poly fit}\footnote{There could be other contributions to our direct integration computation that were missed when fitting the numerical data. They could modify the two-point functions, just like how including the contribution of the Schwarzian mode significantly improved the agreement between exact diagonalization results and large $N$ saddle point results in the SYK model in~\cite{Kobrin:2020xms}.}.      

Note that despite the discrepancies between the direct integration (or saddle point) results and exact diagonalization at intermediate values of $\tau$, they agree at the boundaries $\tau = 0,1$ except for $k = -6,-5,-4$. This is due to the fact that \eqref{Gk direct int result} and \eqref{Gk fin saddle} are consistent with the boundary values \eqref{Gk bdy vals}, as shown in \eqref{Gk direct int 0 bdy}, \eqref{Gk direct int 1 bdy} and \eqref{Gk saddle pt bdy}. There are exceptions for the right boundary values at $\tau = 1$ when $k = -6,-5,-4$, but this is because the ratio $Z_{k+1}/Z_k$ appearing in $G_k(1^-)$ in \eqref{Gk bdy vals} differs greatly between the different sets of data for these value of charges. This can be seen from the plot of $\log (Z_k)/N$ in Figure \ref{fig: Zk result}, where any discrepancy must be multiplied by a large integer $N$ and exponentiated to find $Z_{k+1}/Z_k$.

\begin{figure}[htbp]
\centering
\includegraphics[width=0.8\textwidth]{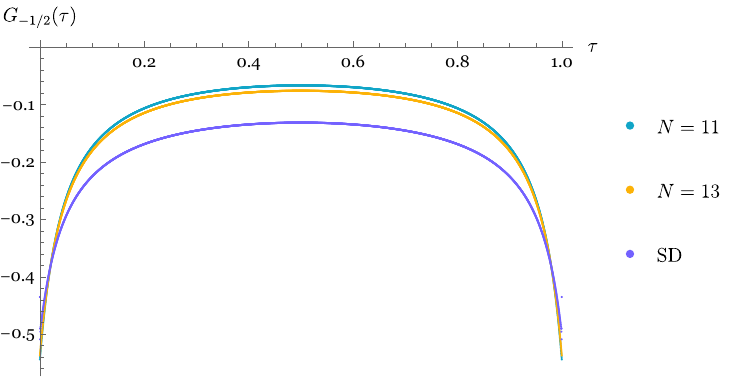}
\caption{Plots of $G_{-\frac{1}{2}}(\tau)$ computed using exact diagonalization at $\beta J = 50$ for $N = 11$ (blue) and $N = 13$ (orange). The data for $N = 11$ is the result of averaging over $1000$ instances, while $100$ instances were used for $N = 13$. The Schwinger-Dyson solution $G_0(\tau;0)$ is plotted in purple.}
\label{fig: kappa0}
\end{figure}

We also test the prediction \eqref{Gk saddle pt fixed kappa} from the saddle point method, which is valid at fixed $\kappa$ and large $N$. For instance, we expect that for $N$ odd,
\be
\lim_{N\rightarrow\infty}G_{-\frac{1}{2}}(\tau) = \lim_{N\rightarrow\infty}G_{N\left(-\frac{1}{2N}\right)}(\tau)=\lim_{N\rightarrow\infty}G_{N\kappa}(\tau)\Big|_{\kappa = 0} = G_0(\tau;0)\,,
\ee
where the last equality follows from \eqref{Gk saddle pt fixed kappa} and the fact that $\beta\mu(0)=0$. We chose to compare the Schwinger-Dyson solution $G_0(\tau;0)$ with the exact diagonalization results of $G_{-\frac{1}{2}}(\tau)$ for large odd integers $N$, instead of $G_{k=0}(\tau)$ at large even $N$. This is because $Z_{-\frac{1}{2}}=Z_{\frac{1}{2}}$ due to charge conjugation invariance, and as a consequence the boundary values of $G_{-\frac{1}{2}}(\tau)$ are equal at $0^+$ and $1^-$, as seen from \eqref{Gk bdy vals}. This is more consistent with the symmetry of the Schwinger-Dyson solution $G_0(\tau;0)$ under the reflection $\tau\mapsto 1-\tau$, but in principle $\lim_{N\rightarrow \infty}G_{k=0}(\tau)$ is equally valid. The asymmetry of $G_{k=0}(\tau)$ vanishes in the large $N$ limit. In Figure \ref{fig: kappa0}, we plot the exact diagonalization results for $G_{-\frac{1}{2}}(\tau)$ at $N = 11, 13$, as well as the Schwinger-Dyson solution $G_0(\tau;0)$. The data is consistent with the expectation that $G_{-\frac{1}{2}}(\tau)$ will slowly converge to the Schwinger-Dyson solution $G_0(\tau;0)$ at large $N$. Since the latter is well approximated by the conformal ansatz \eqref{conf ansatz full} at $\tau\gg(\beta J)^{-1}$, we deduce that the theory is conformal at large $\beta J$ in the zero charge sector.   

\begin{figure}[htbp]
\centering
\includegraphics[width=0.8\textwidth]{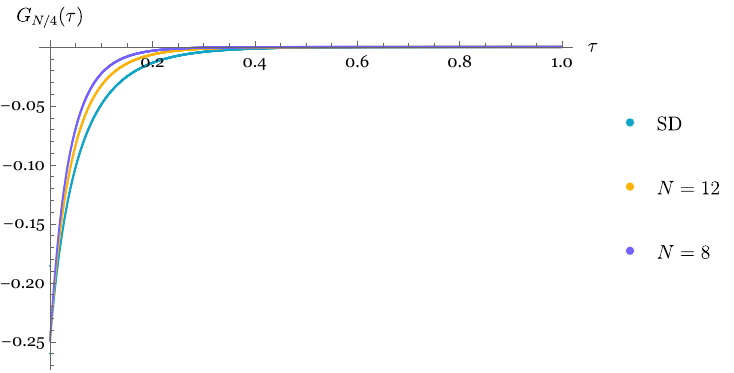}
\caption{Plots of $G_{N\kappa}(\tau)$ computed using exact diagonalization at $\beta J = 50$ for $N = 8$ (blue) and $N = 12$ (orange). The data for $N = 8$ is the result of averaging over $8000$ instances, while $200$ instances were used for $N = 12$. The expected large $N$ limit $e^{-\beta\mu(1/4)}G_0(\tau;1/4)$ is plotted in purple.}
\label{fig: kappa1d4}
\end{figure}

We also test \eqref{Gk saddle pt fixed kappa} for a nonzero value of $\kappa = \frac{1}{4}$. In Figure \ref{fig: kappa1d4}, we plot $G_{N/4}(\tau)$ computed using exact diagonalization for $N = 8$ and $N = 12$, as well as the right hand side of \eqref{Gk saddle pt fixed kappa}, which is the expected large $N$ limit. One again observes that the exact diagonalization results seem to converge slowly to the expected large $N$ limit. 

Unlike the case of $\kappa = 0$, the large $N$ limit of the two-point function at $\kappa = \frac{1}{4}$ is not well approximated by the conformal ansatz. This can be traced to the fact that although $G_0(\tau;1/4)$ in the right hand side of \eqref{Gk saddle pt fixed kappa} is approximately conformal at large $\beta J$, the exponential factor $e^{-\beta\mu(1/4)}$ spoils this. We therefore deduce that in the large $N$ limit, the theory is not conformal in the sector with charge $k=N\kappa$, $\kappa\neq 0$. The fact that the theory is not conformal at nonzero charge can be understood schematically from the observation that starting from the zero charge sector, the operator $\hat{Q}$ has dimension $\sim\frac{2}{q}$, and is therefore not a marginal deformation. We note that this breakdown of conformality is distinct from that in \cite{Azeyanagi:2017drg,Ferrari:2019ogc,Heydeman:2022lse} because the observables in which it is observed is different. In these references, the breakdown of conformality was observed in the bare fermion two-point function, without any Wilson line insertion, which is computed by $G_0$. Here, we are computing the two-point function with a connecting Wilson line, and the gauge field is dynamical. Indeed as mentioned, the contributing solution $G_0$ is approximately conformal and it is the on-shell value of the Wilson line that gives an exponential contribution which spoils conformality.

\section{Operator dimensions}\label{sec: op dim}

Recall that in Section \ref{sec:num}, we found numerically that there are inequivalent solutions $G_0(\tau;u,0)$ (see \eqref{gt sols branch 0}) for the Schwinger-Dyson equations at $q = 4$ and $u\in(-\infty,\infty)$, which are well-approximated by the conformal solution \eqref{conf const fixed}. Therefore, one might be interested in the conformal dimensions of low-lying operators in the presence of the background gauge field $u$. To compute this, we follow the standard procedure used in \cite{Gu:2019jub,Heydeman:2022lse} for the spectrum in the presence of a chemical potential.

We start with the bilocal action $S$ in \eqref{wt S def} and expand it to second order in fluctuations around solutions to the Schwinger-Dyson equations. Namely, writing the bilocal fields as $G = G_0 + \delta G/N^\frac{1}{2}$ and $\Sigma = \Sigma_0 + \delta \Sigma/N^\frac{1}{2}$, we expand $S$ to second order in $\delta G$ and $\delta \Sigma$. The result is\footnote{To find the second order term in $\delta\Sigma$, we have used the identity $\frac{\partial M^{-1}_{mn}}{\partial M_{ij}}=-M^{-1}_{mi}M^{-1}_{jn}$ and the Schwinger-Dyson equations \eqref{full SD eqn}.} 
\bea
&-S[u,G,\Sigma] = -S[u,G_0,\Sigma_0] - \frac{1}{2}\int d\tau_1d\tau_2d\tau_3d\tau_4\,\delta\Sigma(\tau_1,\tau_2)G_0(\tau_{23})G_0(\tau_{41})\delta\Sigma(\tau_3,\tau_4)\\
&+\int d\tau_1d\tau_2\,\delta\Sigma(\tau_1,\tau_2)\delta G(\tau_2,\tau_1)+\frac{(\beta J)^2}{2}\int d\tau_1d\tau_2\,\bigg[\left(\frac{q}{2}-1\right)\delta G(\tau_1,\tau_2)G_0(\tau_{21})^{\frac{q}{2}}(-G_0(\tau_{12}))^{\frac{q}{2}-2}\delta G(\tau_1,\tau_2)\\
&\hspace{4cm}-\frac{q}{2}\delta G(\tau_1,\tau_2)G_0(\tau_{21})^{\frac{q}{2}-1}(-G_0(\tau_{12}))^{\frac{q}{2}-1}\delta G(\tau_2,\tau_1)\bigg]+\cO(N^{-\frac{1}{2}})\,.
\eea
Integrating out $\delta\Sigma$ to leading order in $N$, the action can be expressed in terms of a kernel operator $K$:
\bea\label{S quad expansion}
-&S[u,G,\Sigma] = - S[u,G_0,\Sigma_0] + \frac{1}{2}\int d\tau_1\cdots d\tau_6\,\delta G(\tau_5,\tau_6)G_0^{-1}(\tau_{61})G_0^{-1}(\tau_{25})\left[\delta(\tau_{13})\delta(\tau_{24})\right.\\
&\hspace{7cm}\left.-K(\tau_1,\tau_2;\tau_3,\tau_4)\right]\delta G(\tau_3,\tau_4)+\cO(N^{-1})\,,\\
&K(\tau_1,\tau_2;\tau_3,\tau_4)\equiv (-1)^{\frac{q}{2}-1}(\beta J)^2\bigg[\left(\frac{q}{2}-1\right)G_0(\tau_{14})G_0(\tau_{32})G_0(\tau_{34})^{\frac{q}{2}-2}G_0(\tau_{43})^{\frac{q}{2}}\\
&\hspace{6cm}+\frac{q}{2}G_0(\tau_{13})G_0(\tau_{42})G_0(\tau_{34})^{\frac{q}{2}-1}G_0(\tau_{43})^{\frac{q}{2}-1}\bigg]\,.
\eea
Note that the one-loop determinant $\Delta(G_0)$ appearing in \eqref{SD sum Z G} and in numerous subsequent expressions is precisely the determinant of the operator appearing between $\delta G(\tau_5,\tau_6)$ and $\delta G(\tau_3,\tau_4)$ above. Since we are expanding a gauge invariant action around $G_0$, each term at a fixed order in $\delta G$ must be separately invariant under large gauge transformations, and that includes the quadratic term above. Indeed, it is straightforward to check that \eqref{S quad expansion} is invariant under $G_0(\tau)\mapsto e^{2\pi i n\tau}G_0(\tau)$ and $\delta G(\tau_1,\tau_2)\mapsto e^{2\pi i n\tau_{12}}\delta G(\tau_1,\tau_2)$. Therefore, if a gauge invariant regularization scheme can be found, the determinant $\Delta(G_0)$ will be gauge invariant.

\begin{figure}[htbp]
\centering
\includegraphics[width=0.8\textwidth]{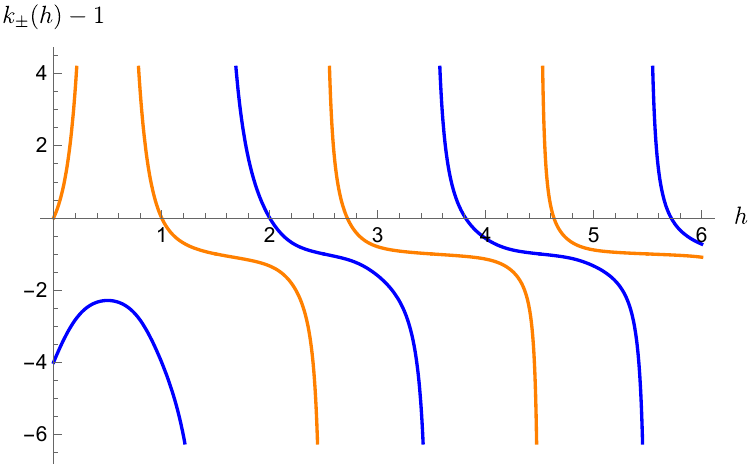}
\caption{Plots of $k_+(h;u_\text{IR})-1$ in blue, and $k_-(h;u_\text{IR})-1$ in orange against $h$, for the case of $q = 4$ and $u_\text{IR}=\frac{\pi}{4}$. The values of $h$ at the horizontal intercepts are the conformal dimensions of various operators at the given conformal solution.}
\label{fig: k minus 1}
\end{figure}

Upon substituting the conformal solution \eqref{conf const fixed}, the kernel becomes  
\bea
&K(\tau_1,\tau_2;\tau_3,\tau_4) = \frac{\pi\left(\frac{q}{2}-1\right)\sin\frac{2\pi}{q}}{\cos u_\text{IR}+\cos\frac{2\pi}{q}}\,e^{iu_\text{IR}(\tau_{12}-\tau_{34})}\left[\frac{1}{2}\frac{l(\tau_{13};u_\text{IR})l(\tau_{42};u_\text{IR})}{\sin^\frac{2}{q}(\pi|\tau_{13}|)\sin^\frac{2}{q}(\pi|\tau_{24}|)\sin^{2-\frac{4}{q}}(\pi|\tau_{34}|)}\right.\\
&\left.-\frac{1}{q}\left(\frac{q}{2}-1\right)\frac{l(\tau_{14};u_\text{IR})l(\tau_{32};u_\text{IR})\Big(\Theta(\tau_{34})e^{iu_\text{IR}}+\Theta(\tau_{43})e^{-iu_\text{IR}}\Big)}{\sin^\frac{2}{q}(\pi|\tau_{14}|)\sin^\frac{2}{q}(\pi|\tau_{23}|)\sin^{2-\frac{4}{q}}(\pi|\tau_{34}|)}\right]\,,\\
&\hspace{5cm}
l(\tau;u)\equiv \Theta(\tau)e^{-\frac{iu}{2}}-\Theta(-\tau)e^{\frac{iu}{2}}\,.
\eea
Operators in the spectrum are in one-to-one correspondence with eigenfunctions of the kernel $K$ with eigenvalue $1$. Due to the conformal symmetry of the kernel, it can be simultaneously diagonalized with the conformal casimir, whose eigenvalues are labelled by the conformal dimensions $h\in\bR$ we seek. At fixed $h$, the eigenfunctions are spanned by  
\be
f_\pm(\tau_1,\tau_2)\equiv e^{iu_\text{IR}\tau_{12}\mp\frac{iu_\text{IR}}{2}}\frac{\Theta(\pm\tau_{12})\cos^{-h}(\pi\tau_1)\cos^{-h}(\pi\tau_2)}{\sin^{\frac{2}{q}-h}(\pi|\tau_{12}|)}\,.
\ee
When computing the action of $K$ on $f_\pm$, it is sufficient to evaluate the integral
\be\label{int K}
\int_{-\frac{1}{2}}^\frac{1}{2}d\tau_3 d\tau_4\frac{l(\tau_{13};u_\text{IR})l(\tau_{42};u_\text{IR})\Theta(\tau_{34})\cos^{-h}(\pi\tau_3)\cos^{-h}(\pi\tau_4)}{\sin^\frac{2}{q}(\pi|\tau_{13}|)\sin^\frac{2}{q}(\pi|\tau_{24}|)\sin^{2-\frac{2}{q}-h}(\pi|\tau_{34}|)}\,,
\ee
or its complex conjugate. The integral can be rewritten to the form computed in~\cite{Peng:2017spg}. Here, for completeness, we write out some details explicitly. Upon the change of variables
\be
\tan\pi\tau_3 = (\tan\pi\tau_2-\tan\pi\tau_1)u + \tan\pi\tau_1\,,\quad \tan\pi\tau_4 = (\tan\pi\tau_2-\tan\pi\tau_1)v + \tan\pi\tau_1\,, 
\ee
the integral becomes
\bea
&\frac{\Theta(\tau_{12})\cos^{-h}(\pi\tau_1)\cos^{-h}(\pi\tau_2)}{\sin^{\frac{2}{q}-h}(\pi|\tau_{12}|)}\frac{1}{\pi^2}\int_{-\infty}^\infty dudv\,\frac{l(u;u_\text{IR})l(1-v;u_\text{IR})\Theta(v-u)}{|u|^\frac{2}{q}|1-v|^\frac{2}{q}|u-v|^{2-\frac{2}{q}-h}}\\
+&\frac{\Theta(\tau_{21})\cos^{-h}(\pi\tau_1)\cos^{-h}(\pi\tau_2)}{\sin^{\frac{2}{q}-h}(\pi|\tau_{12}|)}\frac{1}{\pi^2}\int_{-\infty}^\infty dudv\,\frac{l(u;-u_\text{IR})l(1-v;-u_\text{IR})\Theta(u-v)}{|u|^\frac{2}{q}|1-v|^\frac{2}{q}|u-v|^{2-\frac{2}{q}-h}}\,.
\eea
After substituting the Fourier transform
\be
\frac{\Theta(u-v)}{|u-v|^{2-\frac{2}{q}-h}}=-\int_{-\infty}^\infty\frac{d\omega}{2\pi}e^{i\omega(v-u)}i\sgn(\omega)|\omega|^{1-\frac{2}{q}-h}e^{i\pi\sgn(\omega)\left(\frac{1}{q}+\frac{h}{2}\right)}\Gamma\left(\frac{2}{q}+h-1\right)\,,
\ee
the remaining integrals can be evaluated using further Fourier transforms
\be
\int_{-\infty}^\infty d\tau e^{-i\omega\tau}\frac{l(\tau,u_\text{IR})}{|\tau|^{\frac{2}{q}}}=-2i\sgn(\omega)|\omega|^{\frac{2}{q} - 1}\Gamma\left(1-\frac{2}{q}\right)\cos\left(\frac{\pi}{q}-\sgn(\omega)\frac{u_\text{IR}}{2}\right)\,.
\ee
The result for \eqref{int K} is then
\bea\label{int K res}
&\frac{\cos^{-h}(\pi\tau_1)\cos^{-h}(\pi\tau_2)}{\sin^{\frac{2}{q}-h}(\pi|\tau_{12}|)}\frac{2i}{\pi^3}\Gamma\left(\frac{2}{q}-h\right)\Gamma\left(\frac{2}{q}+h-1\right)\Gamma\left(1-\frac{2}{q}\right)^2\\
\times &\left\{\Theta(\tau_{12})\left[e^{i\pi h}\cos^2\left(\frac{\pi}{q}+\frac{u_\text{IR}}{2}\right)-e^{-i\pi h}\cos^2\left(\frac{\pi}{q}-\frac{u_\text{IR}}{2}\right)\right]\right.\\
&\left.+\Theta(\tau_{21})\left[e^{\frac{2\pi i}{q}}\cos^2\left(\frac{\pi}{q}+\frac{u_\text{IR}}{2}\right)-e^{-\frac{2\pi i}{q}}\cos^2\left(\frac{\pi}{q}-\frac{u_\text{IR}}{2}\right)\right]\right\}\,.
\eea
Using \eqref{int K res}, one finds that the kernel $K$ acts on $f_\pm$ as the $2\times 2$ matrix $M$ defined as follows
\bea
&\hspace{2.7cm}\int d\tau_3d\tau_4 K(\tau_1,\tau_2;\tau_3,\tau_4)\begin{pmatrix}f_+ \\ f_-\end{pmatrix}(\tau_3,\tau_4)=M\cdot\begin{pmatrix}f_+ \\ f_-\end{pmatrix}(\tau_1,\tau_2)\,,\\
&M\equiv \frac{i}{\pi}\frac{1}{\cos u_\text{IR}+\cos\frac{2\pi}{q}}\frac{\Gamma\left(\frac{2}{q}-h\right)\Gamma\left(\frac{2}{q}+h-1\right)\Gamma\left(2-\frac{2}{q}\right)}{\Gamma\left(\frac{2}{q}+1\right)}\begin{pmatrix} 1 & \frac{2}{q}-1 \\ \frac{2}{q}-1 & 1 \end{pmatrix}\begin{pmatrix} A & B \\ - B^* & -A^*\end{pmatrix}\,,\\
&\hspace{2.5cm}A\equiv e^{i\pi h}\cos^2\left(\frac{\pi}{q}+\frac{u_\text{IR}}{2}\right)-e^{-i\pi h}\cos^2\left(\frac{\pi}{q}-\frac{u_\text{IR}}{2}\right)\,, \\
&\hspace{2.5cm}B\equiv e^{\frac{2\pi i}{q}-iu_\text{IR}}\cos^2\left(\frac{\pi}{q}+\frac{u_\text{IR}}{2}\right)-e^{-\frac{2\pi i}{q}-iu_\text{IR}}\cos^2\left(\frac{\pi}{q}-\frac{u_\text{IR}}{2}\right)\,.
\eea
Its eigenvalues are 
\bea\label{kernel eigenvals}
&k_\pm(h;u_\text{IR}) = \frac{1}{\pi}\frac{\Gamma\left(\frac{2}{q}-h\right)\Gamma\left(\frac{2}{q}+h-1\right)\Gamma\left(2-\frac{2}{q}\right)}{\Gamma\left(\frac{2}{q}+1\right)}\left\{\left(1-\frac{2}{q}\right)\sin\frac{2\pi}{q}-\sin\pi h\,\frac{1+\cos u_\text{IR} \cos\frac{2\pi}{q}}{\cos u_\text{IR} + \cos\frac{2\pi}{q}}\right.\\
&\left.\pm\left[\left(\left(1-\frac{2}{q}\right)\sin\frac{2\pi}{q}-\sin\pi h\,\frac{1+\cos u_\text{IR} \cos\frac{2\pi}{q}}{\cos u_\text{IR} + \cos\frac{2\pi}{q}}\right)^2+\frac{4}{q}\left(1-\frac{1}{q}\right)\left(\sin^2\frac{2\pi}{q}-\sin^2\pi h\right)\right]^\frac{1}{2}\right\}\,.
\eea
Graphically, the dimensions of the operators are given by the positions where the graphs of $k_\pm(h;u_\text{IR})-1$ intersect the horizontal axis. For instance, Figure \ref{fig: k minus 1} shows these graphs for the case $q = 4$ and $u_\text{IR} = \frac{\pi}{4}$, from which we can read off the operator dimensions $h = 1,2,2.718,3.813,4.632,5.717,\ldots$.

As $u_\text{IR}$ varies in the allowed range $(0,\frac{\pi}{2})$ for $q = 4$, the two operators at $h = 1,2$ are always present. The $h = 2$ operator corresponds to the Schwarzian mode, which descends from the approximate IR reparametrisation invariance of the Schwinger-Dyson equations, while the $h = 1$ operator descends from the local gauge invariance of the same equations. The fact that they are present for all $q$ can be seen analytically from \eqref{kernel eigenvals}, due to the identities $k_-(1,u_\text{IR})=1$ and $k_+(2,u_\text{IR})=1$. However, it must be emphasized that because the $h = 1$ operator corresponding to the $\U(1)$ axion $\lambda$ has its origins in gauge transformations $G_0(\tau)\mapsto e^{i\lambda(\tau)-i\lambda(0)}G_0(\tau)$, it must be excluded from the spectrum because such transformations spoil the gauge fixing \eqref{gauge fixing} that we have been using all along. The only IR degree of freedom from $\lambda$ that survives the gauge fixing is the winding number $n$. 

The dimensions of the remaining operators vary with $u_\text{IR}$, as shown in Figure \ref{fig: spec v u}. At $u_\text{IR} = 0$, which corresponds to $u = 0$, they agree with the dimensions found in Appendix B of \cite{Gu:2019jub} for vanishing chemical potential. As $u_\text{IR}\rightarrow \frac{\pi}{2}$, which corresponds to $u\rightarrow \infty$, they gradually increase and tend to integer values, as shown by the dotted lines in Figure \ref{fig: spec v u}. This suggests that the theory becomes free as $u\rightarrow\infty$, and the anomalous dimensions of operators with the form $\psi^\dagger\partial_\tau^n\psi$ are vanishing. Since we have not performed the integral in $u$ yet, we can think of $u$ as a modulus of the theory, in which case the observation above is consistent with the swampland conjecture that there should be an infinite tower of massless states at infinity in moduli space~\cite{Ooguri:2006in}. Notice that similar observations have been made in other disordered models~\cite{Peng:2018zap,Ahn:2018sgn,Chang:2021fmd,Chang:2021wbx}. \footnote{Alternatively, near the boundary of moduli space, some models could experience phase transitions and the model could behave significantly differently~\cite{Peng:2017kro, Heydeman:2022lse,Gao:2024lem,Heydeman:2024ohc}.}

\begin{figure}[h]
\centering
\includegraphics[width=0.5\textwidth]{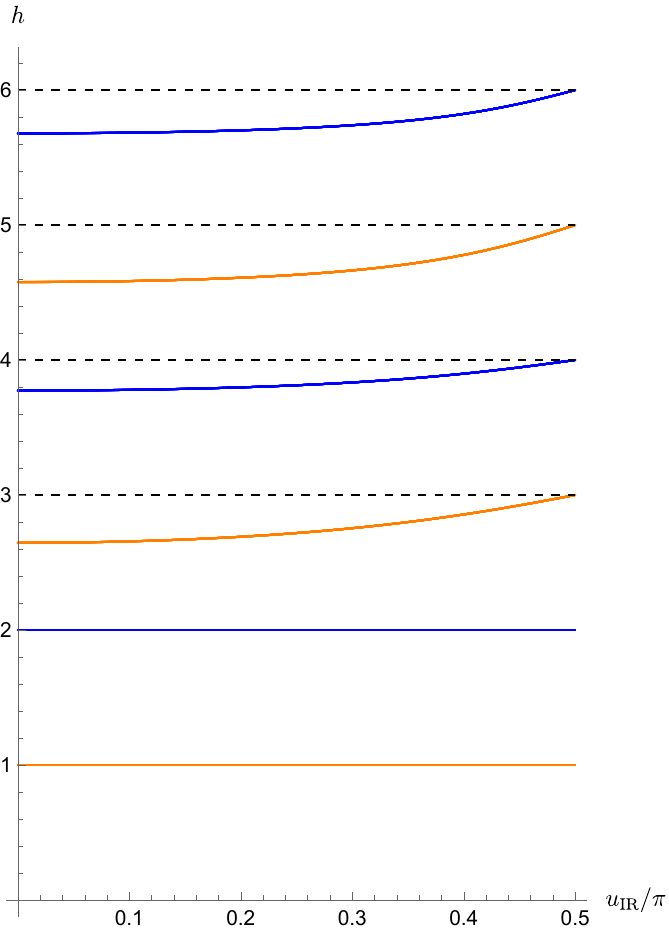}
\caption{The dimensions of several lowest lying operators as functions of $u_\text{IR}/\pi$. A curve is coloured blue or orange depending on whether it plots a solution of $k_+(h,u_\text{IR})=1$ or $k_-(h,u_\text{IR})=1$ respectively.}
\label{fig: spec v u}
\end{figure}

\subsection{The 4-point function}

Similarly to \eqref{full gauge inv 2pt op}, the most general gauge invariant $4$-point insertion which can be expressed in terms of the bilocal fields is\footnote{Naively, there seems to be one other way of obtaining a gauge invariant operator, by connecting $\tau_1$ and $\tau_4$ with a Wilson line, and $\tau_2$ and $\tau_3$ with another. However, this is identical to \eqref{gauge inv 4pt op} since $\exp i\left(\int_{\tau_1}^{\tau_4}+\int_{\tau_3}^{\tau_2}\right)d\tau' A_{\tau'}(\tau')=\exp i\left(\int_{\tau_1}^{\tau_2}+\int_{\tau_2}^{\tau_4}+\int_{\tau_3}^{\tau_4}+\int_{\tau_4}^{\tau_2}\right)d\tau' A_{\tau'}(\tau')=\exp i\left(\int_{\tau_1}^{\tau_2}+\int_{\tau_3}^{\tau_4}\right)d\tau' A_{\tau'}(\tau')$.
} 
\be\label{gauge inv 4pt op}
\frac{1}{N^2}\psi_i^\dagger(\tau_2)\exp \left[i\int_{\tau_1}^{\tau_2} d\tau' A_{\tau'}(\tau')\right]\psi_i(\tau_1)\psi_j^\dagger(\tau_4)\exp \left[i\int_{\tau_3}^{\tau_4} d\tau' A_{\tau'}(\tau')\right]\psi_i(\tau_3)\,.
\ee
After the gauge fixing \eqref{gauge fixing}, we consider the path integral with an insertion of \eqref{gauge inv 4pt op}, normalized by the path integral without insertions. In the Hamiltonian formalism, this computes 
\bea
&\frac{1}{N^2Z_k}\int_{-\pi}^\pi du\,e^{-iu(k+\tau_{12}+\tau_{34})}\Tr_{\cH}\left[e^{-\hat{H}}\hat{\psi}_i^\dagger(\tau_2)\hat{\psi}_i(\tau_1)\hat{\psi}_j^\dagger(\tau_4)\hat{\psi}_j(\tau_3)\right]\\
=\frac{1}{N^2Z_k}\int_{-\pi}^\pi du\,&e^{-iu(k+\tau_{12}+\tau_{34})}\sum_{Q=-\frac{N}{2}}^{\frac{N}{2}}\Tr_{\cH_Q}\left[e^{(\tau_{23}-1)\hat{H}}\hat{\psi}_i^\dagger(0)e^{\tau_{12}\hat{H}}\hat{\psi}_i(0)e^{\tau_{41}\hat{H}}\hat{\psi}_j^\dagger(0)e^{\tau_{34}\hat{H}}\hat{\psi}_j(0)\right]\\
=\frac{1}{N^2Z_k}\sum_{Q=-\frac{N}{2}}^{\frac{N}{2}}&\int_{-\pi}^\pi du\,e^{iu(Q-k)}\Tr_{\cH_Q}\left[e^{(\tau_{23}-1)\hat{H}_0}\hat{\psi}_i^\dagger(0)e^{\tau_{12}\hat{H}_0}\hat{\psi}_i(0)e^{\tau_{41}\hat{H}_0}\hat{\psi}_j^\dagger(0)e^{\tau_{34}\hat{H}_0}\hat{\psi}_j(0)\right]\\
&=\frac{1}{N^2Z_k}\Tr_{\cH_k}\left[e^{(\tau_{23}-1)\hat{H}_0}\hat{\psi}_i^\dagger(0) e^{\tau_{12}\hat{H}_0}\hat{\psi}_i(0)e^{\tau_{41}\hat{H}_0}\hat{\psi}_j^\dagger(0)e^{\tau_{34}\hat{H}_0}\hat{\psi}_j(0)\right]\,.
\eea
We observe that the effect of gauging in the presence of a Wilson line with charge $k$ is simply to restrict the 4-point function of the ungauged theory to the charge $k$ sector, exactly as in \eqref{Gk hamiltonian expr}. Under the large $N$ saddle point approximation, the same path integral is computed as
\bea\label{4 pt u int}
&\hspace{3cm}\frac{1}{Z_k}\int_{-\pi}^\pi du\,e^{-iu(k+\tau_{12}+\tau_{34})}\int\cD[G]\cD[\Sigma]\,G(\tau_{12})G(\tau_{34})\,e^{-S[u,G,\Sigma]}\\
&\approx\frac{1}{Z_k}\int_{-\pi}^\pi du\,e^{-iu(k+\tau_{12}+\tau_{34})}\sum_{n\in\bZ}\int\cD[\delta G]\left(G_0(\tau_{12};u,n)G_0(\tau_{34};u,n)+\frac{1}{N}\delta G(\tau_{12})\delta G(\tau_{34})\right)\\
&\hspace{1cm}\times\exp\left\{-S[u,G_0(u,n),\Sigma_0(u,n)]+\frac{1}{2}\delta G\cdot G_0^{-1}G_0^{-1}(u,n)\cdot(1-K[G_0(u,n)])\cdot\delta G\right\}\\
&=\frac{1}{Z_k}\int_{-\infty}^\infty du\,e^{-iu(k+\tau_{12}+\tau_{34})}\int\cD[\delta G]\left(G_0(\tau_{12};u)G_0(\tau_{34};u)+\frac{1}{N}\delta G(\tau_{12})\delta G(\tau_{34})\right)\\
&\hspace{1cm}\times\exp\left\{-S[u,G_0(u),\Sigma_0(u)]+\frac{1}{2}\delta G\cdot G_0^{-1}G_0^{-1}(u)\cdot(1-K[G_0(u)])\cdot\delta G\right\}\\
&=\frac{1}{Z_k}\int_{-\infty}^\infty du\,\frac{1}{\sqrt{\Delta(G_0(u))}}\,e^{-iu(k+\tau_{12}+\tau_{34})-S[u,G_0(u),\Sigma_0(u)]}\bigg[G_0(\tau_{12};u)G_0(\tau_{34},u)\\
&\hspace{4cm}\left.-\frac{1}{N}(1-K[G_0(u)])^{-1}\cdot G_0G_0(u)(\tau_1,\tau_2;\tau_3,\tau_4)\right]
\eea
In the third line, we have used a shorthand product notation for the integrals in the quadratic action for $\delta G$. Their explicit expression can be found in \eqref{S quad expansion}, and here $\Delta G=\det\left(G_0^{-1}G_0^{-1}(u)\cdot(1-K[G_0(u)])\right)$. To obtain the equality on the fourth line, we have repeated the same steps as in \eqref{Zk u int} and \eqref{Gk int u inf}, which make use of the gauge invariance of the action and operator insertion. For convenience, we package the $\cO(N^{-1})$ contribution appearing in the last line of \eqref{4 pt u int} into
\bea\label{cF def}
\cF(\tau_1,\tau_2;\tau_3,\tau_4) &\equiv \int d\tau_5d\tau_6\,(1-K)^{-1}(\tau_1,\tau_2;\tau_5,\tau_6)\cF_0(\tau_5,\tau_6;\tau_3,\tau_4)\,, \\
\cF_0(\tau_1,\tau_2;\tau_3,\tau_4)&\equiv - G_0(\tau_2,\tau_3)G_0(\tau_4,\tau_1)\,.
\eea
When expanded in powers of $K$, the right hand side of \eqref{cF def} has the interpretation of summing all ladder diagrams contributing to the $4$-point function. Equivalently, $\cF$ satisfies
\be\label{cF SD eqn}
\cF(\tau_1,\tau_2;\tau_3,\tau_4)-\int d\tau_5d\tau_6\, K(\tau_1,\tau_2;\tau_5,\tau_6)\cF(\tau_5,\tau_6;\tau_3,\tau_4) = \cF_0(\tau_1,\tau_2;\tau_3,\tau_4)\,.
\ee

In a sector with nonzero charge, it does not make sense to classify operators in terms of their conformal dimensions, because the conformal symmetry is broken. This is evidenced by the behaviour of the two-point function in Section \ref{subsec: 2pt ED}. Since the result is not expected to have a conformal structure like in \cite{Maldacena:2016hyu}, we will not attempt to compute $\cF$ and perform the integration in $u$ to obtain the four-point function at $\cO(N^{-1})$ in a sector with nonzero charge. In the zero charge sector, the operator dimensions are those at $u = 0$ or zero chemical potential, which have already been computed in \cite{Gu:2019jub}, with the crucial difference that the $h = 1$ operator corresponding to the $\U(1)$ axion is absent due to gauge fixing.

\section{The chaos exponent}\label{sec: chaos exp}

\begin{figure}[htbp]
\centering
\includegraphics[width=0.6\textwidth]{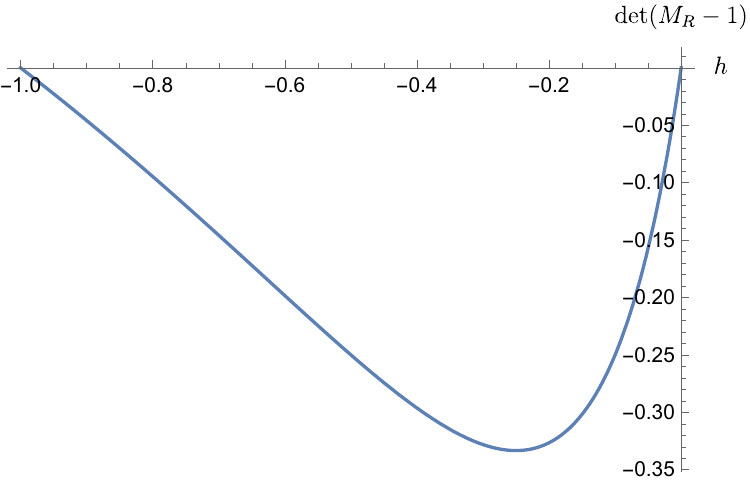}
\caption{Plot of $\det(M_R - 1)$ against $h$ when $q=4$. The plots for $q = 4,6,\ldots$ are qualitatively similar, with exactly one horizontal intercept at $h = -1$ in the allowed range $-1\leq h<0$.}
\label{fig: detkr}
\end{figure}

Following \cite{Murugan:2017eto}, we consider out of time order correlation functions (OTOCs) as a measure of quantum chaos. Specifically, we want to compute the double commutators
\bea\label{double comm def}
W(t_1,t_2)&\equiv-\frac{1}{N^2Z_k}\Tr_{\cH_k}\left[e^{-\hat{H}}\left\{\hat{\psi}_i^\dagger\left(\frac{1}{2}+it_2\right),\hat{\psi}_j^\dagger\left(\frac{1}{2}\right)\right\}\left\{\hat{\psi}_i\left(it_1\right),\hat{\psi}_j\left(0\right)\right\}\right]\,,\\
\wt W(t_1,t_2)&\equiv\frac{1}{N^2Z_k}\Tr_{\cH_k}\left[e^{-\hat{H}}\left\{\hat{\psi}_i\left(\frac{1}{2}+it_2\right),\hat{\psi}_j^\dagger\left(\frac{1}{2}\right)\right\}\left\{\hat{\psi}_i^\dagger\left(it_1\right),\hat{\psi}_j\left(0\right)\right\}\right]\,.
\eea
Using the large $N$ saddle point approximation for the path integral as in \eqref{4 pt u int}, the double commutators can be expressed as 

\begin{equation*}
\begin{aligned}
&W(t_1,t_2) \approx \frac{1}{Z_k}\int_{-\infty}^\infty du\,\frac{1}{\sqrt{\Delta(G_0(u))}}e^{-iu(k + it_{12} - 1)-S[u,G_0(u),\Sigma_0(u)]}\left[-G_0\left(it_1,\frac{1}{2}+it_2\right)G_0\left(0^-,\frac{1}{2}^+\right)\right. \\
&\hspace{3cm}+G_0\left(it_1,\frac{1}{2}+it_2\right)G_0\left(0^+,\frac{1}{2}^+\right) + G_0\left(it_1,\frac{1}{2}+it_2\right)G_0\left(0^-,\frac{1}{2}^-\right)\\
&\hspace{5cm}\left.-G_0\left(it_1,\frac{1}{2}+it_2\right)G_0\left(0^+,\frac{1}{2}^-\right) + \frac{1}{N}\cW(t_1,t_2)\right]\\
&\cW(t_1,t_2)\equiv - \cF\left(it_1,\frac{1}{2}+it_2;0^-,\frac{1}{2}^+\right) + \cF\left(it_1,\frac{1}{2}+it_2;0^+,\frac{1}{2}^+\right) \\
&\hspace{2cm} + \cF\left(it_1,\frac{1}{2}+it_2;0^-,\frac{1}{2}^-\right) - \cF\left(it_1,\frac{1}{2}+it_2;0^+,\frac{1}{2}^-\right)\,,
\end{aligned}    
\end{equation*}
\bea\label{W saddle pt}
&\wt W(t_1,t_2)\approx\frac{1}{Z_k}\int_{-\infty}^\infty du\,\frac{1}{\sqrt{\Delta(G_0(u))}}e^{u(-ik-t_{12})-S[u,G_0(u),\Sigma_0(u)]}\left[-G_0\left(\frac{1}{2}+it_2,it_1\right)G_0\left(0^-,\frac{1}{2}^+\right)\right.\\
&\hspace{3cm}+G_0\left(\frac{1}{2}+it_2,it_1\right)G_0\left(0^+,\frac{1}{2}^+\right)+G_0\left(\frac{1}{2}+it_2,it_1\right)G_0\left(0^-,\frac{1}{2}^-\right)\\
&\hspace{5cm}\left.-G_0\left(\frac{1}{2}+it_2,it_1\right)G_0\left(0^+,\frac{1}{2}^-\right)+\frac{1}{N}\wt\cW(t_1,t_2)\right]\,,\\
&\wt\cW(t_1,t_2)\equiv - \cF\left(\frac{1}{2}+it_2,it_1;0^-,\frac{1}{2}^+\right) + \cF\left(\frac{1}{2}+it_2,it_1;0^+,\frac{1}{2}^+\right) \\
&\hspace{2cm} + \cF\left(\frac{1}{2}+it_2,it_1;0^-,\frac{1}{2}^-\right) - \cF\left(\frac{1}{2}+it_2,it_1;0^+,\frac{1}{2}^-\right)\,.
\eea
Our goal is to examine whether the $\cO(N^{-1})$ contributions $\cW$ and $\wt \cW$ have an exponential growth in $t=t_1=t_2$ at late times $t\gg1$, and if so, determine the exponent of this growth. By considering the analytic continuation of \eqref{cF SD eqn} on the Schwinger-Keldysh contour in \cite{Murugan:2017eto} and substituting the conformal solution \eqref{conf const fixed} into the kernel $K$, one finds that $\cW$ and $\wt \cW$ satisfy 
\bea\label{cW SD eqn}
&\begin{pmatrix}
    \cW \\ \wt\cW
\end{pmatrix}(t_1,t_2) - \int_0^{t_1} dt_3\int_0^{t_2} dt_4 \frac{\pi\left(\frac{q}{2}-1\right)\sin\frac{2\pi}{q}}{\sinh^{\frac{2}{q}}(\pi t_{13})\sinh^{\frac{2}{q}}(\pi t_{24})\cosh^{2-\frac{4}{q}}(\pi t_{34})}\\
&\times\begin{pmatrix}
    e^{u_\text{IR}(-t_{12}+t_{34})} & (\frac{2}{q}-1)e^{-u_\text{IR}(t_{12}+t_{34})} \\ (\frac{2}{q}-1)e^{u_\text{IR}(t_{12}+t_{34})} & e^{u_\text{IR}(t_{12}-t_{34})}
\end{pmatrix}\begin{pmatrix}
    \cW \\ \wt\cW
\end{pmatrix}(t_3,t_4) = \begin{pmatrix}
    \cW_0 \\ \wt\cW_0
\end{pmatrix}(t_1,t_2)\,,\\
&\cW_0 = 0\,,\quad \wt\cW_0 = \left[\frac{\pi\left(1-\frac{2}{q}\right)\sin\frac{2\pi}{q}}{2(\beta J)^2\left(\cos u_\text{IR}+\cos\frac{2\pi}{q}\right)}\right]^\frac{2}{q}\frac{2e^{u_\text{IR}t_{12}}\left(\cos u_\text{IR}+\cos\frac{2\pi}{q}\right)}{\sinh^\frac{2}{q}(\pi t_1)\sinh^\frac{2}{q}(\pi t_2)}\,.
\eea
In the limit of large $t = t_1=t_2$, the right hand side of \eqref{cW SD eqn} is negligible since $\wt \cW_0\sim e^{-\frac{4\pi}{q}t}$ decays exponentially. In addition, assuming that $\cW$ and $\wt\cW$ grow exponentially, the lower limits of integration in \eqref{cW SD eqn} can be changed to $-\infty$ since the error incurred is negligible \cite{Murugan:2017eto}. Therefore, under the assumption of exponential growth and at late times, \eqref{cW SD eqn} simplifies to
\bea\label{cW SD simp}
&\begin{pmatrix}
    \cW \\ \wt\cW
\end{pmatrix}(z_1,z_2) - \int_{-\infty}^{z_1} dz_3\int_{z_2}^{\infty} dz_4 \frac{\left(\frac{q}{2}-1\right)\sin\frac{2\pi}{q}}{\pi|z_{13}|^\frac{2}{q}|z_{24}|^\frac{2}{q}|z_{34}|^{2-\frac{4}{q}}}\\
&\times\begin{pmatrix}
    \frac{|z_1|^{\frac{1}{q}+\frac{u_\text{IR}}{2\pi}}|z_2|^{\frac{1}{q}-\frac{u_\text{IR}}{2\pi}}}{|z_3|^{\frac{1}{q}+\frac{u_\text{IR}}{2\pi}}|z_4|^{\frac{1}{q}-\frac{u_\text{IR}}{2\pi}}} & (\frac{2}{q}-1)\frac{|z_1|^{\frac{1}{q}+\frac{u_\text{IR}}{2\pi}}|z_2|^{\frac{1}{q}-\frac{u_\text{IR}}{2\pi}}}{|z_3|^{\frac{1}{q}-\frac{u_\text{IR}}{2\pi}}|z_4|^{\frac{1}{q}+\frac{u_\text{IR}}{2\pi}}} \\ (\frac{2}{q}-1)\frac{|z_1|^{\frac{1}{q}-\frac{u_\text{IR}}{2\pi}}|z_2|^{\frac{1}{q}+\frac{u_\text{IR}}{2\pi}}}{|z_3|^{\frac{1}{q}+\frac{u_\text{IR}}{2\pi}}|z_4|^{\frac{1}{q}-\frac{u_\text{IR}}{2\pi}}} & \frac{|z_1|^{\frac{1}{q}-\frac{u_\text{IR}}{2\pi}}|z_2|^{\frac{1}{q}+\frac{u_\text{IR}}{2\pi}}}{|z_3|^{\frac{1}{q}-\frac{u_\text{IR}}{2\pi}}|z_4|^{\frac{1}{q}+\frac{u_\text{IR}}{2\pi}}}
\end{pmatrix}\begin{pmatrix}
    \cW \\ \wt\cW
\end{pmatrix}(z_3,z_4) \approx 0\,,
\eea
where we have used the change of variables $z_{1,3}=-e^{-2\pi t_{1,3}}$, $z_{2,4}=e^{-2\pi t_{2,4}}$. Now suppose that $\cW$ and $\wt\cW$ follow the ansatz
\bea\label{cW ansatz}
\cW(z_1,z_2) &= w \frac{|z_1|^{\frac{1}{q}+\frac{u_\text{IR}}{2\pi}}|z_2|^{\frac{1}{q}-\frac{u_\text{IR}}{2\pi}}}{|z_{12}|^{\frac{2}{q}-h}}=w\frac{e^{-2\pi t_1\left(\frac{1}{q}+\frac{u_\text{IR}}{2\pi}\right)}e^{-2\pi t_2\left(\frac{1}{q}-\frac{u_\text{IR}}{2\pi}\right)}}{\left(e^{-2\pi t_1}+e^{-2\pi t_2}\right)^{\frac{2}{q}-h}}\,,\quad w=\text{const.}\\
\wt\cW(z_1,z_2) &= \wt w \frac{|z_1|^{\frac{1}{q}-\frac{u_\text{IR}}{2\pi}}|z_2|^{\frac{1}{q}+\frac{u_\text{IR}}{2\pi}}}{|z_{12}|^{\frac{2}{q}-h}}=\wt w\frac{e^{-2\pi t_1\left(\frac{1}{q}-\frac{u_\text{IR}}{2\pi}\right)}e^{-2\pi t_2\left(\frac{1}{q}+\frac{u_\text{IR}}{2\pi}\right)}}{\left(e^{-2\pi t_1}+e^{-2\pi t_2}\right)^{\frac{2}{q}-h}}\,,\quad \wt w=\text{const.}
\eea
When $t=t_1=t_2$, this ansatz becomes proportional to $e^{-2\pi h t}$. Therefore the chaos exponent is $\lambda=-2\pi h$, where $h$ is determined by solving \eqref{cW SD simp}. Due to the bound $0<\lambda\leq2\pi$ derived in \cite{Maldacena:2015waa}, we must have $-1 \leq h < 0$. Using the identity
\be
\int_{-\infty}^{z_1}dz_3\int_{z_2}^\infty dz_4\frac{1}{|z_{13}|^\frac{2}{q}|z_{24}|^\frac{2}{q}|z_{34}|^{2-\frac{2}{q}-h}}=\frac{\Gamma\left(1-\frac{2}{q}\right)^2\Gamma\left(\frac{2}{q}-h\right)}{\Gamma\left(2-\frac{2}{q}-h\right)}\frac{1}{|z_{12}|^{\frac{2}{q}-h}}\,,
\ee
one finds that on the ansatz \eqref{cW ansatz}, \eqref{cW SD simp} becomes the eigenvalue equation
\be
M_R\cdot\begin{pmatrix}
    w \\ \wt w
\end{pmatrix} = \begin{pmatrix}
    w \\ \wt w
\end{pmatrix}\,,\quad M_R\equiv \frac{\Gamma\left(2-\frac{2}{q}\right)\Gamma\left(\frac{2}{q}-h\right)}{\Gamma\left(1+\frac{2}{q}\right)\Gamma\left(2-\frac{2}{q}-h\right)}\begin{pmatrix} 
1 & \frac{2}{q} -1 \\ \frac{2}{q} -1 & 1     
\end{pmatrix}\,.
\ee
Note that this is independent of $u_\text{IR}$ or the background gauge field. We seek the values of $h$ for which $M_R$ has $1$ as an eigenvalue. Graphically, these values can be found by reading off the horizontal intercepts of the graph of $\det(M_R - 1)$ against $h$. In this way, we find that $h = -1$ is the only intercept in the allowed range $-1 \leq h < 0$, and the chaos exponent is maximal. For instance, Figure \ref{fig: detkr} shows the graph of $\det(M_R - 1)$ at $q = 4$.  

In fact, $\det(M_R - 1)= 0$ is an identity at $h = -1$, indicating the presence of maximal chaos regardless of $q$ and $u_\text{IR}$. If we plug this result for $\cW$ and $\wt \cW$ at $t_1=t_2=t$ back into \eqref{W saddle pt}, it would imply that there is maximal chaotic growth in the double commutators \eqref{double comm def} at any charge $k$. \footnote{One subtlety in this computation is the following. Notice that (7.7) is a linear equation, meaning that the eigenvector $(w, \wt w)^T$ can be multiplied by an arbitrary function of $u$ and it would still be a solution of (7.7). When computing $\cW$ and $\wt\cW$, we have made several approximations which are only valid at late times, such as neglecting $\cW_0$ and $\wt\cW_0$ in (7.3), which have nontrivial $u$ dependence. Perhaps it is only by solving the unapproximated equation that we can fix the $u$-dependent factor multiplying the eigenvector. Perhaps this $u$-dependence will destroy the chaotic behaviour for $\kappa\neq 0$, once the integral in $u$ is performed. But notice that this is nothing but the fact that if we can solve~\eqref{cW SD eqn} exactly without dropping the bare terms, we should be able to see the raising of the OTOC as time increases. We expect that at relatively large $t$, this improved ``exact" treatment should lead to the same Lyapunov  as the current result which assumes large $t$ in the first step of solveing~\eqref{cW SD eqn}.} Note that at $t_1=t_2=t$, the only $t$ dependence at $\cO(N^{-1})$ is inherited from $\cW$ and $\wt \cW$.  On a technical level, this independence of $q$ and $u_{\text{IR}}$ is because in the large $t=t_1=t_2$ limit, the effect of the Wilson line drop off from all exponential factors and hence does not affect the value of the Lyapunov exponent. Intuitively we have the picture that at late time, the mode that causes the chaotic behavior is the scramblon, see e.g.~\cite{Gu:2021xaj},   which is charge neutral and hence can mediate a growing OTOC in any charge sector with the same maximal Lyapunov exponent.

\section*{Acknowledgements}
We thank Zhenbin Yang, Pengfei Zhang for stimulating discussions. CP and ZZ are supported by NSFC NO. 12175237, NSFC NO. 12447108 and NSFC NO. 12247103, the Fundamental Research Funds for the Central Universities, and funds from the Chinese Academy of Sciences.

\appendix

\section{Conformal ansatz}\label{app: conformal ansatz deriv}

We want to constrain the 2-point function $\wt G$ using $\text{PSL}(2,\bR)$ conformal invariance in the presence of the constant gauge field $u_\text{IR}$, and thereby also constrain solutions $G_0$ to the Schwinger-Dyson equations. This discussion is an adaptation of \cite{Davison:2016ngz,Benini:2023xxx}. From an ``active" point of view, a conformal transformation acts as
\bea\label{conf transf}
&\tau\mapsto\tau'\,,\quad \tan\left(\pi\tau'\right)=\frac{a \tan\left(\pi\tau\right)+b}{c \tan\left(\pi\tau\right)+d}\,,\quad a,b,c,d\in\bR\,,\quad ad-bc=1\,,\\
&\psi_i(\tau)\mapsto\psi_i'(\tau')=\left(\frac{d\tau'}{d\tau}\right)^{-\Delta}\psi_i(\tau)\,,\quad A_\tau(\tau)\mapsto A_{\tau'}(\tau')=\frac{d\tau}{d\tau'}A_\tau (\tau)\,.
\eea
This coordinate transformation is the pullback to the interval $(-1/2,1/2)$ of the fractional linear action of $\text{PSL}(2,\bR)$ on $\bR\cup\{\infty\}$ via the map $\tau\mapsto\tan(\pi\tau)$. Here $\bR\cup\{\infty\}$ denotes the one point compactification of $\bR$ where $\pm\infty$ are identified, corresponding to the identification of $\pm 1/2$ in the interval. Since $\tan^{-1}$ is ambiguous mod $\pi$, the branch of $\tau'$ should be chosen such that the function $\tau'(\tau)$ is continuous. As a diffeomorphism, $\tau'(\tau)$ covers $S^1$ exactly once (winding number 1) and $\tau'\mapsto\tau'+\beta$ as $\tau\mapsto\tau+\beta$. The important point to note here is that the gauge field after the conformal transformation is not what we started with: $u_\text{IR}$ is mapped to $u_\text{IR}\frac{d\tau}{d\tau'}$. To bring the gauge field back to its original value, we need to perform a compensating $\U(1)$ gauge transformation with parameter
\be
\alpha(\tau')=u_\text{IR}(\tau'-\tau(\tau'))\,.
\ee
This transformation is single-valued on $S^1$ since $\alpha(\tau'+1)=\alpha(\tau')$, from the fact that conformal transformations have winding number $1$. Putting the conformal and compensating gauge transformations together, the fields transform as
\be
\psi_i(\tau)\mapsto\psi_i'(\tau')=e^{iu_\text{IR}(\tau'-\tau)}\left(\frac{d\tau'}{d\tau}\right)^{-\Delta}\psi_i(\tau)\,.
\ee
The corresponding Ward identity is
\be\label{conf WI w gauge}
\wt G(\tau'(\tau_1),\tau'(\tau_2))=e^{iu_\text{IR}(\tau'(\tau_1)-\tau'(\tau_2)-\tau_1+\tau_2)}\left[\frac{d\tau'}{d\tau}(\tau_1)\frac{d\tau'}{d\tau}(\tau_2)\right]^{-\Delta}\wt G(\tau_1,\tau_2)\,.
\ee
For all $\text{PSL}(2,\bR)$ transformations in \eqref{conf transf}, one can check that
\bea
\frac{d\tau'}{d\tau}(\tau_1)\frac{d\tau'}{d\tau}(\tau_2)&=\frac{\sec^2(\pi\tau_1)\sec^2(\pi\tau_2)}{\Big[\big(a\tan(\pi\tau_1)+b\big)^2+\big(c\tan(\pi\tau_1)+d\big)^2\Big]\Big[\big(a\tan(\pi\tau_2)+b\big)^2+\big(c\tan(\pi\tau_2)+d\big)^2\Big]}\\
&=\frac{\sin^2(\pi\tau'_{12})}{\sin^2(\pi\tau_{12}\big)}\,,\quad \tau'_{12}\equiv\tau'(\tau_1)-\tau'(\tau_2)\,. 
\eea
The Ward identity \eqref{conf WI w gauge} can then be written as
\be
\wt G(\tau'_{12})=\frac{e^{iu_\text{IR}\tau'_{12}}\sin^{-2\Delta}(\pi\tau'_{12})}{e^{iu_\text{IR}\tau_{12}}\sin^{-2\Delta}(\pi\tau_{12})}\,\wt G(\tau_{12})\,.
\ee
In the large $N$ saddle point expansion \eqref{SD sum Z G}, the $\tau$ dependence of $\wt G$ must be carried by the solutions $G_0$ to the Schwinger-Dyson equations, which must therefore satisfy the same conformal Ward identities. It is clear that the identities are solved by the ansatz
\be\label{ansatz imposing conf}
G_0(\tau)=e^{iu_\text{IR}\tau}\bigg|\frac{\pi}{\sin(\pi\tau)}\bigg|^{2\Delta}\left[A\Theta(\tau)+B\Theta(-\tau)\right]\,,\quad A,B\in\bC\,,
\ee
if $\sgn(\tau'_{12})=\sgn(\tau_{12})$. In order to be consistent with the KMS condition \eqref{KMS condition sd}, it is essential to allow for a discontinuity at $\tau=0$ where the operators coincide. Also, this ansatz is only valid for $\tau\in(-1,1)$. For other values of $\tau$, additional pieces proportional to $\Theta(\pm\tau-n)$, where $n\in\bZ_{\geq 1}$, should be added to be consistent with the KMS condition.

Now, imposing the KMS condition on \eqref{ansatz imposing conf} gives
\be
Ae^{iu_\text{IR}}+B=0\,,
\ee
which is solved by
\be
A=-ge^{-\frac{iu_\text{IR}}{2}}\,,\quad B=ge^{\frac{iu_\text{IR}}{2}}\,,\quad g\in\bC\,.
\ee
 
Applying the reality condition \eqref{G cc relation} to the conformal ansatz implies that the coefficient $g$ is real. Thus far, the ansatz is constrained to be
\be\label{conf ansatz g real}
G_0(\tau)=-ge^{iu_\text{IR}\tau}\bigg|\frac{\pi}{\sin(\pi\tau)}\bigg|^{2\Delta}\left[e^{-\frac{iu_\text{IR}}{2}}\Theta(\tau)-e^\frac{iu_\text{IR}}{2}\Theta(-\tau)\right]\,,\quad g(u)\,. 
\ee
A final constraint can be put on the sign of $g$ using \eqref{reG bdy}. The boundary value of $G_0$ in the conformal ansatz is
\be
\lim_{\epsilon\rightarrow 0^+}\re G_0(\epsilon)=-\lim_{\epsilon\rightarrow 0^+}\left(\frac{g\cos\frac{u_\text{IR}}{2}}{\epsilon^{2\Delta}}+\cO(\epsilon^{1-2\Delta})\right)\,.
\ee
In the actual solution, the divergent term proportional to $\epsilon^{2\Delta}$ will be regularized to something close to the finite value of $-\frac{1}{2}$. For this regularization to be smooth, the coefficient of the divergent term must be negative. This implies   
\be\label{sgn g req}
\sgn(g)=\sgn\left(\cos\frac{u_\text{IR}}{2}\right)=(-1)^{\lfloor\frac{u_\text{IR}+\pi}{2\pi}\rfloor}\,.
\ee

\bibliographystyle{ytphys}
\bibliography{BHEntropy}
\end{document}